 \documentclass[twocolumn,trackchanges]{aastex631}
 
\usepackage{amsmath}

\def\FeII{Fe\,{\sc ii}}

\def \OIII {[O\,{\sc iii}]}

\begin{document}

\title{Revisiting the dust torus size -- luminosity relation based on a uniform reverberation mapping analysis}

\author[0000-0001-9957-6349]{Amit Kumar Mandal}
\affiliation{Department of Physics $\&$ Astronomy, Seoul National University\\
Seoul 08826, Republic of Korea, jhwoo@snu.ac.kr}

\author[0000-0002-8055-5465]{Jong-Hak Woo}
\affiliation{Department of Physics $\&$ Astronomy, Seoul National University\\
Seoul 08826, Republic of Korea, jhwoo@snu.ac.kr}

\author[0000-0002-2052-6400]{Shu Wang}
\affiliation{Department of Physics $\&$ Astronomy, Seoul National University\\
Seoul 08826, Republic of Korea, jhwoo@snu.ac.kr}

\author[0000-0002-8377-9667]{Suvendu Rakshit}
\affiliation{Aryabhatta Research Institute of Observational Sciences\\
Manora Peak, Nainital-263001, Uttarakhand, India}

\author[0000-0003-2010-8521]{Hojin Cho}
\affiliation{Department of Physics $\&$ Astronomy, Seoul National University\\
Seoul 08826, Republic of Korea, jhwoo@snu.ac.kr}

\author[0000-0002-4704-3230]{Donghoon Son}
\affiliation{Department of Physics $\&$ Astronomy, Seoul National University\\
Seoul 08826, Republic of Korea, jhwoo@snu.ac.kr}

\author[0000-0002-4998-1861]{C. S. Stalin}
\affiliation{Indian Institute of Astrophysics\\
Block II, Koramangala, Bangalore, 560 034, India}

%\author{August Muench}
%\affiliation{American Astronomical Society \\
%1667 K Street NW, Suite 800 \\

%\collaboration{20}{(AAS Journals Data Editors)}

%\author{F.X Timmes}
%\affiliation{Arizona State University}
%\affiliation{AAS Journals Associate Editor-in-Chief}

%\altaffiliation{AASTeX v6+ programmer}

%\affiliation{AAS Director of Publishing}
%\affiliation{American Astronomical Society \\
%1667 K Street NW, Suite 800 \\




\begin{abstract}

We investigate the torus size -- luminosity relation of Type 1 AGNs based on the reverberation-mapping analysis using the light curves of the optical continuum and the IR continuum obtained with the W1 and W2- bands of the Wide-field Infrared Survey Explorer (WISE) survey. The final sample consists of 446 and 416 AGNs, respectively, for W1 and W2- band light curves, covering a large dynamic range of bolometric luminosity from $10^{43.4}$ to $10^{47.6}$ $erg \, s^{-1}$, which show reliable lag measurements based on our quality assessment analysis. After correcting for the accretion disk contamination in the observed IR flux, we constrain the torus size ($R_{dust}$) and AGN bolometric luminosity ($L_{bol}$) relationship with the best-fit slope of 0.39 (0.33) for the W1- (W2) band, which is shallower than expected from the dust radiation equilibrium model. By combining the previous K- band lag measurements, we find that the measured torus size depends on the observed wavelength of the dust radiation, as  $R_{dust,K}:R_{dust,W1}:R_{dust,W2}$ = 1.0:1.5:1.8 ($R_{dust} \, \propto \, \lambda^{0.80}$) at $L_{bol}$ = $10^{46} \, erg \, s^{-1}$, confirming a stratified structure of the torus, where wavelength-dependent emissions originate from distinct regions of the torus. By investigating the deviation from the best-fit torus size -- luminosity relation, we find a moderate correlation between the offset from the $R_{dust}$--$L_{bol}$ relation and Eddington ratio. This suggests a possible influence of the Eddington ratio on the observed flattening of the $R_{dust}$--$L_{bol}$ relationship.

\end{abstract}



\section{Introduction} \label{sec:intro}

Active galactic nuclei (AGNs) are amongst the most luminous objects in the Universe, powered by the accretion of matter onto a supermassive black hole (SMBH) and emit radiation across the entire electromagnetic spectrum \citep{1984ARA&A..22..471R}. The unification model of AGN posits the presence of a dust torus surrounding the broad emission line region (BLR). Based on different obscuration by the torus, AGNs are classified into two main categories, namely Type 1 and Type 2, characterized by the presence or absence of broad emission lines in their optical/infrared spectrum, respectively \citep{1943ApJ....97...28S, 1995PASP..107..803U}. 

In AGNs with UV luminosities ranging from $10^{42}$ to $10^{44}$ $erg \, s^{-1}$, the BLR typically resides at a distance of  $\sim$ 0.003 to 0.02 pc from the accretion disk (AD), while the surrounding dust torus can extend from 0.01 to 0.1 pc. Consequently, the central regions of AGNs are compact and pose challenges for direct imaging techniques. Nevertheless, the advancements in near-infrared (NIR) and MIR interferometry have enabled the measurement of torus sizes in more than 40 sources, primarily focusing on nearby and bright AGNs \citep{2009A&A...507L..57K, 2011A&A...527A.121K, 2011A&A...536A..78K, 2013A&A...558A.149B, 2020A&A...635A..92G}.

An alternative way to resolve the central part of AGNs is the technique of reverberation mapping \citep[RM;][]{1982ApJ...255..419B, 1993PASP..105..247P}. The broad-band spectral energy distribution (SED) of an AGN presents an excess emission in the IR, which is attributed to the thermal emission from the dusty torus \citep{1964ApJ...140..796S, 1989ApJ...347...29S, 1994ApJS...95....1E}. By measuring the delayed response ($\tau$) of the reprocessed IR continuum relative to the ionizing UV/optical continuum from the AD, the torus size can be obtained as $R_{torus} \, < \, c \times \tau$, where c is the speed of light.

 To measure the inner extent of the torus, the dust RM (DRM) between the optical and the NIR continuum in the K- band has been applied to $\sim$ 40 AGNs over a relatively narrow range of luminosity, i.e., V- band luminosity $L_{V} \, < \, 10^{45} \, erg \, s^{-1}$
 %using ground based telescopes
 \citep{2006ApJ...639...46S, 2014ApJ...788..159K, 2018MNRAS.475.5330M, 2019ApJ...886..150M, 2021MNRAS.501.3905M}. \citet{2014ApJ...788..159K} found that the dust torus radius ($R_{dust}$) correlated with the optical luminosity ($L$) as $R_{dust} \, \propto \, L^{\alpha}$ with the exponent $\alpha$ = 0.5, which is consistent with the expectation when the dust particles are in radiative equilibrium with the dust sublimation temperature \citep[$\mathrm{T_{sub}=1500K}$;][]{1987ApJ...320..537B, 2007A&A...476..713K}. However, \citet{2019ApJ...886..150M} reported a somewhat shallower slope with $\alpha = 0.424$, proposing that the anisotropic illumination of the AD leads to a smaller dust-sublimation radius in the equatorial plane compared to the polar directions. They also suggested that the dust torus does not respond instantaneously to the changes in the luminosity of AD, leading to larger torus sizes being more frequently observed in less luminous AGNs, which show relatively small timescales of flux variation as the AD of low-luminosity AGNs has  dimmed \citep{2009ApJ...700L.109K}. Consequently, the average inner radius of the dust torus tends to be larger for less luminous AGNs. In contrast, the torus radius from the K-band interferometric measurements is roughy a factor of two larger than the dust reverberation radius measured in the same band. This difference could be attributed to the distinction between the flux-weighted and response-weighted radii of the innermost dust torus obtained from interferometry and DRM, respectively \citep{2011A&A...536A..78K, 2014ApJ...788..159K}. These results indicate that the torus size and luminosity relation is complex, requiring more detailed investigations.

\begin{table*}
\centering
\caption{Details of the surveys used in this work}
\label{tab:survey}
%\resizebox{10cm}{!}{
%\small
%\setlength{\tabcolsep}{1pt}

\begin{tabular}{llccclr} \hline
Name & Filter & Time Span & Cadence & $\mathrm{N_{epoch}}$ & Coverage & mag \\
 &  &  & days ($\#$/season) &  & $\mathrm{deg^2}$ &  \\
(1) & (2) & (3) & (4) & (5) & (6) & (7) 
\\ \hline

  &  &   & Sample 1 &  &   & \\ 

CRTS  & unfiltered & 2005--2013  & $\sim$ 20  & 15--120  & 33,000 $\mathrm{deg^2}$ & 19--21\\  
PTF  & $g^{\prime}$, R & 2009--2014  & $\sim$ 5--15  & 5--100  & 11,233 $\mathrm{deg^2}$ & 21.3($g^{\prime}$), 20.6 (R)\\ 
ASAS--SN  & V & 2012--2019  & $\sim$ 3--4  & 150--350  & All sky & 17\\
ZTF  & ZTF-$g$, ZTF-$r$ & 2018--2021  & $\sim$ 3--4  & 20--350  & 25,000--30,000 $\mathrm{deg^2}$ & 20.8($g$), 20.6($r$)\\
WISE/NEOWISE  & W1, W2 & 2010--2020  & $\sim$ 180 & 15--17 & All sky  & 16.6 (W1), 16.0 (W2)\\ \hline

  &  &   & Sample 2 &  &   & \\ 

SDSS(S82)  & $g$ & 1998--2007  & $\sim$ 5  & $\sim$ 60  & 300 $\mathrm{deg^2}$ & 22.2\\
CRTS  & unfiltered & 2005--2013  &  $\sim$ 21  & $\sim$ 40  & 33,000 $\mathrm{deg^2}$ & 19--21\\
PTF  & $g^{\prime}$ & 2009--2014  & $\sim$ 5  & $\sim$ 4  & 11,233 $\mathrm{deg^2}$ & 21.3\\ 
PS1  & $g (r, i)$ & 2011--2014  & 2/season & $\sim$ 10  & 3$\pi$ & 22.0\\
ASAS--SN  & V & 2012--2019  & $\sim$ 3--4  & $\sim$ 200  & All sky & 17\\
DES  & $g (r, i)$ & 2013--2018  & 1--4/season & $\sim$ 10  & 5100 $\mathrm{deg^2}$ & 23.6\\
ZTF  & ZTF-$g$ & 2018  & $\sim$ 3--4  & $\sim$ 30  & 25,000--30,000 $\mathrm{deg^2}$ & 20.8 \\
WISE/NEOWISE  & W1, W2 & 2010--2020  & $\sim$ 180 & 12--16 & All sky  & 16.6 (W1), 16.0 (W2)\\

\hline
\end{tabular}
%}
%\vspace{2.9cm}
\raggedright Note:  Columns are (1) name of the survey, (2) name of the filter, (3) observing time period of the light curves, (4) cadence of the light curves, (5) number of data points present in the light curves, (6) sky coverage of the corresponding survey, and (7) limiting magnitude in mag unit.

\end{table*}

\begin{table}
\centering
\caption{Details of the initially selected samples with optical and IR (W1, W2- band) light curve data}
\label{tab:sample}
%\resizebox{10cm}{!}{
%\small
\setlength{\tabcolsep}{2pt}

\begin{tabular}{lcccr} \hline
Sample & $\mathrm{N_{targets}}$  & redshift  & $L_{bol}$  & WISE-band  \\
 &  & $z$ & $erg \, s^{-1}$ & \\

 (1) & (2) & (3) & (4) & (5)

\\ \hline

Sample 1 & 142 & 0.02 -- 0.34 & $10^{43.4}$ -- $10^{46.5}$ & W1, W2 \\

Sample 2 & 587 & 0.3 -- 2.0 & $10^{44.5}$ -- $10^{47.6}$ & W1, W2 \\

\hline

\end{tabular}
%}

\raggedright Note:  Columns are (1) name of the sample, (2) initial number of AGNs present in the sample, (3) redshift range of the sample, (4) AGN bolometric luminosity range of the sample, and (5) name of the WISE bands used.

\end{table}

By expanding the K-band studies, \citet{2019ApJ...886...33L} and \citet{2020ApJ...900...58Y} performed RM studies by utilizing various optical surveys and IR imaging data from the all-sky Wide-field Infrared Survey Explorer (WISE) \citep{2010AJ....140.1868W, 2014ApJ...792...30M}. They focused on high-luminosity AGNs with bolometric luminosities ($L_{bol}$) exceeding $10^{45} \, erg \, s^{-1}$, effectively excluding the low-luminosity counterparts. Using the WISE W1 and W2-band lag analysis, \citet{2019ApJ...886...33L} derived reliable torus sizes  for 67 among 87 Palomar–Green (PG) quasars at redshifts $z \leq 0.5$, reporting that the torus size correlates with AGN bolometric luminosity with a slope of 0.47 (0.45) for W1- (W2) band lags. Using a large sample of 587 AGNs within the SDSS stripe 82 region, \citet{2020ApJ...900...58Y} performed  a DRM study, although they did not report an independent torus size--luminosity relationship. More recently, \citet{2023MNRAS.tmp.1073C} conducted DRM analysis for 78 AGNs, for which $H\beta$ lag measurements are available, reporting a much shallower slope, $\alpha = 0.37$ between the torus size and bolometric luminosity. They also uncovered a correlation between the deviations from the best-fit torus size--luminosity relation and the accretion rate \citep{2014ApJ...782...45D}, concluding AGNs with higher mass accretion tend to have a more shortened torus size because of the self-shadowing effect of the slim disk \citep{2011ApJ...737..105K, 2014ApJ...797...65W}.

While the thermal radiation from the dust torus dominates the IR flux of Type 1 AGNs, it includes emission from the AD. Consequently, the presence of AD  contamination in the observed IR light curves leads to derived dust lags appearing shorter than the actual lag \citep{2014ApJ...788..159K, 2021MNRAS.501.3905M}, because the continuum emission from the AD in the IR is nearly synchronous with the optical emission compared to the lag of the torus IR emission \citep{2006ApJ...652L..13T, 2008Natur.454..492K, 2011MNRAS.415.1290L}. Therefore, it is necessary to correct for AD contamination in the observed IR emission in order to reliably measure the dust lag. Notably, while \citet{2019ApJ...886...33L} performed AD contamination correction in the observed IR fluxes, \citet{2023MNRAS.tmp.1073C} and \citet{2020ApJ...900...58Y} did not consider this correction, potentially leading to an underestimation of their obtained torus sizes.

In a different approach, \citet{2023arXiv230212437L} employed an ensemble of structure functions (SFs) based on optical and IR light curves in W1- band of 587 AGNs in the SDSS Stripe 82 region. They determined the torus size--luminosity relation with an exponent of $\alpha = 0.51$ by binning the luminosity range into nine sub-samples and calculating the average effective torus size from the ensemble SF analysis for each sub-sample. Unlike the RM technique, their method relies on fitting parameters, such as the inclination angle ($\theta_{inc}$), inner radius ($R_{in}$), outer-to-inner radius ratio ($Y$), torus half-opening angle ($\theta_{opn}$), and radial power-law index ($p$).

The previous studies of DRM showed a notable diversity in analysis method and the slope of the correlation between torus size and luminosity varies among these studies \citep[See][]{2019ApJ...886...33L, 2020ApJ...900...58Y, 2023arXiv230212437L, 2023MNRAS.tmp.1073C}. Thus, it is necessary to further investigate the torus size -- luminosity relation using a uniform and consistent analysis method over a large dynamic range of  AGN luminosity. Such investigations would also contribute to establishing AGNs as standard candles for constraining cosmological parameters \citep{2014ApJ...784L..11Y, 2017MNRAS.464.1693H}.

In this paper, we revisit the torus size--luminosity ($R_{dust}$--$L_{AGN}$) relation with a uniform analysis of the torus lag measurements, using of a large sample of AGNs. First, our sample covers a large dynamic range in luminosities ($10^{43.4}$ to $10^{47.6}$ $erg \, s^{-1}$) with a redshift ranging from 0.02 to 2.
Second, we employ a uniform method for cross correlation analysis and perform quality assessment, ensuring consistent and reliable lag measurements. Third, we investigate the effect of the AD contamination on the observed IR fluxes in W1 and W2- bands, and correct for the contamination in order to determine the $R_{dust}$--$L_{AGN}$ relationship.
The presentation of the paper is as follows. In Section \ref{sec:sample}, we describe the  sample and data acquired from different surveys and literature. We discuss the construction of the optical and IR light curves in Section \ref{lc_con}. In Section \ref{analysis}, we describe the time series analysis. The results are presented in Section \ref{result}. We discuss the wavelength dependent lag and scatter in our $R_{dust}$--$L_{AGN}$ relation in Section \ref{discussion} and a summary in Section \ref{summary}.

.

\begin{figure*}
\centering
\includegraphics[scale=0.68]{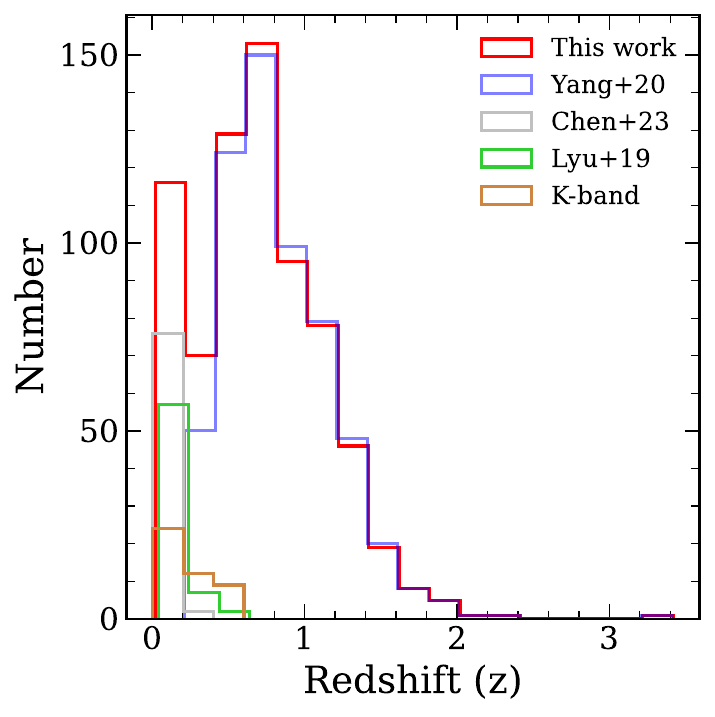}
\includegraphics[scale=0.68]{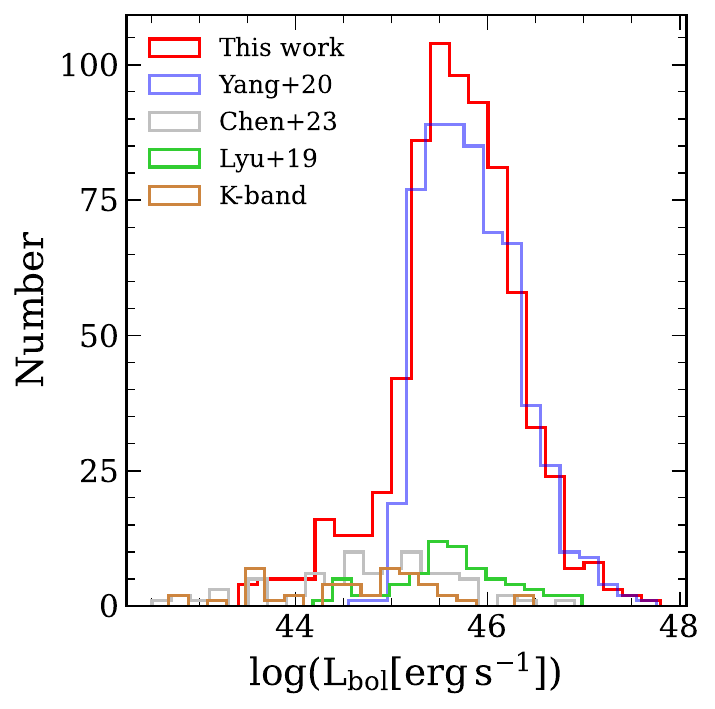}
%\resizebox{6cm}{5cm}{\includegraphics{grt_curve.pdf}}
%\resizebox{9cm}{8cm}{\includegraphics{fvar1.pdf}}
\caption{Distribution of redshift (left) and bolometric luminosity (right) for different samples of AGNs with IR lag measurements.}
\label{fig:dist}
\end{figure*}

\section{Sample and Data} \label{sec:sample}

In order to investigate the relationship between the torus size and AGN luminosity over a wide range of luminosities, we conducted a  selection process of Type 1 AGNs with available optical and IR light curve data. We selected targets from two distinct categories based on their redshifts and luminosities.

\subsection{Sample 1 $(z < 0.35)$}\label{sec:sample1}

To explore the low luminosity end of the $R_{dust}$--$L_{AGN}$ relationship, we selected targets from the catalog of \citet{2019ApJS..243...21L}. This catalog comprises a uniformly defined sample of 14,584  Type 1 AGNs compiled from SDSS DR7 at $z$ $<$ 0.35 and covering a range of AGN bolometric luminosities between $10^{41.5}$ and $10^{46.6}$ $erg \, s^{-1}$. Since the AGNs in this catalog are selected based on the robustness of the broad components of the Balmer lines instead of imposing a lower limit on the FWHM of emission line, the catalog contains many low BH mass ($M_{BH}$) AGNs with lower luminosity than other existing Type 1 AGN catalogs. Thus, we can explore the low luminosity end of the $R_{dust}$--$L_{AGN}$ relation using this catalog. The authors carefully removed the stellar continuum by fitting their SDSS spectra. We used other key parameters (i.e., optical luminosity at 5100 {\AA}, $H\beta$, {\OIII} and {\FeII} luminosities, FWHM of $H\beta$, Eddington ratio etc.) from the catalog.

\subsection{Sample 2 $(z > 0.3)$}\label{sec:sample2}

To cover a moderate to high luminosity range in the $R_{dust}$--$L_{AGN}$ relationship, we used the SDSS Stripe 82 (S82) region \citep{2012ApJ...753..106M}, where a total of approximately 9258 spectroscopically confirmed broad-line AGNs are available. These AGNs span a wide range of redshifts and luminosities, useful for investigating luminous AGN population in the $R_{dust}$--$L_{AGN}$ relation. \citet{2020ApJ...900...58Y} performed a DRM study of S82 AGNs using $\sim$ 20 yr ground-based optical light curves and 10 yr IR light curves in the W1- band from the WISE. They constructed optical light curves using data from the SDSS, Pan-STARRS \citep[PS1;][]{2016arXiv161205560C}, Dark Energy Survey \citep[DES;][]{2018ApJS..239...18A}, Catalina Real-Time Transient Survey \citep[CRTS;][]{2009ApJ...696..870D}, Palomar Transient Factory  \citep[PTF;][]{2009PASP..121.1395L}, Zwicky Transient Facility \citep[ZTF;][]{2019PASP..131a8002B} and All-Sky Automated Survey for Supernovae \citep[ASAS-SN;][]{2014AAS...22323603S}. They were able to find lag between the optical and W1- bands for 587 AGNs with redshift in the range 0.3 $<$ $z$ $<$ 2, and luminosity in the range $10^{44.5}$ to $10^{47.6}$ $erg \, s^{-1}$. They  employed {\tt JAVELIN}  \citep{2011ApJ...735...80Z} to search for the lags between the optical and W1- band light curves. Thus, Sample 2 comprises a total of 587 AGNs from S82 region. We used the same object ID number as assigned by \citet{2020ApJ...900...58Y} for Sample 2.

\subsection{Light curves data}\label{sec:lc_data}

We gathered optical and IR data from both ground-based and space-based telescopes, as well as from the relevant literature for our target AGNs.

\subsubsection{Optical data from ground based surveys}
For Sample 1, we compiled the available optical photometric data from different ground-based transient surveys. Initially, we conducted a search for data from the ASAS-SN survey, as it overlaps with the major part of the IR light curves from WISE and NEOWISE. Since 2012, ASAS-SN monitored nearly the entire visible sky to a V- band depth of 17 mag with a cadence of 2--3 days using twenty telescopes each of 14 cm diameter. Therefore, we selected AGNs with $g<$ 17.5 mag from our sample of 14,584 Type 1 AGNs, which resulted in 732 targets. We searched light curve data from the publicly available  ASAS-SN light curve servers, i.e., ASAS-SN Photometry Database \footnote{\url{https://asas-sn.osu.edu/photometry}} \citep{2014AAS...22323603S, 2019MNRAS.485..961J}. Among the 732 targets, we found ASAS-SN light curve data for 142 AGNs. Thus in Sample 1, we arrived at an initial number of 142 sources.

We also collected V- band photometric light curve data for the 142 targets from the CRTS. This survey uses the 0.7 m Catalina Schmidt Telescope in Tucson, Arizona, the 1.5 m telescope on Mt. Lemmon, Arizona and the 0.5 m Siding Springs Survey (SSS) telescope in Australia to search for optical transients with timescales of minutes to yr in the northern sky. CRTS uses {\tt SExtractor} \citep{1996A&AS..117..393B} to process the images following standard techniques. The limiting V- band magnitude for the 1.5 m and 0.7 m are 21.5 mag and 19.5 mag, respectively. The photometry for sources brighter than 13 V-mag is problematic because of saturation and non-linearity response of the CCD.

In addition, we obtained photometric light curve data from the PTF Data Release 3 (DR3) when available. This survey used the 48 inch Samuel Oschin Telescope at the Palomar Observatory in northern San Diego County, California. The photometry was performed in two broadband filters, SDSS $g$$^\prime$ and Mould $R$.  The survey provides a depth of $m_g$ $\sim$ 21.3 mag and $m_R$ $\sim$ 20.6 mag with an exposure of 60 s. The PTF data were used to fill the gap between the CRTS and ASAS-SN light curves.

Finally, we gathered light curve data from the ZTF Data Release 14 (DR14). ZTF is a wide-field time-domain survey, that uses the 1.2 m Schmidt telescope at the Palomar Observatory with 47 deg$^2$ field of view. It provides photometric light curve data with a cadence of 3--4 days in the $g r i$ bands since 2018 March (MJD $\sim$ 58194). It scans the entire northern visible sky to median depths of $g$ $\sim$ 20.8 mag and $r$ $\sim$ 20.6 mag (AB, 5$\sigma$) with a 30 s exposure \citep{2019PASP..131a8003M}. ZTF uses point-source-function-fit photometry to build the light curves. We selected ZTF data with the condition {\bf catflags=0} to avoid the data contaminated by clouds, moonlight or bad seeing. 

In summary, we collected optical photometric light curve data for 142 Type 1 AGNs in Sample 1 with a baseline from 2005 to 2021. This sample consists of targets with bolometric luminosity within the range of $\mathrm{10^{43.4}}$ to $\mathrm{10^{46.5}}$ $\mathrm{erg \, s^{-1}}$ at $z < 0.3$. 

We utilized the publicly available light curve data in the optical band from \citet{2020ApJ...900...58Y} for 587 Type 1 AGNs in Sample 2 for our own analysis.

\begin{figure*}
\includegraphics[scale=0.5]{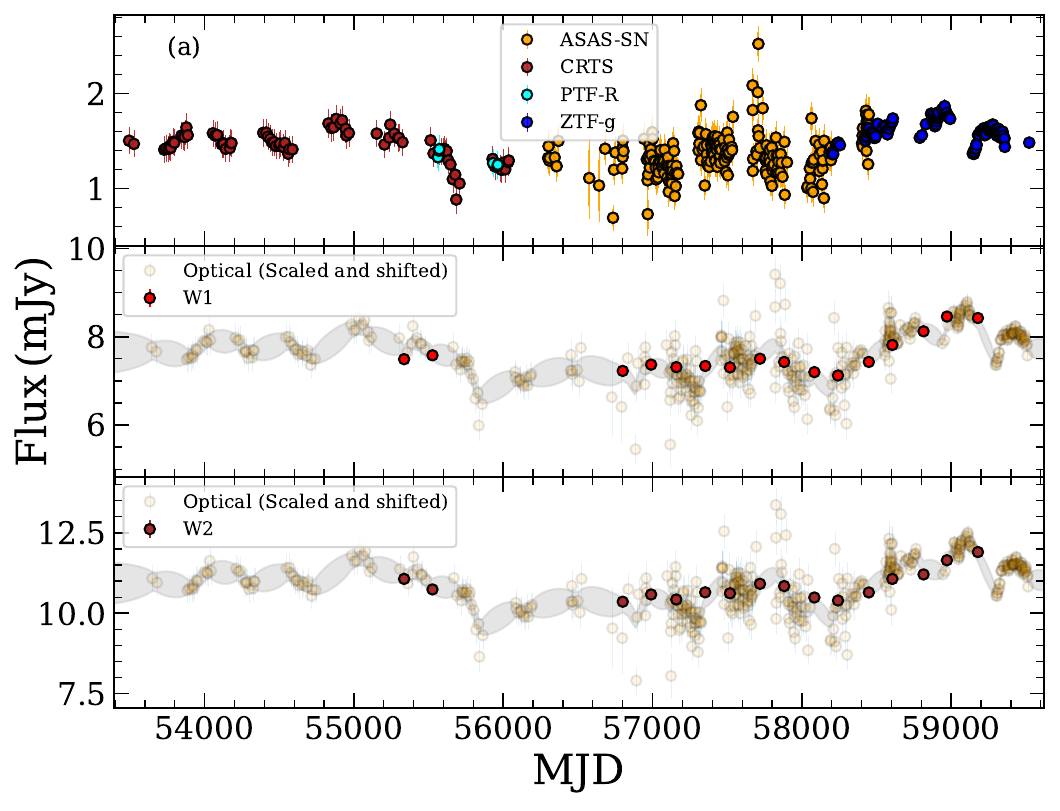}
\includegraphics[scale=0.5]{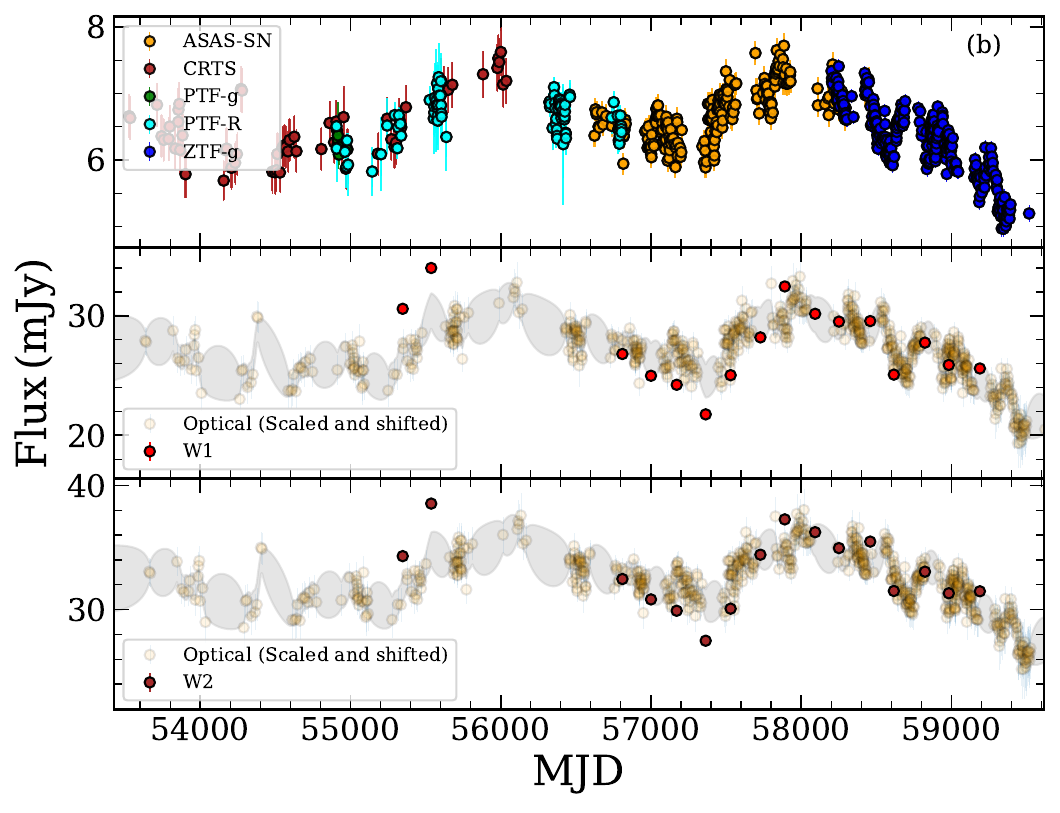}
\includegraphics[scale=0.5]{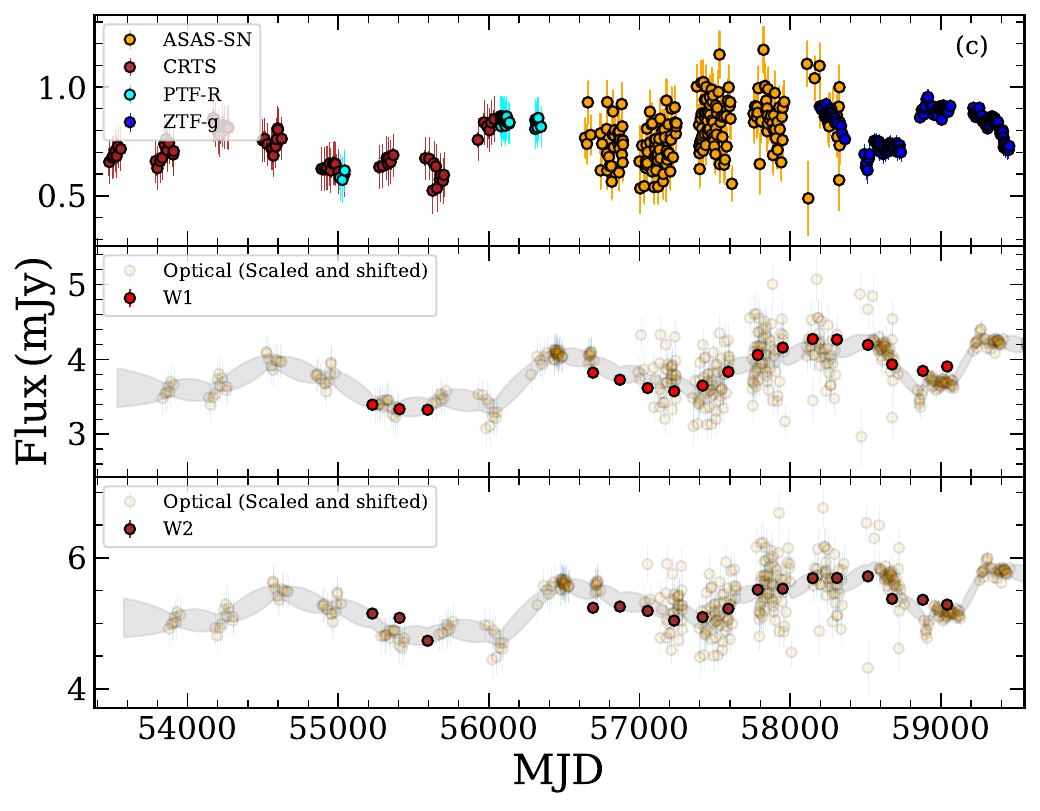}
\includegraphics[scale=0.5]{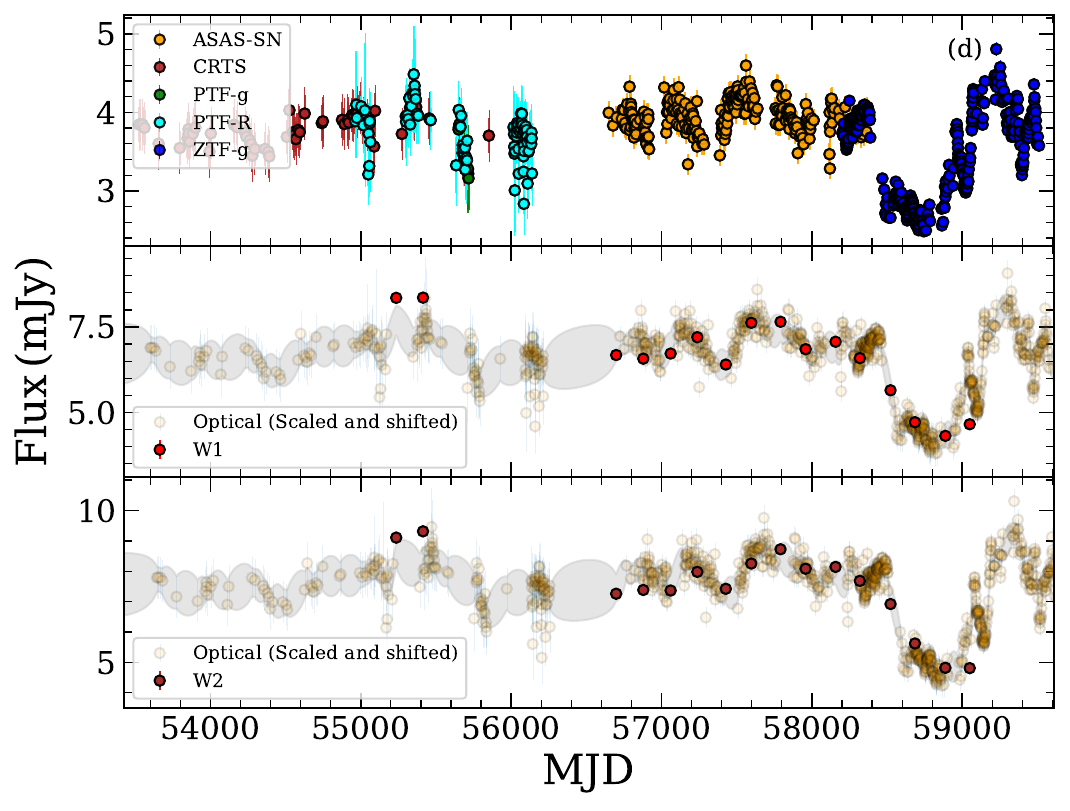}
\includegraphics[scale=0.49]{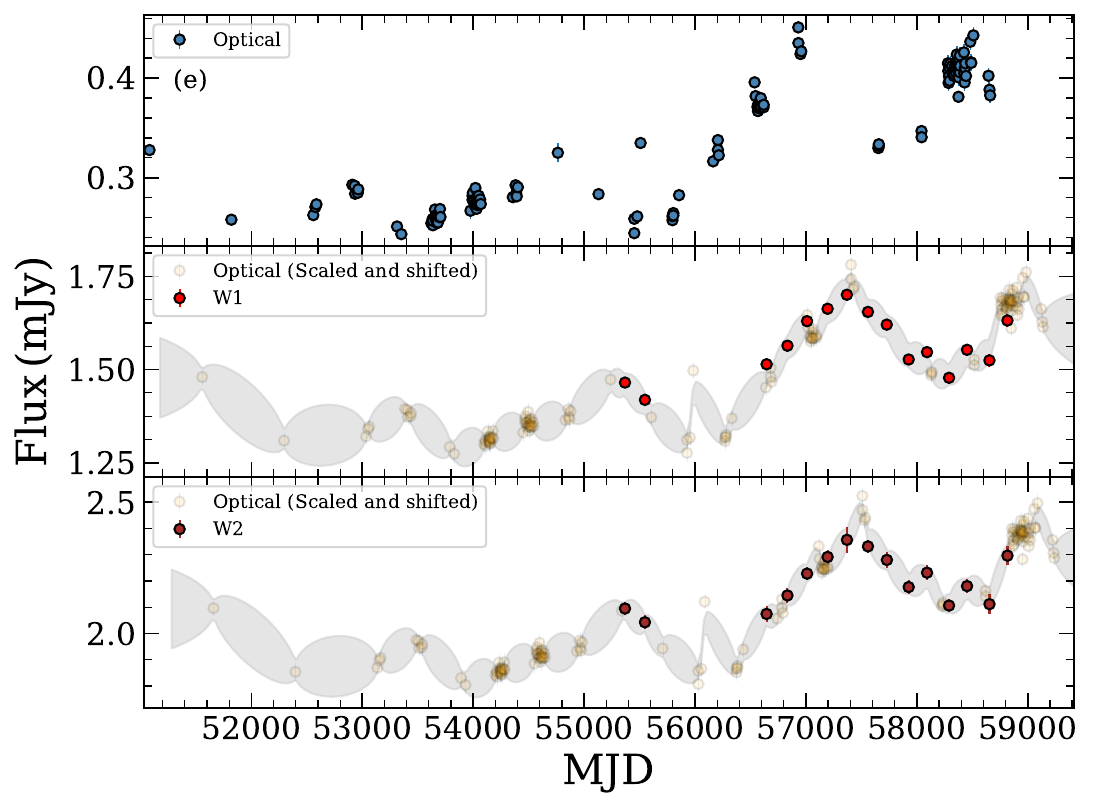}
\includegraphics[scale=0.49]{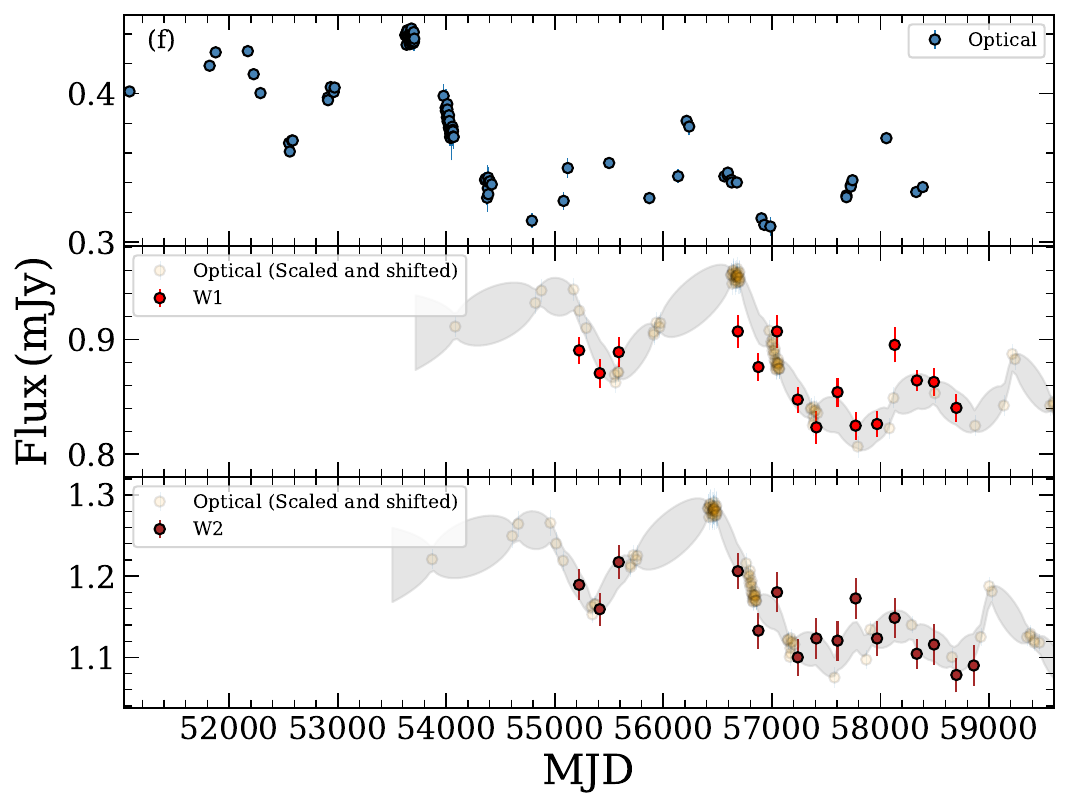}

\caption{From top left to bottom right: optical and IR  W1 and W2- bands light curves of the targets (a) OB21, (b) OB44, (c) OB64, and  (d) OB76 from Sample 1, as well as (e) ID24646, and (f) ID2497212 from Sample 2. The final calibrated optical light curves including data from CRTS, ASAS-SN, PTF and ZTF from Sample 1 are shown in the top panel, while for Sample 2 finally calibrated DES $g$- band optical light curves are presented. For Sample 1, the measurement errors (thick bars) and the total errors including the systematic uncertainty of the offset from  {\tt PyCALI} are presented.
%with and without caps represent photometric errors associated with the data points, and total errors including systematic errors computed by {\tt PyCALI}, respectively, for Sample 1.}  
The bottom two panels represent the IR light curves in W1 and W2- bands.  The W1 and W2- bands light curves shown in the figure are not corrected for the AD contamination. The shifted (by the lag from ICCF) and scaled version of the optical light curves (yellow circles) are also shown with the observed IR light curves, where the shade represents the $1\sigma$  uncertainty obtained in the scaling process.}
\label{fig:light}
\end{figure*}

\subsubsection{IR data from WISE}

WISE surveys the full sky at wavelengths centered at 3.4 (W1), 4.6 (W2), 12 (W3) and 22 $\mu m$ (W4) using a 40 cm diameter infrared telescope since 2010. We collected profile-fit photometric light curve data of our targets in the W1 and W2-- bands from the Near-Earth Object Wide-field Infrared Survey Explorer Reactivation Database (NEOWISE-R), WISE All-Sky and WISE Post-Cryo Database from the NASA/IPAC Infrared Science Archive \footnote{\url{https://irsa.ipac.caltech.edu/cgi-bin/Gator/nph-scan?mission=irsa&submit=Select&projshort=WISE}}. The retrieved WISE data covers a time period from 2010  to 2020. There is a gap of 3 yr between 2011 and 2014. We only selected good-quality photometric measurements following the suggestions given at \url{https://wise2.ipac.caltech.edu/docs/release/neowise/expsup/sec2_3.html}.

For Sample 2, we acquired W1- band light curve data from \citet{2020ApJ...900...58Y}, while the light curve data in the W2- band was retrieved from the WISE database.

The survey details utilized in this study are summarized in Table \ref{tab:survey}, while an overview of the initially selected sample properties can be found in Table \ref{tab:sample}. Figure \ref{fig:dist} presents a comparison between the redshift and luminosity distributions of our initially selected sample and those reported in the literature for DRM. Sample 1 covers a range of luminosity with redshifts below 0.33, while Sample 2 mainly represents high-luminosity and high-redshift AGNs. Therefore, our total sample encompasses the range of luminosity and redshift that has been used to measure the torus size in AGNs to date.

\section{Light Curves Construction} \label{lc_con}
\subsection{Photometric Calibration of the optical light curves}

The optical light curves originate from different surveys, each observed with distinct telescopes and filter properties. For instance, the CRTS observations are unfiltered. We converted the CRTS magnitudes to V- band fluxes following \citet{2014MNRAS.439..690H}. While ASAS-SN provides light curve data in V- band, PTF have data either or both in SDSS-$g$ and Mould R- bands. The ASAS-SN and PTF magnitudes were converted to fluxes using the conversion factors given in \citet{1998A&A...333..231B} and \citet{1996AJ....111.1748F} in their respective filters. We also utilized ZTF-$g$ band data to construct our optical light curves and used ZTF-$r$ when ZTF-$g$ is not available. The ZTF magnitudes were converted into fluxes using the zero-points provided by the Spanish Virtual Observatory (SVO) filter profile service \footnote{\url{http://svo2.cab.inta-csic.es/theory/fps/index.php?mode=browse&gname=Palomar&gname2=ZTF&asttype=}}. We took the average of the fluxes if more than one observation were taken in a single night.

 Various optical monitoring surveys utilize different filters and adopt various aperture schemes in photometry, generating systematic offset among the continuum light curves. Thus, it is necessary to inter-calibrate heterogeneous data sets and include systematic errors.
We inter-calibrated the light curves from different surveys with various optical bands using the python-based software {\tt PyCALI}\footnote{\url{https://github.com/LiyrAstroph/PyCALI}}\citep{2014ApJ...786L...6L}. {\tt PyCALI} uses Bayesian statistics to determine the best scaling factors for inter-calibration by modeling each light curve using damped random walk model \citep[DRW;][]{2009ApJ...698..895K, 2010ApJ...721.1014M} for describing AGN variability. In this process, we used ASAS-SN light curve as the reference to scale all the other light curves, because ASAS-SN covers a significant portion of the overlapping region between the optical and WISE light curves. In this inter-calibration we ignored the small optical inter-band continuum lags \citep[accretion disk RM;][]{2017ApJ...836..186J, 2018ApJ...862..123M, 2022ApJ...940...20G}, which are expected to be a few days \citep[$<$ 10 days;][]{2022MNRAS.511.3005J}, negligible compared to IR time scale.  In our subsequent analysis,  we adopted the total error computed by {\tt PyCALI}, which accounts for measurement and systematic errors. For Sample 2, we used DES $g$- band light curves that were finally calibrated using various optical data sources from \citet{2020ApJ...900...58Y}.

\subsection{WISE Light Curves}

The WISE magnitudes are obtained using profile-fitting photometry and circular aperture photometry \footnote{\url{https://wise2.ipac.caltech.edu/docs/release/allsky/expsup/sec4_4c.html}}. WISE Vega magnitudes were then converted to AB magnitudes using $m_{AB} \, = \, m_{Vega} \, + \, \Delta m$, where $\Delta m$ = 2.699, 3.339, 5.174, and 6.620 for the W1, W2, W3, and W4- bands, respectively \citep{2011ApJ...735..112J}, and finally into flux densities following \citet{2011ApJ...735..112J}. WISE observed each object for 15--17 epochs, with 10--20 frames within a 36 hour window in each epoch during the period 2010--2020. Thus, we calculated the mean value of the IR fluxes within a 6-month window for both W1 and W2- band light curves.

The light curves were not corrected for the host galaxy contribution considering a constant host star-light contamination, which makes a negligible effect on the lag measurements. We obtained the errors in the flux measurements through the standard propagation of errors. In Figure \ref{fig:light}, we show light curves for 6 targets in the optical and IR  bands. The displayed light curves reveal a correlated flux variations between the optical and IR- (W1, W2) bands for the selected AGNs. A few more light curves for objects with various lags and luminosities are shown in Figures \ref{fig:light_app} and \ref{fig:light_appS2} for Sample 1 and Sample 2, respectively. The light curve data are presented in Table \ref{tab:lc}.

\section{Analysis}\label{analysis}

The final calibrated optical and WISE light curves, with a typical cadence of 5 days in the optical and 180 days in the WISE light curves, span over 16--20, and 10 yr, respectively. In this section we describe our lag measurements based on these light curves, quality assessment, and correction for $z$ and AD contamination in WISE light curves.

%To enhance the constraint on the $\mathrm{R_{dust} - L_{bol}}$ relationship, we expanded the MIR lag measurements across a broad luminosity range ($\mathrm{10^{43.4} - 10^{47.6} \, erg \, s^{-1}}$) by incorporating AGNs of varying luminosities into our analysis. This allowed us to include both low and high luminosity AGNs and achieve a more comprehensive understanding of the relationship.

\begin{table}
\centering
 \movetableright= -15mm
\caption{Optical and IR photometric light curves. The full table is available for Sample 1 and Sample 2 separately in a machine-readable form online.}
\label{tab:lc}
\centering
\begin{tabular}{ccccc} \hline

% ID & & & \multicolumn{2}{c}{ICCF} & \multicolumn{2}{c}{\textsc{Javelin}} \\

ID no & MJD& $F$ & $F_{err}$ & Survey \\
 &  & mJy & mJy &  \\
(1) & (2) & (3) & (4) & (5)
\\ \hline

OB1 & 53627.295443 & 1.2872 & 0.1976 & CRTS \\
OB1 & 53637.455742 & 1.2723 & 0.1978 & CRTS \\
OB1 & 53646.362378 & 1.2807 & 0.1978 & CRTS \\

... & ... & ... & ... & ... \\
... & ... & ... & ... & ... \\

\hline

\end{tabular}
\vspace{0.04cm}

\raggedright Note: Columns are (1) object ID number, (2) Modified Julian Date, (3) flux, (4) error in flux, and (5) name of the imaging survey. Data table for the observed  and AD contamination corrected IR light curves for Sample 1 and Sample 2 will be provided separately with the same format.
\end{table}

\begin{figure*}
\resizebox{9.3cm}{4.8cm}{\includegraphics{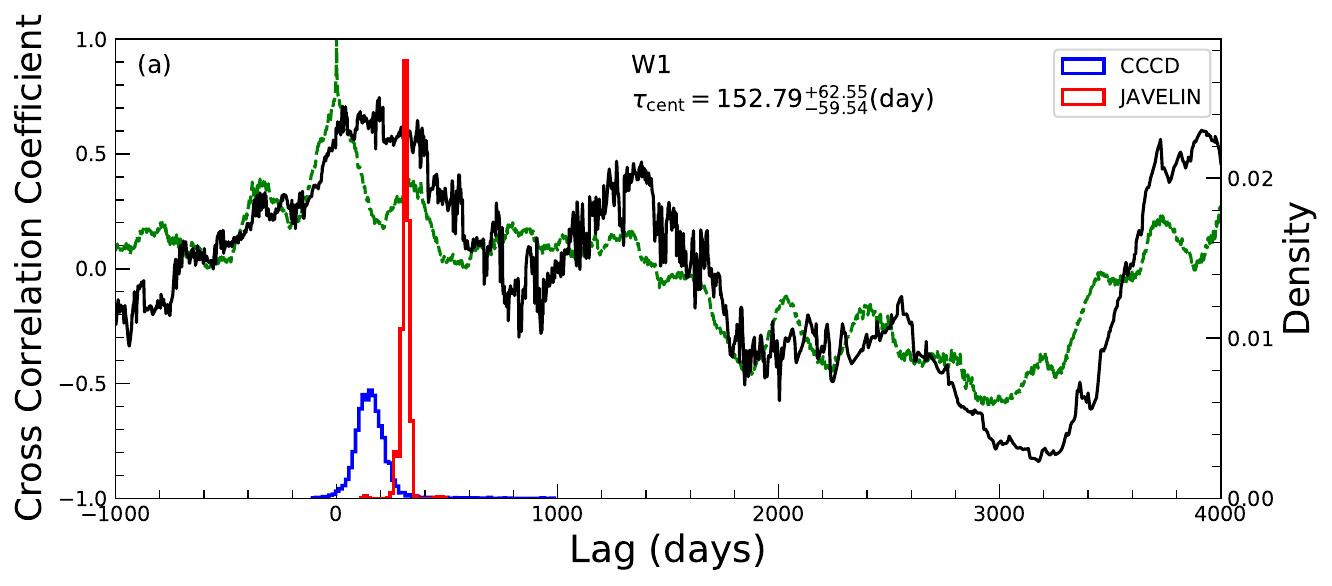}}
\resizebox{9.3cm}{4.8cm}{\includegraphics{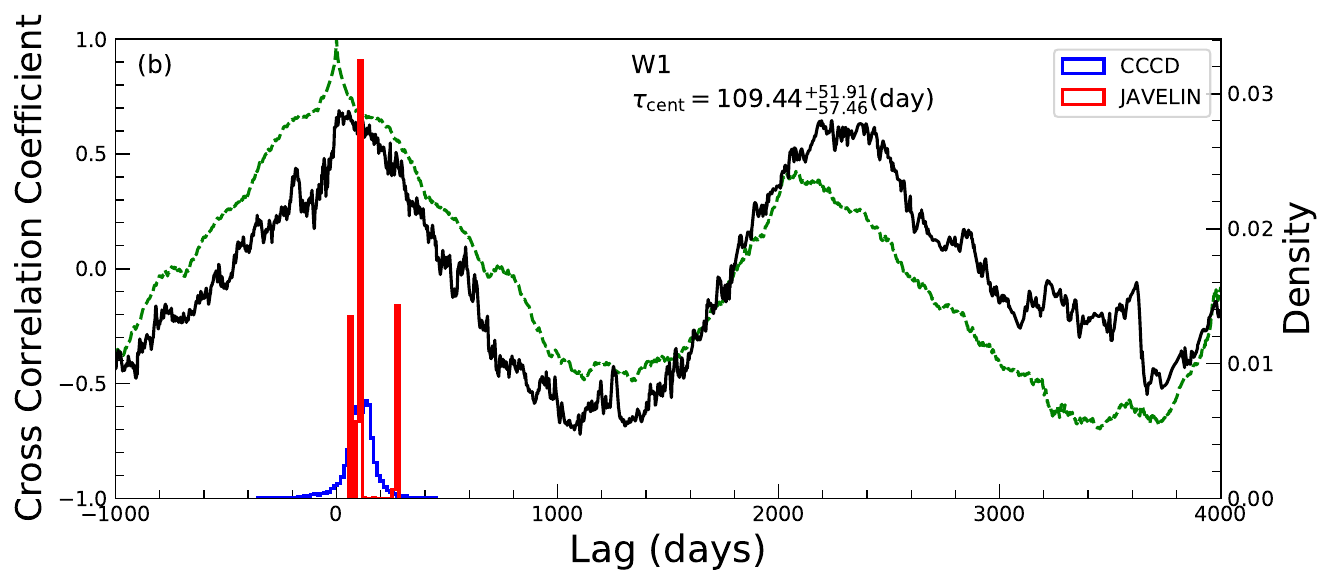}}
\resizebox{9.3cm}{4.8cm}{\includegraphics{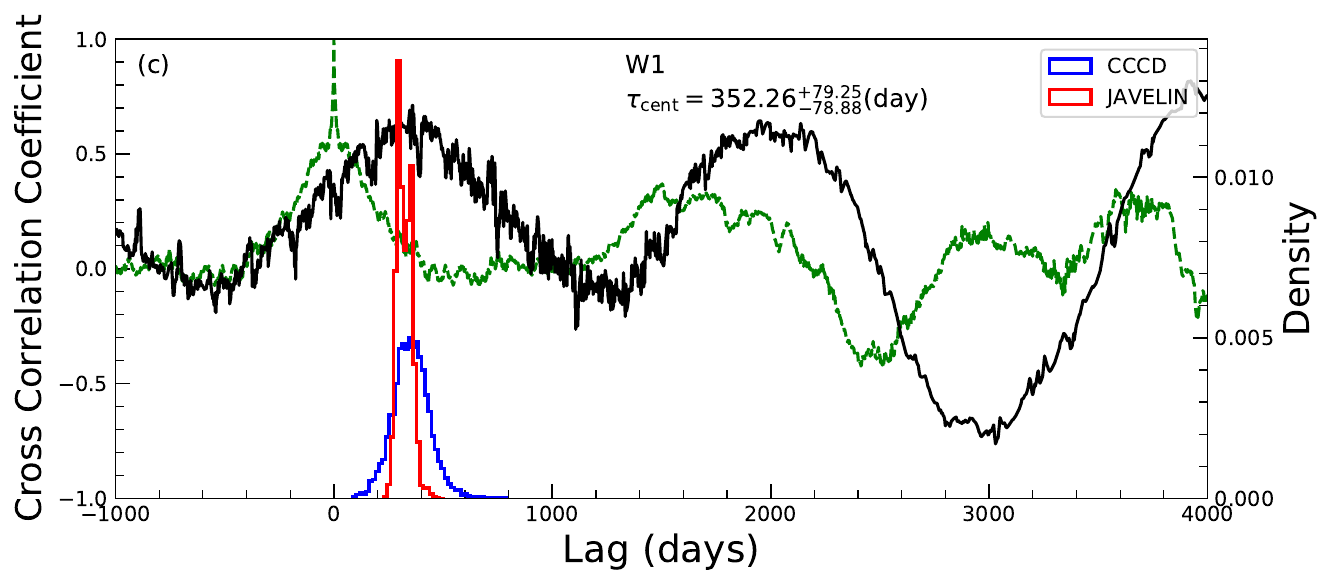}}
\resizebox{9.3cm}{4.8cm}{\includegraphics{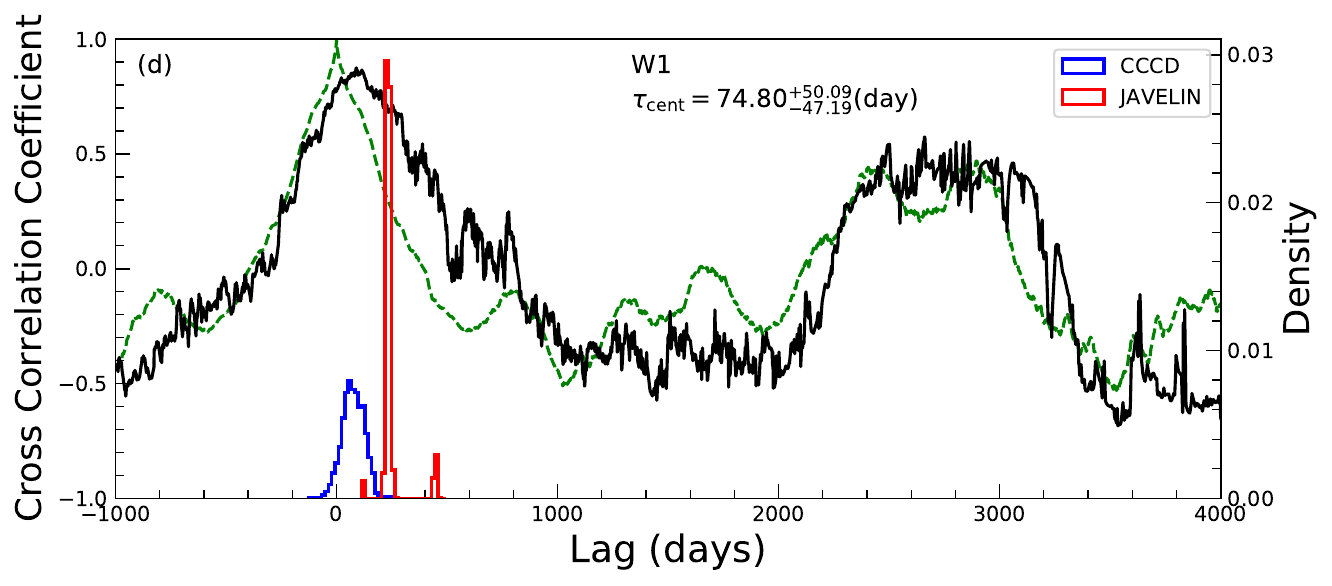}}
\resizebox{9.3cm}{4.8cm}{\includegraphics{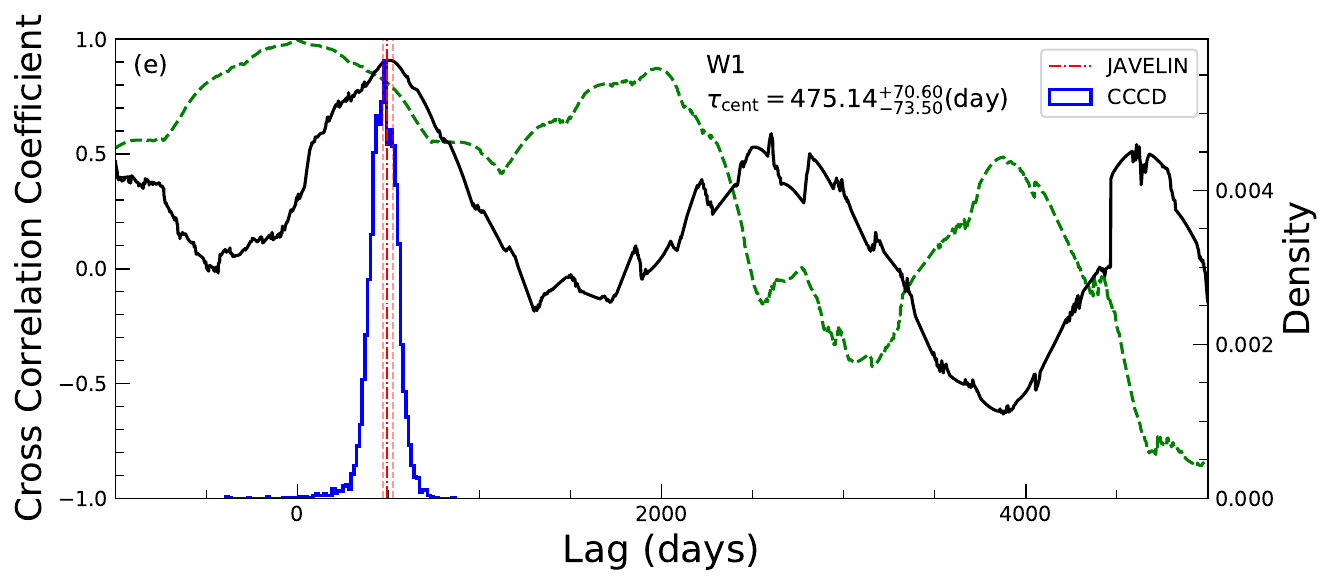}}
\resizebox{9.3cm}{4.8cm}{\includegraphics{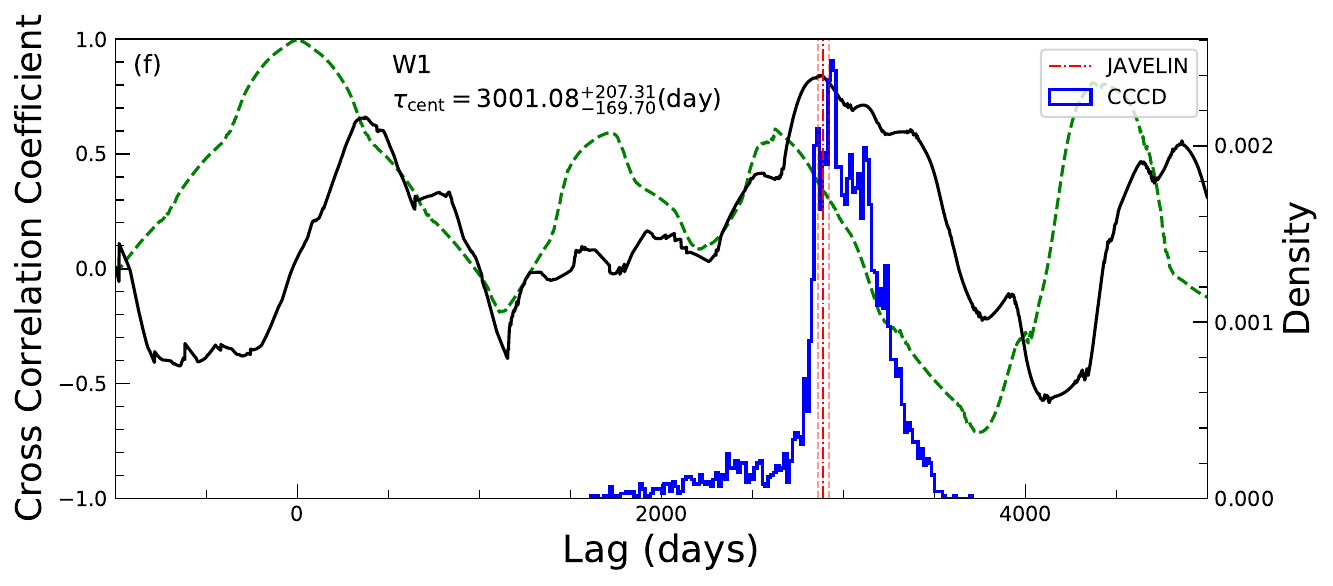}}

\caption{From top left to bottom right: the ICCFs and {\tt JAVELIN} lag analysis between optical and W1- band light curves (not corrected for AD contamination) for the targets (a) OB21, (b) OB44, (c) OB64, and (d) OB76, respectively, from Sample 1. Black solid line represents the ICCF between the optical and W1- band light curves and dashed green line is the ACF of the optical light curve. The CCCD obtained from ICCF is shown by the blue histogram, whereas the red histogram represents the lag probability distribution of the time delay from {\tt JAVELIN}. The last two in the bottom panel are the ICCFs for the targets (e) ID24646, and (f) ID2497212 from Sample 2, where the vertical red dashed lines represent the {\tt JAVELIN} lag measurements from \citet{2020ApJ...900...58Y} with 1$\sigma$ uncertainties shown by the red dotted lines. The observed-frame time lags obtained from ICCF with their uncertainties are also mentioned.}
\label{fig:ccf}
\end{figure*}

\subsection{Lag measurements between the optical and IR light curves}

All the AGNs selected for this study exhibit synchronized flux variations over time in both the optical and IR W1 and W2- bands, with the optical observations consistently leading the IR measurements. To determine the time lag between the optical and IR continua, we employed the Interpolated Cross-Correlation Function  \citep[ICCF;][]{1986ApJ...305..175G, 1987ApJS...65....1G} as the primary method. It determines time lags by linearly interpolating two light curves and calculating the cross-correlation coefficient (r). We evaluated CCF by setting the lag search range to $\sim$ 70 \% of the baseline defined by the optical light curve, which is sufficient for robust time lag measurements \citep{2017ApJ...851...21G, 2019ApJ...887...38G, 2020ApJ...900...58Y}. We applied the ICCF method to measure time lags in 142 AGNs from Sample 1 and 587 AGNs from Sample 2, respectively. We first located the most significant peak within the search range and the lag was estimated from the centroid ($\tau_{cent}$) of the CCF around the most significant peak, defined as \citep{1998PASP..110..660P}

\begin{equation}
\tau_{cent} = \frac{\Sigma_j \, \tau_j CCF_j} {\Sigma_j \, CCF_j}
\end{equation}
where, $\tau_{cent}$ was obtained by considering all the points within 80 \%  of the maximum of the CCF ($r_{max}$). The uncertainties in the measured lags were estimated using a model-independent Monte Carlo simulation based on flux randomization (FR) and random subset selection (RSS) as described in \citet{1998PASP..110..660P, 1999ApJ...526..579W}; and  \citet{2004ApJ...613..682P}. This process was repeated for 5000 iterations, and the centroid of the lag was determined in each iteration. The cross-correlation centroid distribution (CCCD) was then constructed. As the CCCD was not Gaussian, the lag was evaluated from the median of the distribution and the lower and upper uncertainties in the measured lag are that values at the 15.9 and 84.1 percentile of the distribution. This is equivalent to $1\sigma$ error for a Gaussian distribution.

We also used {\tt JAVELIN} to find the time delays between the optical and IR light curves and validated the ICCF measurements. In {\tt JAVELIN} the driving optical continuum was first modeled using the DRW model with two free parameters; amplitude ($\sigma_d$) and time scale of variability ($\tau_d$). To recover the IR continuum light curve a top-hat transfer function was convolved with the driving optical continuum light curve. We used the same search range as used for ICCF to find the lag, and adopted Markov Chain Monte Carlo approach to find the best-fitting model maximizing the likelihood. However, we observed that {\tt JAVELIN} tends to produce strong aliases, particularly near the boundaries of the lag search window when a large lag search range was applied, leading to multiple peaks in the lag posterior distribution. This is a feature of aliasing in the light curves resulting from sparsely sampled multi-year data with seasonal gaps \citep{2019ApJ...887...38G, 2020ApJ...901...55H}. Thus, we excluded the false peaks in the posterior distribution. First, we identified the most significant peak around the location of the ICCF peak, and subsequently excluded false peaks located outside of the $\pm 1000$ days range from the ICCF peak in the posterior distribution. The final lag was determined by calculating the median of the lag probability distribution around the most significant peak in the distribution.

Figure \ref{fig:ccf} shows examples of the results from ICCF and {\tt JAVELIN} for the targets shown in Figure \ref{fig:light}. For a few targets, along with a primary peak at smaller lag, there is a secondary peak at a larger lag (for examples, see the  CCFs of OB44 and OB64 in Figure \ref{fig:ccf}). While the primary peak represents the actual lag between the optical and W1- band light curves, the other peak arises mainly because  of the shape and aliasing in the light curves \citep{2022MNRAS.516.4898G}. For example, in Figure \ref{fig:ccf}, the CCF of the target OB44 shows a primary peak at around 110 days, and a secondary peak around 2200 days. The corresponding Auto Correlation Function (ACF) of the optical light curve has also two peaks, one around 0 lag and another secondary peak around 2090 days. Thus, the IR W1 light curve will also correlate with the optical light curve at the lag $\tau$=109 days and at a latter time $\tau$  + 2090 days, resulting two peaks in the CCF. The same is applicable for the target OB64 shown in Figure \ref{fig:ccf}. Thus, we calculated lags around the primary peaks in these cases. We obtained te lags between the optical and WISE W1 and W2- band light curves for all of those 729 (142+587) AGNs.

\begin{figure}
\centering
\includegraphics[scale=0.46]{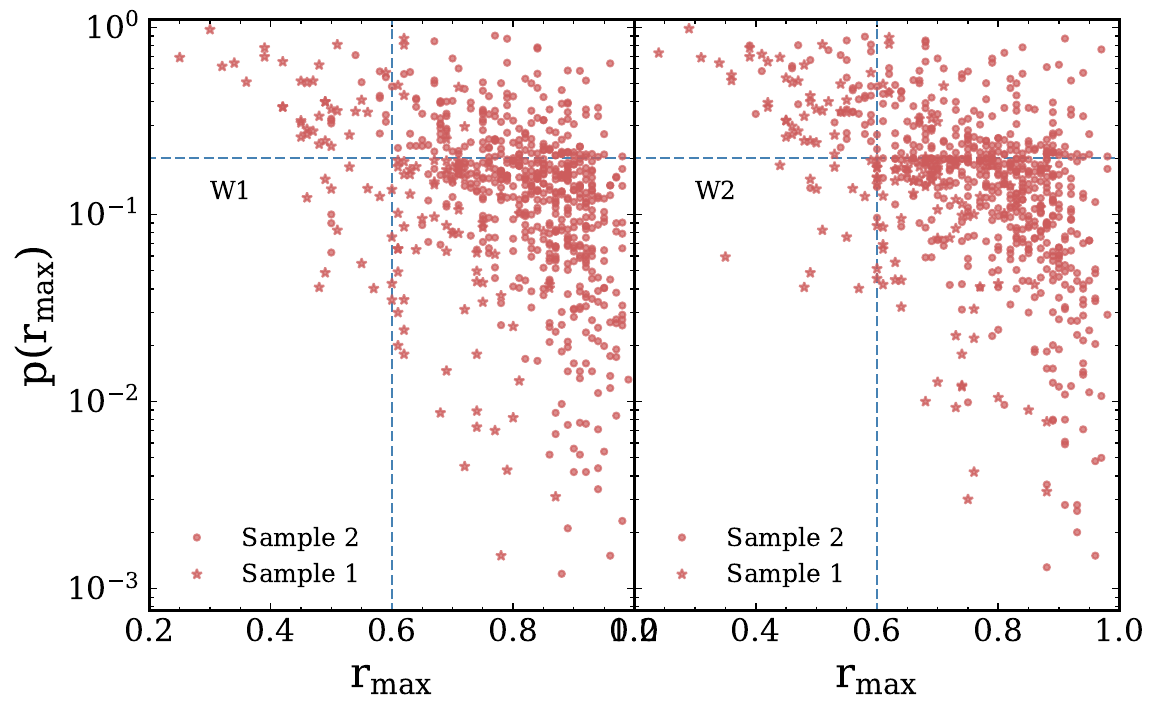}
%\resizebox{9cm}{6cm}{\includegraphics{pval_final.pdf}}
\caption{ The probability $p(r_{max})$ represents the likelihood of uncorrelated light curves producing a cross-correlation coefficient equal to or exceeding $r_{max}$. This probability is presented as a function of the maximum cross-correlation coefficient $r_{max}$ in the left and right panels for W1 and W2- bands, respectively. The vertical dashed line represents $r_{max}$=0.6, whereas $p(r_{max})$=0.2 is shown by the horizontal dashed line. The reliable time lags are defined with $r_{max}$ $\geq$ 0.6, and $p(r_{max})$ $<$ 0.2.}
\label{fig:pval}
\end{figure}

\begin{figure*}

\resizebox{9.3cm}{4.8cm}{\includegraphics{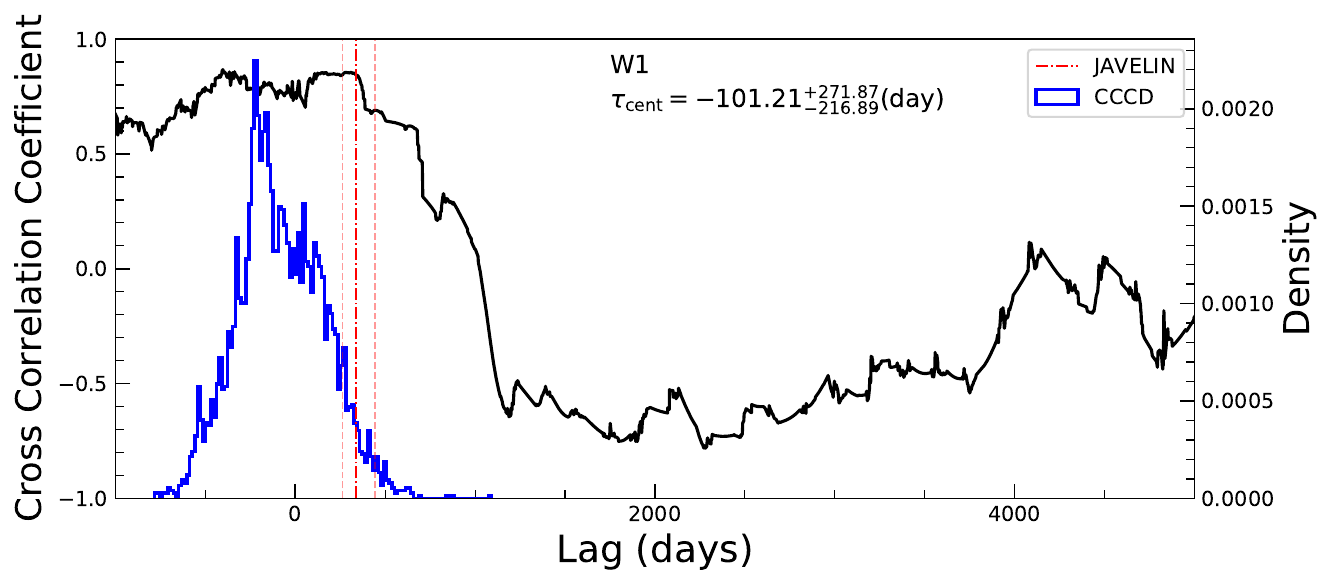}}
\resizebox{9.3cm}{4.8cm}{\includegraphics{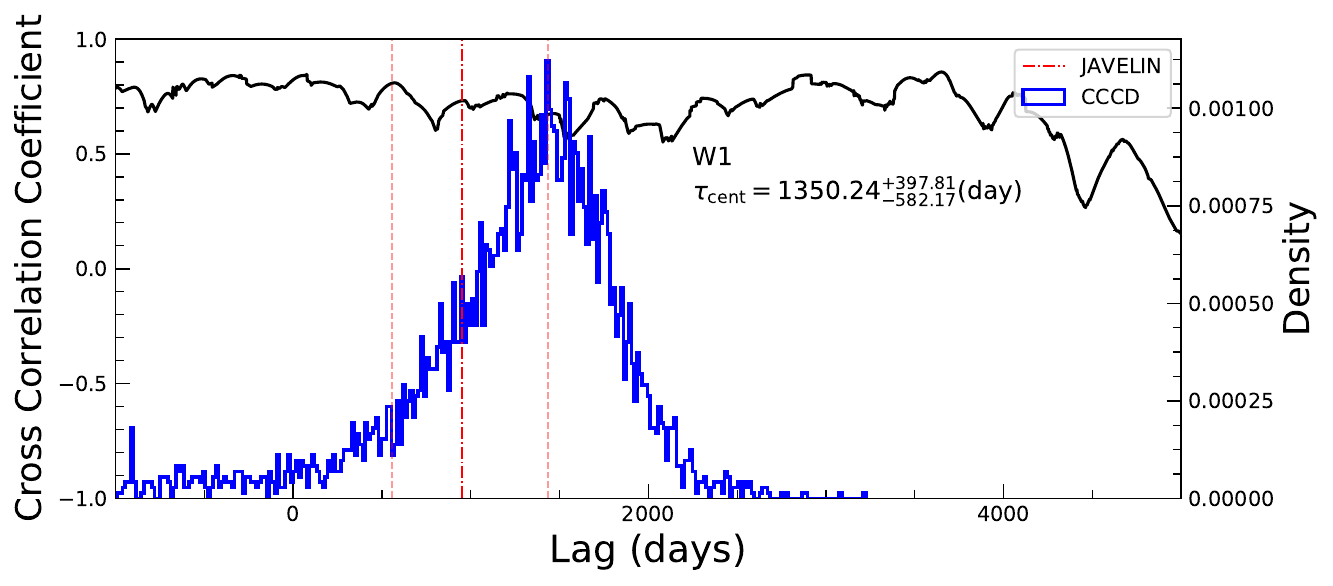}}

\caption{Examples of rejected targets based on ICCFs between the optical and W1- band light curves.  The two targets ID7907462 (left) and  ID954476 (right) from Sample 2 are the examples with negative lag and a flat CCF, respectively. The {\tt JAVELIN} lag measurements from \citet{2020ApJ...900...58Y} for these two targets are shown by the dashed red lines. The observed-frame time lags obtained from ICCF with their uncertainties are also mentioned at each panel.}
\label{fig:rjccf}
\end{figure*}

\begin{figure}
\includegraphics[scale=0.6]{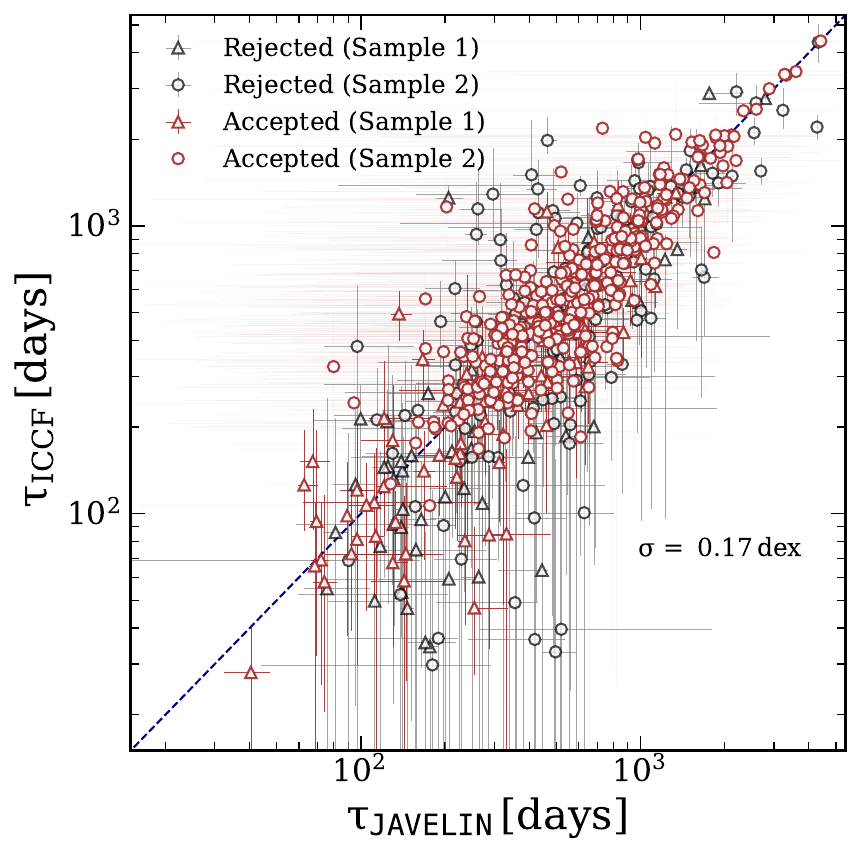}
%\resizebox{6cm}{5cm}{\includegraphics{grt_curve.pdf}}

\caption{Comparison of the time lags in the W1- band obtained through ICCF and the {\tt JAVELIN} method in the observed-frame of the targets. The dashed blue line is the 1:1 relation. The brown triangles and circles represent the accepted targets from Sample 1 and Sample 2, respectively, for use in the RM analysis. The black points are the rejected targets. For Sample 2, the {\tt JAVELIN} lags are obtained from \citet{2020ApJ...900...58Y}. Most of the rejected targets deviate from the 1:1 relation.}
\label{fig:comp}
\end{figure}

 \subsection{Quality assessment} \label{qual}

In this section, we assess the dependability of the lag measurements obtained through cross-correlation analysis. The accuracy of the recovered time lags between the optical light curves, derived from various surveys, and sparsely sampled IR light curves is influenced by different sources of uncertainty. The presence of substantial data gaps in the IR light curves, coupled with a limited number of data points in the optical-IR light curve pairs, can lead to false positive peaks near the boundaries of the lag search window when using the ICCF method. These factors collectively impact the reliability of the obtained time lags. Consequently, it is crucial to exercise caution when selecting targets with significant and reliable lags.

 The maximum cross-correlation coefficient ($r_{max}$) is a key factor to assess the reliability in lag measurements. This coefficient signifies the strength of the correlation between two light curves. Previous DRM studies have often employed specific thresholds for $r_{max}$, such as $r_{max}$ $>$ 0.5 \citep{2020ApJ...900...58Y} or $r_{max}$ $>$ 0.4 \citep{2023MNRAS.tmp.1073C}, as a means to determine the reliability of lag measurements.

However, it is crucial to acknowledge that relying solely on $r_{max}$ is insufficient for guaranteeing lag reliability. There are situations where $r_{max}$ might be relatively high even when the light curves possess only a limited number of data points but exhibit strong linear variation. Even small variability amplitude compared to the flux uncertainties in the light curves may result high $r_{max}$ as long as certain features in the light curves align well. Moreover, it is crucial to emphasize that the calculations for $r_{max}$ do not incorporate the uncertainties in flux.

To address these issues and evaluate the reliability of lag measurements, a new method called {$\tt PyI^2CCF$} \footnote{\url{https://github.com/legolason/PyIICCF}} has been introduced \citep[][Guo et al., in prep]{2021AAS...23722608G, 2022ApJ...925...52U}. This method employs a null-hypothesis approach to assess whether the observed cross-correlation could be equally derived from two uncorrelated red-noise light curves. To put this method into practice, we utilized the publicly available {$\tt PyI^2CCF$} tool to investigate the reliability of our lag measurements. Our methodology involved generating $10^3$ mock light curve pairs for each object, maintaining the same sampling and cadence as the observed light curves while utilizing DRW models. Subsequently, we calculated the lag significance indicator, referred to as $p(r_{max})_{\tau>0}$, by tallying the occurrences of positive lags ($\tau \, >$  0) where the $r_{max}$ value exceeded the observed $r_{max}$ across all simulations. It is important to mention that we retained the same searching window that was initially used during the lag calculations. Finally, we apply a set of criteria to exclude measurements deemed unreliable, such as:

 \begin{enumerate}
      \item  We define the lag measurements as reliable if $r_{max}$  $\geq$ 0.6 and $p(r_{max})_{\tau>0}$ $<$ 0.2. These thresholds are adopted following \citet{2022ApJ...925...52U} and \citet{2024ApJ...962...67W} (See Figure \ref{fig:pval}). Using these criteria, we identify a total of 85 targets in the W1- band and 79 in the W2- band within Sample 1 as having reliable lags. Similarly, for Sample 2, out of 587 targets, we select 399 in the W1- band and 372 in the W2- band.

      \item  However, we discard 4 and 3 targets from Sample 1 in the W1 and W2- bands, respectively, due to negative lags within a 1$\sigma$ uncertainty range, as negative lags are physically unacceptable (See Figure \ref{fig:rjccf}, left). Similarly, from Sample 2, 16 and 14 targets are rejected with negative lags within 1$\sigma$ uncertainty in W1 and W2- bands, respectively. Additionally, we eliminate targets displaying a flat CCF spanning over thousands of days, as this condition results in an average lag computed over the entire search range (refer to Figure \ref{fig:rjccf}, right). Consequently, 18 targets from Sample 2 in both W1 and W2- bands are excluded from the analysis.

 \end{enumerate}

After applying the criteria, we narrowed down from 142 targets to 81 targets in the W1- band and 76 targets in the W2- band in Sample 1. In the case of Sample 2, we obtained 365 AGNs with W1- band light curves and 340 with W2- band light curves, out of the initial sample of 587 targets. In summary, our final sample consists of 446 (416) AGNs with a reliable and significant dust lag measurements with W1- (W2) band light curves. Note that if we ignore the systematic uncertainty of the light curve calibration between two different telescope data sets (see Section \ref{lc_con}), the error of the photometry decreases by 30-40\%. This reduction of the photometry errors in the light curves will slightly increase the quality of the cross-correlation. To evaluate the effect of the error treatment on the quality assessment, we remeasure the  $p(r_{max})_{\tau>0}$ for Sample 1 after removing the systematic error from the total error. We find that the change of the sample based on the quality assessment (see Figure \ref{fig:pval}) is insignificant. 

In Figure \ref{fig:comp}, we present a comparison of the W1- band lags in the observed-frame between the  ICCF and {\tt JAVELIN} methods. The final sample is represented by brown data points, while the rejected AGNs are shown in black. The majority of the rejected targets deviate from the expected 1:1 relation. In contrast, the lags obtained for the final sample demonstrate a good consistency, with a scatter of 0.17 dex. We present a few examples of the selected ICCF and {\tt JAVELIN} analyses for Sample 1 and Sample 2 in Figures \ref{fig:ccf_app} and \ref{fig:ccf_appS2}, respectively.

We checked that shifting the optical light curve by the median lag value ($\tau_{cent}$) obtained from ICCF resulted in a good alignment with the IR light curves, as demonstrated in Figures \ref{fig:light}, \ref{fig:light_app}, and \ref{fig:light_appS2}. Moreover, the error estimation in ICCF analysis is more conservative \citep{2022ApJ...940...20G}, and the ICCF measurements incorporate a cross-correlation reliability test. Consequently, in all subsequent discussions, our primary focus will be on the lags derived from the ICCF analysis.

\begin{table*}
\centering
 \movetableright= -75mm
%\movetabledown=60mm
%\begin{rotatetable}
%\scriptsize pval$\textunderscore$W1

\caption{Details of the final sample and the results of their time lag measurements for W1 and W2- bands are provided. The complete table is available online in a machine-readable format, separately for Sample 1 and Sample 2.}
\label{tab:lag}
\resizebox{19cm}{!}{
\fontsize{21pt}{21pt}\selectfont
\begin{tabular}{llccccccccccccr} \hline
ID no & ID & $z$ & log($L_{bol}$) & log(Edd) & $r_{max, W1}$ & pval\textunderscore W1 & $\tau_{W1}$  & $\tau_{W1,ADC}$  & $\tau_{W1,ADC}$ (JAV) & $r_{max, W2}$ & pval\textunderscore W2 & $\tau_{W2}$  & $\tau_{W2,ADC}$ & $\tau_{W2,ADC}$ (JAV)  

\\ & & & $\mathrm{erg \, s^{-1}}$ & & & & days & days & days & & & days & days & days  \\
(1) & (2) & (3) & (4) & (5) & (6) & (7) & (8) & (9) & (10) & (11) & (12) & (13) & (14) & (15)

\\ \\ \hline

 &  &  &  &  &  & Sample 1 &  &  &  &  &  &  \\

OB1 & $J014017.07-00503.0$ & 0.334 & $46.14 \pm 0.01$ & $-0.68 \pm 0.14$ & 0.85 & 0.0811 & $617^{+105}_{- 99}$ & $773^{+189}_{-178}$ & $1059^{+97}_{-188}$ & 0.82 & 0.1421 & $619^{+117}_{-113}$ & $795^{+209}_{-208}$ & $1076^{+104}_{-211}$  \\
OB2 & $J031027.83-004950.8$ & 0.080 & $45.05 \pm 0.01$ & $-0.83 \pm 0.14$ & 0.74 & 0.0089 & $73^{+51}_{-33}$ & $133^{+91}_{-78}$ & $163^{+103}_{-91}$ & 0.73 & 0.0093 & $98^{+74}_{-39}$ & $165^{+87}_{-95}$ & $179^{+102}_{-63}$ \\
OB3 & $J032213.90+005513.5$ & 0.185 & $45.69 \pm 0.01$ & $-0.23 \pm 0.14$ & 0.71 & 0.1689 & $375^{+110}_{-150}$ & $400^{+116}_{-145}$ & $616^{+29}_{-27}$ & 0.70 & 0.1953 & $422^{+130}_{-164}$ & $434^{+130}_{-172}$ & $619^{+23}_{-13}$ \\

... & ... & ... & ... & ... & ... & ... & .. & ... & ... & ... & ... & ... & ... & ... \\
... & ... & ... & ... & ... & ... & ... & .. & ... & ... & ... & ... & ... & ... & ... \\

\hline

 &  &  &  &  &  & Sample 2 &  &  &  &  &  &  \\

5871 & $J000535.23+000021.6$ & 0.736 & $45.28 \pm 0.01$ & $-0.82 \pm 0.57$ & 0.78 & 0.1716 & $594^{+204}_{-267}$ & $707^{+291}_{-353}$ & $985^{+179}_{-188}$  & ... & ... & ... & ... & ... \\
2401 & $J000654.10-001533.5$ & 1.725 & $46.99 \pm 0.01$ & ... & ... & ... & ... & ... & ... & 0.83 & 0.0760 & $2571^{+287}_{-233}$ & $2500^{+464}_{-511}$ & $1510^{+1072}_{-653}$ \\ 
9836 & $J001735.69-011325.0$ & 0.805 & $45.89 \pm 0.01$ & $-0.27 \pm 0.18$ & 0.87 & 0.0829 & $437^{+162}_{-178}$ & $585^{+208}_{-259}$ & $795^{+200}_{-243}$ & 0.75 & 0.1909 & $782^{+290}_{-453}$ & $914^{+354}_{-424}$ & $971^{+252}_{-231}$ \\

 ... & ... & ... & ... & ... & ... & ... & .. & ... & ... & ... & ... & ... & ... & ... \\
... & ... & ... & ... & ... & ... & ... & .. & ... & ... & ... & ... & ... & ... & ... \\

 \hline
\end{tabular}

}

\vspace{0.3cm}
\raggedright Note: Columns are (1) object ID number, (2) SDSS object ID, (3) redshift, (4) AGN bolometric luminosity, (5) Eddington ratio obtained from this work, (6) peak correlation coefficient from ICCF between optical and W1- band light curve, (7) $p(r_{max})_{\tau>0}$ value obtained between optical and W1- band light curve, (8) observed-frame W1- band time lag from ICCF, (9) observed-frame AD contamination corrected W1- band time lag from ICCF, (10)  observed-frame AD contamination corrected W1- band time lag from {\tt JAVELIN}, (11) peak correlation coefficient from ICCF between optical and W2- band light curve, (12) $p(r_{max})_{\tau>0}$ value obtained between optical and W2- band light curve, (13) observed-frame W2- band time lag from ICCF, (14) observed-frame AD contamination corrected W2- band time lag from ICCF, and (15)  observed-frame AD contamination corrected W2- band time lag from {\tt JAVELIN}.

%\end{rotatetable}
\end{table*}

%\begin{table*}
%\caption{Format for the final Sample 2 and its lag measurements.}
%\label{tab:lag2}

%\centering
%\resizebox{14cm}{!}{
%\begin{tabular}{lccc} \hline

%% ID & & & \multicolumn{2}{c}{ICCF} & \multicolumn{2}{c}{\textsc{Javelin}} \\

%Column & Format & Unit & Description  \\
%\\ \hline
%\hline
%ID & String & & Object ID \\
%RA & Double & deg & J2000 \\
%DEC & Double & deg & J2000 \\
%REDSHIFT & Double &  &  \\
%LOGLBOL & Double & $\mathrm{erg \, s^{-1}}$ & AGN bolometric luminosity   \\

% & & & obtained from \citet{2020ApJ...900...58Y} \\

%LOGLBOL$\textunderscore$ERR & Double & $\mathrm{erg \, s^{-1}}$ & Error in LOGLBOL \\

%LOGEDD$\textunderscore$RATIO & Double &  & Eddington ratio from \citet{2020ApJ...900...58Y}   \\

%RMAX & Double &  &  Maximum correlation coefficient $\mathrm{r_{max}}$ from ICCF  \\

%TAUW1$\textunderscore$ICCF & Double &  &  Observed-frame W1 time lag from ICCF \\

%TAUW1$\textunderscore$ICCFERR1 & Double &  &  Lower uncertainty in TAUW1$\textunderscore$ICCF \\

%TAUW1$\textunderscore$ICCFERR2 & Double &  &  Upper uncertainty in TAUW1$\textunderscore$ICCF \\

%\hline
%\end{tabular}
%}
%\vspace{0.1cm}

%\raggedright The {\tt JAVELIN} lag measurements for this sample is available in \citet{2020ApJ...900...58Y}. This table will be made available in the electronic version of the article.
%\end{table*}

\begin{figure}
\centering
\includegraphics[scale=0.65]{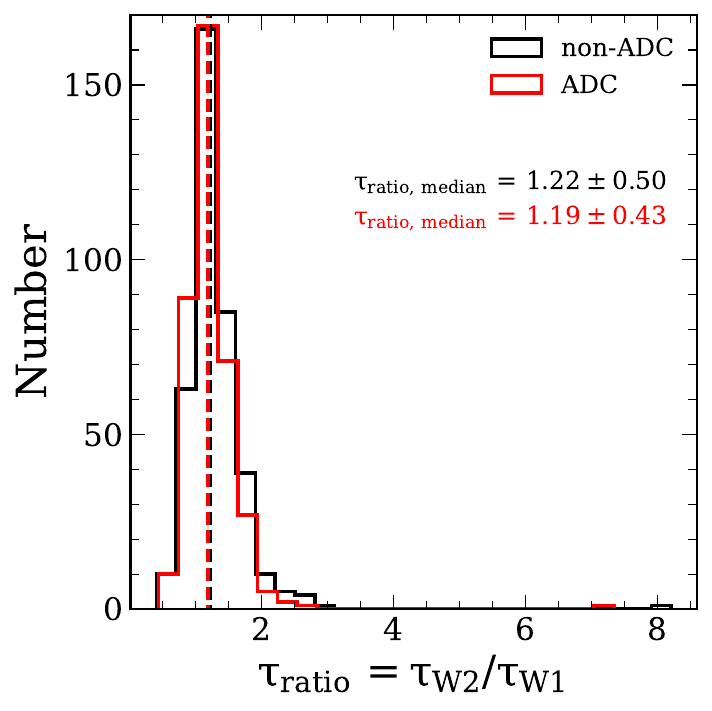}

\caption{ The distribution of the lag-ratio ($\mathrm{\tau_{ratio}}$ = $\tau_{W2}/\tau_{W1}$) between the dust lags in W2 and W1- bands with and without AD contamination correction shown by red and black lines, respectively. The median values with their standard deviations are mentioned in the figure.}
\label{fig:comp_lagratio}
\end{figure}

\subsection{Correcting for redshift effects on dust lags} \label{cor_redshift}

The AGNs under investigation in this study span a wide range of redshifts as illustrated in Figure \ref{fig:dist}.  Because of the wavelength-dependent nature of dust lag, objects at higher redshifts will exhibit shorter rest-frame time lags in specific wavelength bands, than the same luminosity AGNs at lower $z$ \citep{1993ApJ...409..139S, 2001ASPC..224..149O, 2014ApJ...784L..11Y, 2019ApJ...886..150M, 2020AJ....159..259S, 2023MNRAS.tmp.1073C}. This redshift effect must be corrected, especially when dealing with a sample over a large redshift range.  

%Some studies have identified slight but meaningful distinctions in dust lags across various NIR bands among AGNs \citep{2006ApJ...652L..13T, 2011MNRAS.415.1290L, 2015ApJ...814L..12J, 2019ApJ...886...33L, 2023MNRAS.tmp.1073C}, while others have found negligible differences in lag times  \citep{2004MNRAS.350.1049G, 2014A&A...561L...8P, 2018MNRAS.475.5330M, 2021MNRAS.501.3905M}. In contrast, some AGNs exhibited no significant difference in dust lags \citep{2015MNRAS.454..368L, 2018A&A...620A.137R, 2019AstL...45..197O}, without consideration for the correction of AD contamination in NIR bands. As a result, it is reasonable to expect a wavelength-dependent pattern in dust lags in AGNs.

We used the correction factor in the form of (1+$z$)$^\gamma$, as introduced by \citet{2014ApJ...784L..11Y} and \citet{2019ApJ...886..150M}, to account for wavelength-dependent dust lags in order to remove the redshift effect. First, we determined the lag ratio, $\mathrm{\tau_{ratio}}$ = $\tau_{W2}/\tau_{W1}$, between dust lags in W2 and W1- bands for our selected final sample and computed $\gamma$ using the formula, $\gamma$ = log($\tau_{W2}/\tau_{W1}$)/log($\lambda_{W2}/\lambda_{W1}$). Figure \ref{fig:comp_lagratio} represents the distribution of  $\mathrm{\tau_{ratio}}$, both with and without AD contamination correction (see Section \ref{sec:AD_cor} for details of AD contamination correction) obtained from ICCF analysis. Our analysis reveals median values of $\mathrm{\tau_{ratio}}$ to be  $1.19\pm0.43$ ($1.16\pm 0.55$) and $1.22\pm 0.50$ ($1.26\pm 0.85$) for the AD contamination corrected and uncorrected cases, respectively, obtained from ICCF ({\tt JAVELIN}) analysis. These values closely align with the  median values of $\mathrm{\tau_{ratio}}$ = 1.15 reported by \citet{2019ApJ...886...33L} and 1.26 reported by \citet{2023MNRAS.tmp.1073C} within the $1\sigma$ range of the distribution.

By adopting a median $\mathrm{\tau_{ratio}}$ of 1.20 for both the AD contamination corrected and uncorrected cases, 
we find that $\gamma \, \sim 0.62$, leading to a correction factor of $(1+z)^{0.62}$. 

 According to the dust radiation equilibrium model, the evaporation radius from the central ionizing source varies with the UV continuum luminosity and dust grain temperature, as $R_{dust} \, \propto \, L_{UV}^{0.5}T^{-2.8}$ \citep{1987ApJ...320..537B}. Since the dominant source of NIR emission from the torus is attributed to the thermal emission from dust with a temperature T, the wavelength dependency of the dust lag is given by $\tau \, \propto \, \lambda^{2.8}$ \citep{1993ApJ...409..139S, 2001ASPC..224..149O}. The correction factor for the redshift effect is expected to be $(1+z)^{2.8}$. This theoretical prediction is significatnly different from our empirically derived  factor. The discrepancy is partly due to the fact that the dust lag correlates with AGN luminosity with a much shallower slope than the expected value of 0.5 \citep{2019ApJ...886..150M, 2023MNRAS.tmp.1073C}. 
%As a result, this disparity will lead to a distinct dependency of dust lags on temperature and, consequently, yield a redshift correction factor differing from the theoretical prediction.
Additionally, \citet{2001ASPC..224..149O} reported that observed dust lags of certain AGNs are notably smaller than theoretical predictions. This discrepancy is attributed to the presence of thin ring or disk-like dust distributions, diverging from the spherically symmetric distribution of dust particles assumed in the theoretical framework. For empirical constraints, \citet{2019ApJ...886..150M} determined $\gamma$ as 1.18 based on the $K$- and $H$- band lag measurements in six nearby Seyfert galaxies. In the case of W1 and W2-band lag measurements, \citet{2023MNRAS.tmp.1073C} reported a $\gamma$ $\sim$ 0.76 using a sample of 78 AGNs. These results indicate that the power-law index $\gamma$ of the redshift correction factor may vary depending on the observed wavelength for dust lag measurements \citep{2023MNRAS.tmp.1073C}.

Combined with the cosmological time dilation correction factor of (1+$z$)$^{-1}$, the total correction factor for the observed-frame time lag becomes approximately (1+$z$)$^{\gamma-1}$ $\sim$ (1+$z$)$^{-0.38}$. We uniformly applied this total correction factor to the measured time lag in the observed-frame, thereby obtaining the rest-frame time lags for the total sample.

\begin{figure*}
\centering
\includegraphics[scale=0.65]{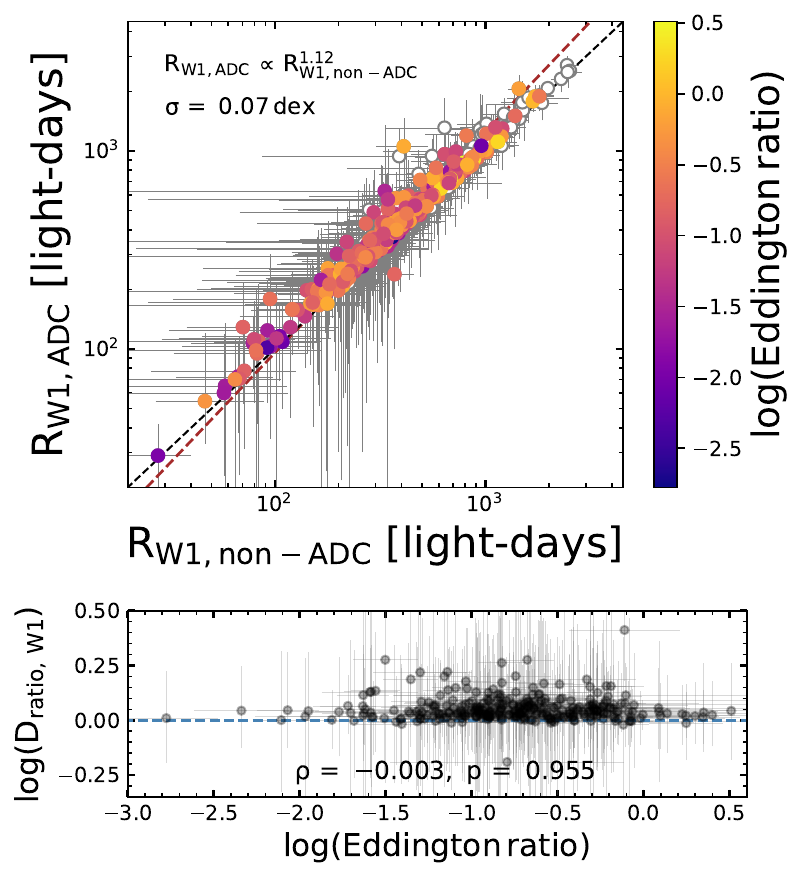}
\includegraphics[scale=0.65]{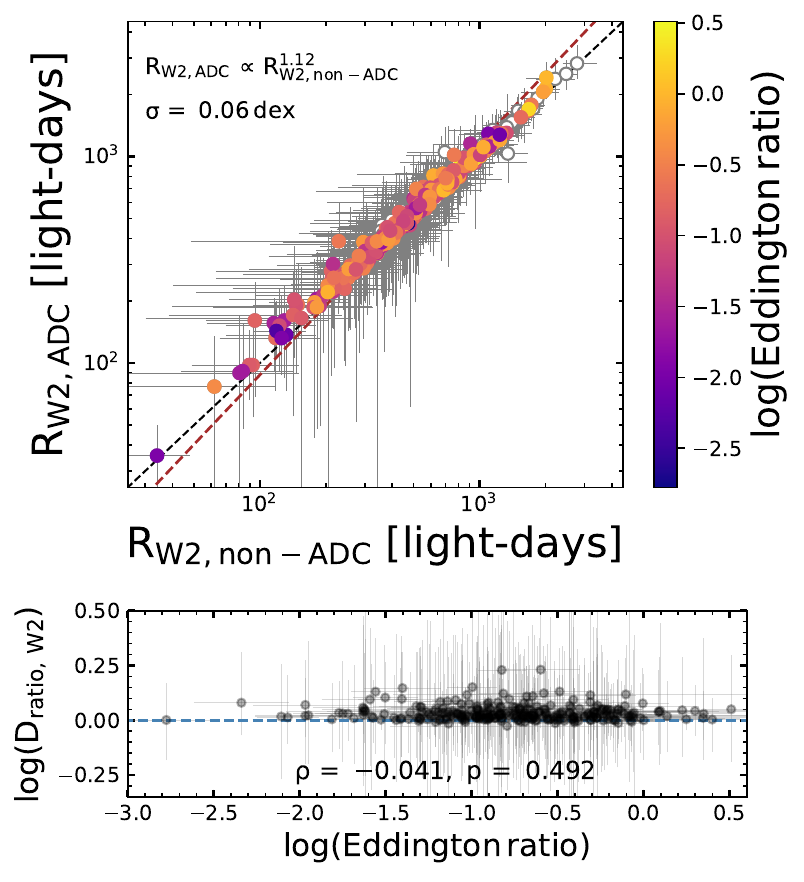}

\caption{Comparisons of torus sizes with and without AD contamination correction are presented in W1 (left) and W2- (right) bands, color-coded by Eddington ratio. The AGNs which do not have Eddington ratio measurements are shown by the open grey circles. The best fits to all the data points are shown by the dashed brown lines, while the dashed black lines represent the 1:1 relation. The intrinsic scatter from the best fit is also mentioned at each panel.  In the bottom panels, we represent the logarithm of the size ratio as a function of Eddington ratio.}

\label{fig:comp_ad}
\end{figure*}

\subsection{Correction for the accretion disk (AD) contamination in the IR W1 and W2- band light curves} \label{sec:AD_cor}

The observed IR fluxes in W1 and W2- bands contain contribution from the torus as well as from the AD \citep{2006ApJ...652L..13T, 2008Natur.454..492K, 2011MNRAS.415.1290L}, the latter of which makes the delay time between the optical and IR light curves shorter than the actual lag. Thus, the AD contamination needs to be removed from the observed IR fluxes to get the actual time lag. The AD emission in the IR can be estimated by considering a power-law spectrum of the accretion disk \citep{2014ApJ...788..159K} as follows

\begin{equation}
F_{IR}^{AD}(t) = F_{OP}(t) \left(\frac{\nu_{IR}}{\nu_{OP}}\right)^{a}
\end{equation}   
where, $F_{IR}^{AD}(t)$ and $F_{OP}(t)$ are the AD component of the IR flux and the optical flux in V- band ($g$- band for Sample 2) at time t, respectively. $\nu_{OP}$ and 
$\nu_{IR}$ are the effective frequencies of optical in V- band ($g$- band for Sample 2) and  IR (W1, W2) bands, respectively, and $a$ is the power-law index. We assumed a constant power-law index of $a = 1/3$ as suggested by the standard disk model \citep{1973A&A....24..337S} to calculate $F_{IR}^{AD}(t)$. Note that the polarized light spectra in the NIR strongly suggest intrinsic AD spectra with $a = 1/3$ \citep{2008Natur.454..492K}. In contrast, \citet{2014ApJ...788..159K} utilized not only $a = 1/3$ but also $a = 0.0$, taking into account the potential maximal contribution of AD contamination in the IR bands, while \citet{2019ApJ...886..150M} used $a = 0.1$ following \citet{2014ApJ...784L..11Y}. If $a$ is smaller than $a=1/3$, the time lag generally becomes larger, because the AD contamination in the IR-band flux is larger (i.e., a larger correction). To empirically test the effect, we performed the ADC with $a = 0.0$ using a representative subsample of 35
AGNs, which covers the entire range of luminosity and Eddington ratio of the total sample. We obtained insignificant effect, with a 0.03 dex increase in both W1 and W2 lag measurements \citep[see also][]{2014ApJ...788..159K, 2019ApJ...886..150M}. 
%Consequently, we adopt $a = 1/3$ for our subsequent analysis.}

We estimated  $F_{IR}^{AD}(t)$ for each epoch of IR (W1 and W2) observations using the average flux values of the quasi-simultaneous epochs (within $\pm$ 5 days) in the optical light curve. If for any epoch in IR light curve, there is no quasi-simultaneous observations in the optical, we obtained the corresponding optical flux from the DRW modeled optical light curve using {\tt JAVELIN}. It is worth noting that the influence of host galaxy flux contamination in the optical band is more evident in lower redshift AGNs. This may introduce a spurious redshift dependence in the dust lags. However, the effect of the host galaxy contribution to the ADC would be negligible since host galaxy flux remains constant and the host-galaxy contribution to the total optical flux would not affect the lag measurements while host galaxy contribution reduces the variability amplitude. Finally, $F_{IR}^{AD}(t)$ was subtracted from the observed IR (W1, W2) flux ($F_{IR,obs}$) to get the IR flux, which is coming from the torus ($F_{IR,dust}$) as

\begin{equation}
  F_{IR,dust}(t) = F_{IR,obs}(t) - F_{IR}^{AD}(t)
\end{equation}

Following this analysis, our next objective is to investigate how the AD contamination corrected and uncorrected dust lags are influenced by AGN luminosity and the Eddington ratio for the selected AGNs in our study. The AGN luminosity can be contaminated by stellar-light from the host galaxies. \citet{2019ApJS..243...21L} modeled and subtracted the stellar continua from the spectra and derived the continuum luminosity at rest-frame 5100 {\AA} ($L_{5100 {\AA}}$). We used their estimated $L_{5100 {\AA}}$ for Sample 1 ($z<$ 0.33) and obtained the AGN bolometric luminosity ($L_{bol}$) by using a bolometric correction factor of 9.26 \citep{2006ApJS..166..470R}.  The AGN bolometric luminosities for Sample 2 were collected from \citet{2020ApJ...900...58Y}. The same bolometric correction factor was used for Sample 2. In order to obtain the Eddington ratios for our own sample, we retrieved FWHM of $H\beta$ values from \citet{2019ApJS..243...21L} and \citet{2011ApJS..194...45S}. Subsequently, the $H\beta$ BLR size ($R_{H\beta}$) were estimated by utilizing the $R_{H\beta}$--$L_{5100 {\AA}}$ relation from \citet{2024ApJ...962...67W}. Then, we employed the virial equation \citep{2002ApJ...579..530W}, incorporating the FWHM of $H\beta$ and a virial factor of $f_{BLR} = 1.12$ \citep{2015ApJ...801...38W}, to derive the single epoch $M_{BH}$. Finally, we obtained the Eddington ratio ($L_{bol}/L_{Edd}$) using Eddington luminosity $L_{Edd} = 1.26 \times 10^{38} M_{BH}/M_{\odot}$. These Eddington ratios were determined for approximately $\sim$72$\%$ of the AGNs in our final sample, specifically those with available FWHM values for $H\beta$.

In Figure \ref{fig:comp_ad}, we present a comparison between the redshift-corrected rest-frame torus sizes, with and without AD contamination correction, denoted as $R_{W1,ADC}$ ($R_{W2,ADC}$) and $R_{W1,non-ADC}$ ($R_{W2,non-ADC}$),  respectively, for the W1- (W2) band. A linear fit to the torus sizes obtained with/without correction for AD contamination results

\begin{equation}
\begin{split}
\label{adeq1}
&\parbox{3\linewidth}{%
 \text{log}($R_{W1, \text{ADC}}/400 \, \text{light-day}$) =
 ($1.12 \pm 0.02$) \text{log}($R_{W1, \text{non-ADC}}$  \\ /
  400 \, \text{light-day})+ ($0.05 \pm 0.01$) }
\end{split}
\end{equation}
and
\begin{equation}
\begin{split}
\label{adeq2}
&\parbox{3\linewidth}{%
 \text{log}($R_{W2, \text{ADC}}$/400\,\text{light-day}) =
 ($1.12 \pm 0.02$) \text{log}($R_{W2, \text{non-ADC}}$ \\  /  
  400 \, \text{light-day})+ ($0.02 \pm 0.01$) }
\end{split}
\end{equation}

While \citet{2019ApJ...886...33L} performed a correction for the AD contamination in the IR light curves, \citet{2020ApJ...900...58Y} and \citet{2023MNRAS.tmp.1073C} did not take this corrective approach into account. Consequently, in our study, we implemented this correction for the W1 and W2- band light curves. The key findings from this analysis are as follows: (1) correction of AD contamination leads to larger torus sizes in AGNs, (2) because of the positive correlation between torus size and AGN luminosity (see Section \ref{result} for details), slopes in Equations \ref{adeq1} and \ref{adeq2} greater than unity suggest that AD contamination becomes more prominent as luminosity increases, (3) the size ratio $\mathrm{D_{ratio}}$ ($\mathrm{R_{IR, ADC}/R_{IR, non-ADC}}$) between the size with/without the AD contamination correction shows a negligible correlation with the Eddington ratio: Spearman's rank correlation coefficient $\mathrm{\rho = -0.003}$ and a p-value of 0.955 ($\mathrm{\rho = -0.041}$, p=0.492) based on W1 (W2)- band lags, implying no significant dependence on Eddington ratio, and (4) intercept values for the W1- band are larger than those for the W2- band, exhibiting a similar slope of $\sim$ 1.12. This indicates that AD contamination is more significant in the W1- band, aligning with expectation as AD contaminates more at shorter IR wavelengths. All of these findings indicate that it is necessary to 
correct for AD contribution to the observed IR fluxes.

We compare our obtained redshift and AD contamination corrected torus sizes with that available in the literature in Section \ref{lit_com}, finding good agreement with \citet{2019ApJ...886...33L}, \citet{2023MNRAS.tmp.1073C}, and \citet{2020ApJ...900...58Y} (See Figure \ref{fig:com_TL}). Table \ref{tab:lag}  describes the details of the finally selected targets and results of their lag-analysis.

\section{Results}\label{result}
\subsection{The torus size and AGN luminosity correlation} \label{lm_result}

In this section, we investigate the correlation between the measured torus size based on W1 and W2- band flux variations with AGN luminosity. 
First we will use bolometric luminosity of each AGN, and then we will also investigate the relation using the torus luminosity measured at W2- band. Note that we use the W1 and W2- band lag measurements after correcting for AD contamination and redshift effect as described in Sections \ref{sec:AD_cor}, and \ref{cor_redshift}, respectively.

We use the following linear equation to parameterize the $R_{dust,IR}$--$L_{bol}$ relationship \\

\begin{equation}
    \text{log}(R_{dust,IR}/1 \, \text{light-day}) = \beta \, + \,  \alpha \, \text{log}\left(\frac{L_{AGN}}{L_{0}}\right)
    \label{eq_rL}
\end{equation}
where $\alpha$ and $\beta$ are the slope and intercept, respectively.
$L_{0}$ is the reference point, which is set close to the median of AGN luminosity $L_{AGN}$ used in the fitting. In fitting Equation \ref{eq_rL}, we employ the Bivariate Correlated Errors and intrinsic Scatter \citep[{\tt BCES};][]{1996ApJ...470..706A}. This method is particularly advantageous as it accommodates measurement errors with heterogeneity in variance, incorporates intrinsic scatter, and accounts for correlations within the measurement errors.

\begin{figure*}
\centering
\includegraphics[scale=0.6]{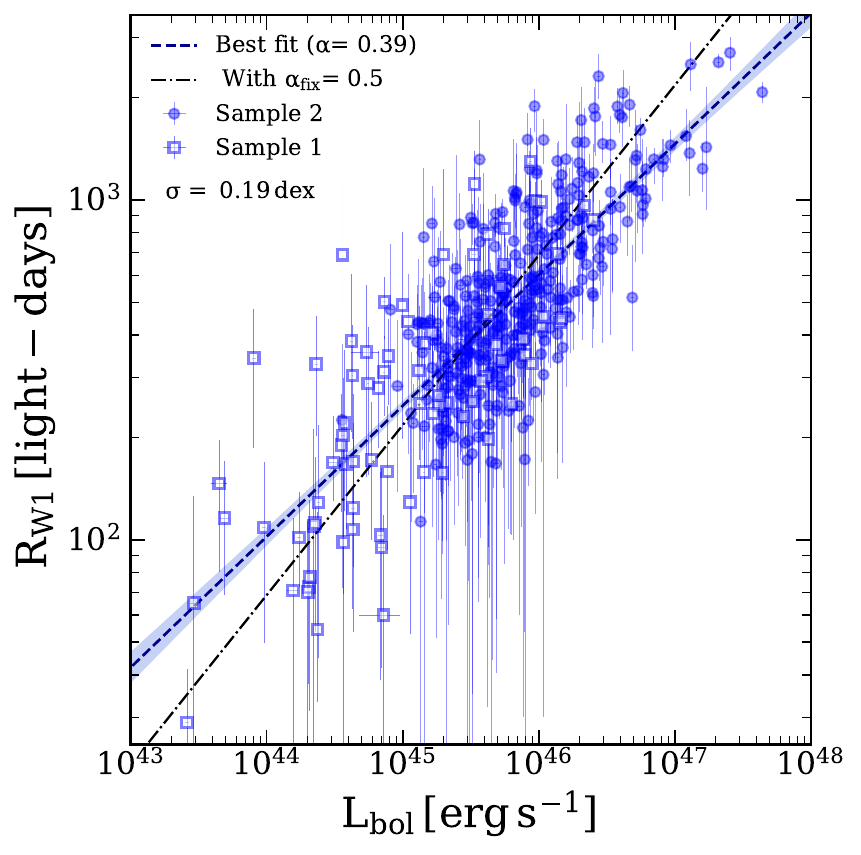}
\includegraphics[scale=0.6]{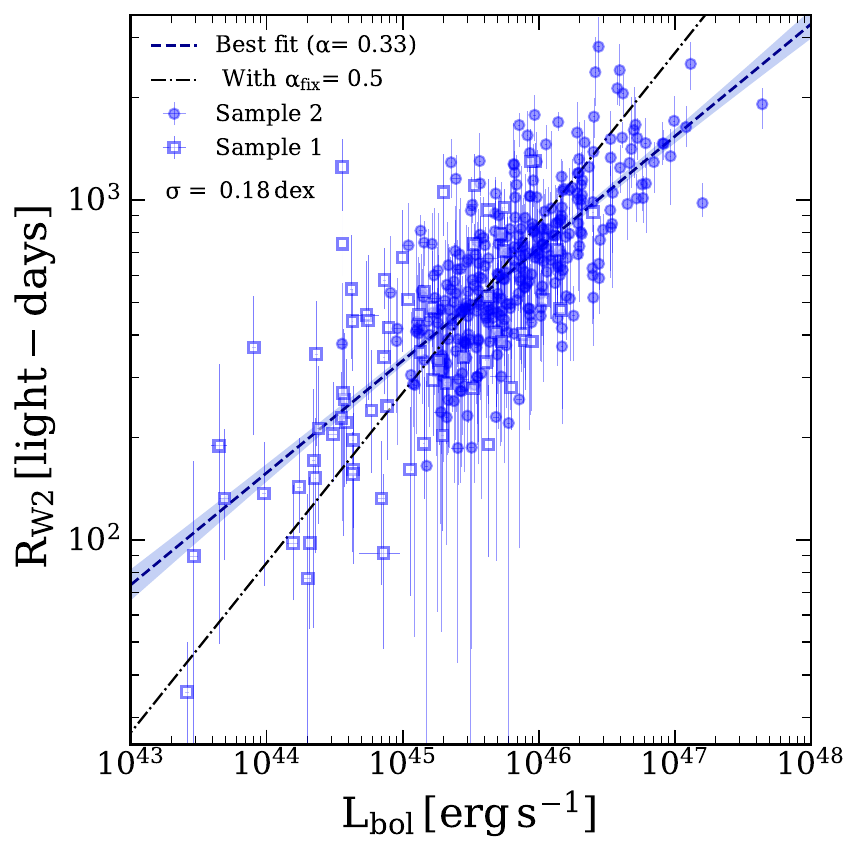}

\caption{Correlation between the torus size and AGN bolometric luminosity, measured in W1 (left) and W2- (right) bands after correcting for the AD contamination. The dashed blue line represents the best fit to the data points and the shaded blue region is the range driven by the uncertainties on the best fit. The intrinsic scatter is also shown at each panel in the figure. The dotted-dashed black colored line shows the best-fits for $\alpha_{fix}=0.5$.}
\label{fig:RW1_Lacn}
\end{figure*}

\subsubsection{The torus size correlation with AGN bolometric luminosity}

In Figure \ref{fig:RW1_Lacn} we present the best-fit relation between torus size and AGN bolometric luminosity ($L_{bol}$) for W1(left) and W2-(right) bands, where $L_{0}$ is set to $10^{45.7} \, erg \, s^{-1}$. The torus size scales with the bolometric luminosity as, $R_{dust,W1} \, \propto \, L_{bol}^{0.39}$, and $R_{dust,W2} \, \propto \, L_{bol}^{0.33}$ with an intrinsic scatter of 0.19 and 0.18 dex, in W1- and W2- bands, respectively. 
%As the dust radiation equilibrium theory \citep{1987ApJ...320..537B} predicts a slope of 0.5, we also fitted the data by fixing the slope value at $\alpha_{fix}$=0.5.
Note that if we use the lag measurements without correcting for AD contamination, we obtain similar slopes of  $\alpha$ =0.38 and 0.35  within the uncertainties for W1 and W2- bands, respectively, but with a smaller intercept ($\beta$) values and slightly larger intrinsic scatters. This result indicates that the presence of AD contamination in the observed IR light curves can cause an underestimation of the torus size. 
Therefore, correcting for AD contamination is essential to accurately recover the true torus sizes in AGNs.

To test the potential difference of the results due to the fitting method \citep{2012ApJS..203....6P}, we employ an additional approach: Bayesian linear regression using the {\tt LINMIX$\textunderscore$ERR} technique \citep{2007ApJ...665.1489K}. This method excels in handling intrinsic scatter within the relationship, addressing measurement errors in both the independent and dependent variables, and effectively managing correlations among the measurement errors. We find that these different fitting methods yield consistent results with a similar slope and intercept within the uncertainties. We also note that two different lag analysis methods, i.e., ICCF and {\tt JAVELIN}, provide consistent results.

 Furthermore, depending on the scatter in the observed $R_{dust,IR}$--$L_{bol}$ relation and the varied uncertainties in luminosities and torus sizes, we perform inverse linear regression fitting, i.e., AGN bolometric luminosity as a function of torus sizes. The inverse linear fitting results indicate a relationship: $L_{bol} \, \propto \, R_{W1}^{1.91\pm0.08}$, and $L_{bol} \, \propto \, R_{W2}^{2.04\pm0.11}$ derived from the AD contamination corrected W1 and W2- band lags, respectively, with a considerably larger intrinsic scatter of $\sigma \sim 0.4$ dex. From these findings, we derive dependencies of torus sizes on $L_{bol}$ as $R_{W1} \, \propto \, L_{bol}^{0.52}$ and $R_{W2} \, \propto \, L_{bol}^{0.49}$, closely aligning with the expected value of 0.5 from the dust-sublimation equilibrium model. It is important to note that the determination of the dust torus size occurs at a distance from the accretion disk, where dust particles evaporate because of the central ionizing luminosity. Hence, the AGN luminosity determines the torus size in AGN. Consequently, in alignment with our scientific objectives, we assert that our choice of the forward linear regression fitting between torus size as a function of AGN luminosity is more appropriate and is consequently employed for further analysis (refer to \citet{1990ApJ...364..104I} and \citet{1992ApJ...397...55F} for the choice of fitting method).

\begin{figure*}
\centering
\includegraphics[scale=0.6]{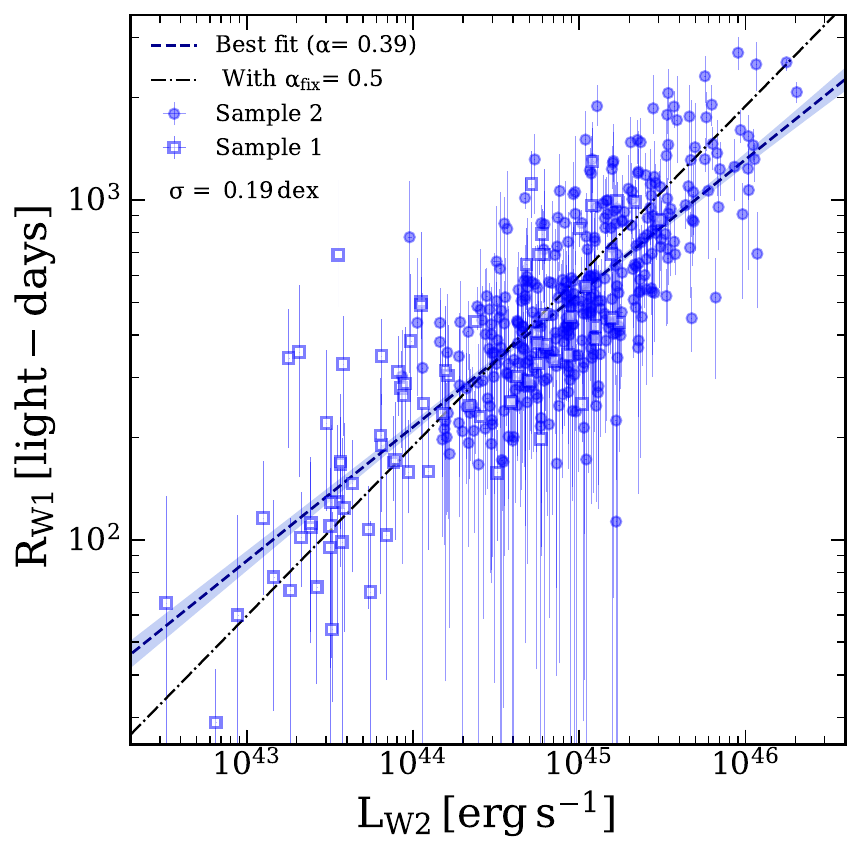}
\includegraphics[scale=0.6]{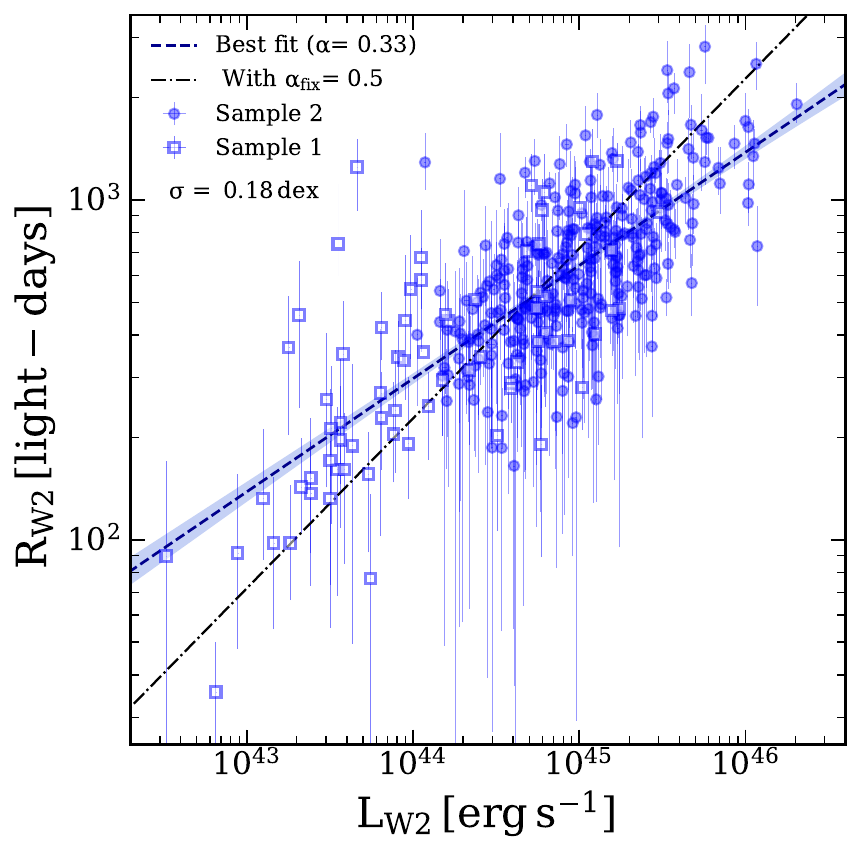}

\caption{Torus sizes in W1 and W2- bands after correcting for the AD contamination plotted against the AGN IR luminosity in W2- band in the left and right panels, respectively. The dashed blue line represents the best fit to the data points and the shaded blue region is the range driven by the uncertainties on the best fit. The intrinsic scatter is also shown at each panel in the figure. The dotted-dashed black colored line shows the best-fits for $\alpha_{fix}=0.5$.}
\label{fig:RW1_Liracn}
\end{figure*}

The slopes ($\alpha$) we obtained in our analysis exhibit values ranging from 0.33 to 0.39. These slopes are notably shallower than the expected value of 0.5 predicted by the dust sublimation model. In contrast, \citet{2019ApJ...886...33L} and \citet{2023arXiv230212437L} reported $R_{dust,W1} \, \propto \, L_{bol}^{0.47}$ ($R_{dust,W2} \, \propto \, L_{bol}^{0.45}$), and $R_{dust,W1} \, \propto \, L_{bol}^{0.51}$, respectively,  in their findings,  which differ considerably from our results. However, it is noteworthy that \citet{2023MNRAS.tmp.1073C} reported a value of $\alpha = 0.37\pm 0.03$ for both W1- and W2- bands based on the study of DRM in AGNs with $H\beta$ RM data. Our obtained slopes align with their findings within $1\sigma$ uncertainty. Furthermore, \citet{2019ApJ...886..150M} also reported a consistent slope of 0.42 from DRM based on K- band dust lags. On the other hand, recent work by \citet{2023A&A...669A..14G} mentioned a much shallower slope of approximately $\alpha = 0.4$ from optical/NIR interferometry. Consequently, the observation of smaller slope values in the relation between torus size and luminosity in DRM is consistent with these findings and is not unexpected.

%This discrepancy in slope values can be attributed to various factors. According to \citet{2011A&A...536A..78K}, higher luminosity objects tend to be more compact, and the physical size of IR emission does not vary with luminosity as $L^{0.5}$, but instead increases much more slowly than predicted. Their IR interferometric observations led to the conclusion that the half-light radius ($R_{1/2}$), which serves as a representative size of the torus, varies with luminosity as $R_{1/2} , \propto , L^{0.21}$ at 8.2 $\mu m$ and becomes nearly constant at 13 $\mu m$. 

Our results indicate that the slope is likely shallower than the expected value of 0.5 from the dust radiation equilibrium model. This deviation could be attributed to the possibility that the optical luminosity at 5100 {\AA}, which is used to obtain AGN bolometric luminosity, may not accurately represent the ionizing luminosity \citep{2019ApJ...886...42D}.

\subsubsection{The torus size correlation with AGN IR luminosity}

To further investigate the potential discrepancy in the observed $R_{dust,IR}$--$L_{bol}$ relationship compared to the expected relation, we obtain a torus size--luminosity relationship ($R_{dust,IR}$--$L_{W2}$) based on the luminosity of the infrared continuum emission in the W2- band, as it serves as luminosity indicators for AGNs, less obscured by dust. Thus, these relations can be applicable for obscured AGNs. We employ Equation \ref{eq_rL} to characterize the relationship between $R_{dust,IR}$ and $L_{W2}$, where $L_{0}$ is set to $10^{45.0} \, erg \, s^{-1}$. $L_{W2}$ is luminosity measured with W2- band, which was derived from the average flux values of the W2- band light curves studied in this work, without correction for host contamination or galactic extinction.   In Figure \ref{fig:RW1_Liracn}, we present the $R_{dust,IR}$--$L_{W2}$ relationships for AD contamination corrected W1 and W2- band light curves in the left and right panels, respectively. All the fitting results are summarized in Table \ref{tab:fit_result}.

Interestingly, we observe similar slopes of $\sim$ 0.39 (W1) and 0.33 (W2) in the $R_{dust,IR}$--$L_{W2}$ relations as those obtained from $R_{dust,IR}$--$L_{bol}$ within the uncertainties,  irrespective of whether we apply corrections for AD contamination or not. Our obtained slope differs considerably from the value of 0.5 reported in the torus size -- IR luminosity relation at 12 $\mu m$, as per the K- band DRM measurements involving only 10 AGNs by \citet{2019ApJ...886..150M}. This discrepancy is likely due to the limited quantity of data and luminosity range in their study.

The similarity in shallower slopes observed in the torus size -- luminosity relations using both optical and IR luminosities can be attributed to the self-shadowing effect of the slim disk model  \citep{2011ApJ...737..105K, 2014ApJ...797...65W, 2018ApJ...856....6D}. Since dust particles within the torus reach thermal equilibrium with the IR radiation re-emitted by dust exposed to ionizing UV radiation, we would expect a slope of 0.5 in the $R_{dust,IR}$--$L_{W2}$ relationship \citep{1987ApJ...320..537B}. However, the self-shadowing effect of slim accretion disks leads to a contraction of the ionization front of the BLR and the torus, resulting in a reduction in torus size with increasing accretion rate. This effect yields the shallower slopes observed in both the $R_{dust,IR}$--$L_{bol}$ and $R_{dust,IR}$--$L_{W2}$ relationships. For further details, please refer to Section \ref{sec:scatter}.

\begin{table*}
\centering
\movetableright= -28mm
\caption{Linear regression fitting results of the torus size -- luminosity relation.}
\label{tab:fit_result}
\resizebox{19cm}{!}{
\begin{tabular}{lccccccccr} \hline

 & Fitting method:  & {\tt BCES} & \multicolumn{3}{c}{ICCF}  & & \multicolumn{3}{c}{{\tt JAVELIN}}  \\

$L_{AGN}$ & Case & lag  & $\alpha$  & $\beta$  & $\sigma$ &  $\beta_{fix}$($\alpha_{fix}$=0.5) &  $\alpha$  & $\beta$ & $\sigma$   \\

(1) & (2) & (3) & (4) & (5) & (6) & (7) & (8)  & (9) & (10)
\\ \hline

$L_{bol}$ & NON ADC & optical--W1   & 0.38 $\pm$ 0.02 &  2.61 $\pm$ 0.01 & 0.20 & 2.66 $\pm$ 0.01   & 0.34 $\pm$ 0.02  & 2.62 $\pm$ 0.01 & 0.20 \\

 &  & optical--W2  & 0.35 $\pm$ 0.02 & 2.72 $\pm$ 0.01 & 0.20 & 2.76 $\pm$ 0.01  & 0.30 $\pm$ 0.02  & 2.76 $\pm$ 0.01 & 0.19 \\
\hline

$L_{bol}$ & ADC & optical--W1  & 0.39 $\pm$ 0.02 & 2.66 $\pm$ 0.01 & 0.19 & 2.69 $\pm$ 0.01  & 0.35 $\pm$ 0.02  & 2.69 $\pm$ 0.01 & 0.19  \\
 &  & optical--W2   & 0.33 $\pm$ 0.02 & 2.76 $\pm$ 0.01 & 0.18 & 2.78 $\pm$ 0.01  & 0.30 $\pm$ 0.02  & 2.78 $\pm$ 0.01 & 0.19  \\ \\

\hline

 & Fitting method:  & {\tt LINMIX} & \multicolumn{3}{c}{ICCF}  & & \multicolumn{3}{c}{{\tt JAVELIN}}  \\

$L_{AGN}$ & Case & lag  & $\alpha$  & $\beta$  & $\sigma$ &  $\beta_{fix}$($\alpha_{fix}$=0.5) &  $\alpha$  & $\beta$ & $\sigma$   \\

%(1) & (2) & (3) & (4) & (5) & (6) & (7) & (8)
\\ \hline

$L_{bol}$ & NON ADC & optical--W1   & 0.37 $\pm$ 0.02 &  2.63 $\pm$ 0.01 & 0.20 & 2.68 $\pm$ 0.01   & 0.35 $\pm$ 0.02 & 2.62 $\pm$ 0.01 & 0.20 \\

  &  & optical--W2   & 0.34 $\pm$ 0.02 & 2.74 $\pm$ 0.01 & 0.20 & 2.78 $\pm$ 0.01  & 0.30 $\pm$ 0.02  & 2.76 $\pm$ 0.01 & 0.19 \\
\hline
$L_{bol}$ & ADC & optical--W1   & 0.38 $\pm$ 0.02 & 2.69 $\pm$ 0.01 & 0.19 & 2.72 $\pm$ 0.01  & 0.35 $\pm$ 0.02  & 2.70 $\pm$ 0.01 & 0.19 \\
&  & optical--W2  & 0.32 $\pm$ 0.02 & 2.78 $\pm$ 0.01 & 0.18 & 2.81 $\pm$ 0.01  & 0.30 $\pm$ 0.02  & 2.79 $\pm$ 0.01 & 0.19 \\ \\

\hline

 & Fitting method:  & {\tt BCES} & \multicolumn{3}{c}{ICCF}  & & \multicolumn{3}{c}{{\tt JAVELIN}}  \\

$L_{AGN}$ & Case & lag  & $\alpha$  & $\beta$  & $\sigma$ &  $\beta_{fix}$($\alpha_{fix}$=0.5) &  $\alpha$  & $\beta$ & $\sigma$   \\

%(1) & (2) & (3) & (4) & (5) & (6) & (7) & (8)  & (9) & (10)
\\ \hline

$L_{W2}$ & NON ADC & optical--W1   & 0.40 $\pm$ 0.02 &  2.67 $\pm$ 0.01 & 0.20 & 2.75 $\pm$ 0.01   & 0.36 $\pm$ 0.02  & 2.67 $\pm$ 0.01 & 0.19 \\

 &  & optical--W2  & 0.36 $\pm$ 0.02 & 2.77 $\pm$ 0.01 & 0.19 & 2.83 $\pm$ 0.01  & 0.30 $\pm$ 0.02  & 2.80 $\pm$ 0.01 & 0.20 \\
\hline

$L_{W2}$ & ADC & optical--W1  & 0.39 $\pm$ 0.02 & 2.73 $\pm$ 0.01 & 0.19 & 2.78 $\pm$ 0.01  & 0.36 $\pm$ 0.02  & 2.75 $\pm$ 0.01 & 0.19  \\
 &  & optical--W2   & 0.33 $\pm$ 0.02 & 2.81 $\pm$ 0.01 & 0.18 & 2.86 $\pm$ 0.01  & 0.30 $\pm$ 0.02  & 2.83 $\pm$ 0.01 & 0.19  \\ \\

\hline

 & Fitting method:  & {\tt LINMIX} & \multicolumn{3}{c}{ICCF}  & & \multicolumn{3}{c}{{\tt JAVELIN}}  \\

$L_{AGN}$ & Case & lag  & $\alpha$  & $\beta$  & $\sigma$ &  $\beta_{fix}$($\alpha_{fix}$=0.5) &  $\alpha$  & $\beta$ & $\sigma$   \\

%(1) & (2) & (3) & (4) & (5) & (6) & (7) & (8)
\\ \hline

$L_{W2}$ & NON ADC & optical--W1   & 0.40 $\pm$ 0.02 &  2.69 $\pm$ 0.01 & 0.20 & 2.79 $\pm$ 0.01   & 0.36 $\pm$ 0.02 & 2.68 $\pm$ 0.01 & 0.19 \\

  &  & optical--W2   & 0.35 $\pm$ 0.02 & 2.79 $\pm$ 0.01 & 0.19 & 2.89 $\pm$ 0.01  & 0.30 $\pm$ 0.02  & 2.81 $\pm$ 0.01 & 0.20 \\
  
\hline

$L_{W2}$ & ADC & optical--W1   & 0.40 $\pm$ 0.02 & 2.75 $\pm$ 0.01 & 0.19 & 2.86 $\pm$ 0.01  & 0.36 $\pm$ 0.02  & 2.76 $\pm$ 0.01 & 0.19 \\
&  & optical--W2  & 0.33 $\pm$ 0.02 & 2.81 $\pm$ 0.01 & 0.18 & 2.93 $\pm$ 0.01  & 0.30 $\pm$ 0.02  & 2.83 $\pm$ 0.01 & 0.19 \\ \\

\hline

\end{tabular}
}

\vspace{0.18cm}

\raggedright Note: Columns are (1) luminosity used in fitting, (2) different cases, NON ADC: without AD contamination correction, ADC: correction for AD contamination applied, (3) name of the bands used to measure the lags, (4) slope of the best-fits with $\alpha$ as free parameter using lags from ICCF, (5)  intercept of the best-fits with $\alpha$ as free parameter using lags from ICCF, (6) intrinsic scatter in dex, (7)  intercept of the best-fits with $\alpha$ fixed to 0.5 using lags from ICCF, (8) slope of the best-fits with $\alpha$ as free parameter using lags from {\tt JAVELIN}, (9) intercept of the best-fits with $\alpha$ as free parameter using lags from {\tt JAVELIN}, and (10) intrinsic scatter in dex.

\end{table*}

\begin{figure}
\includegraphics[scale=0.6]{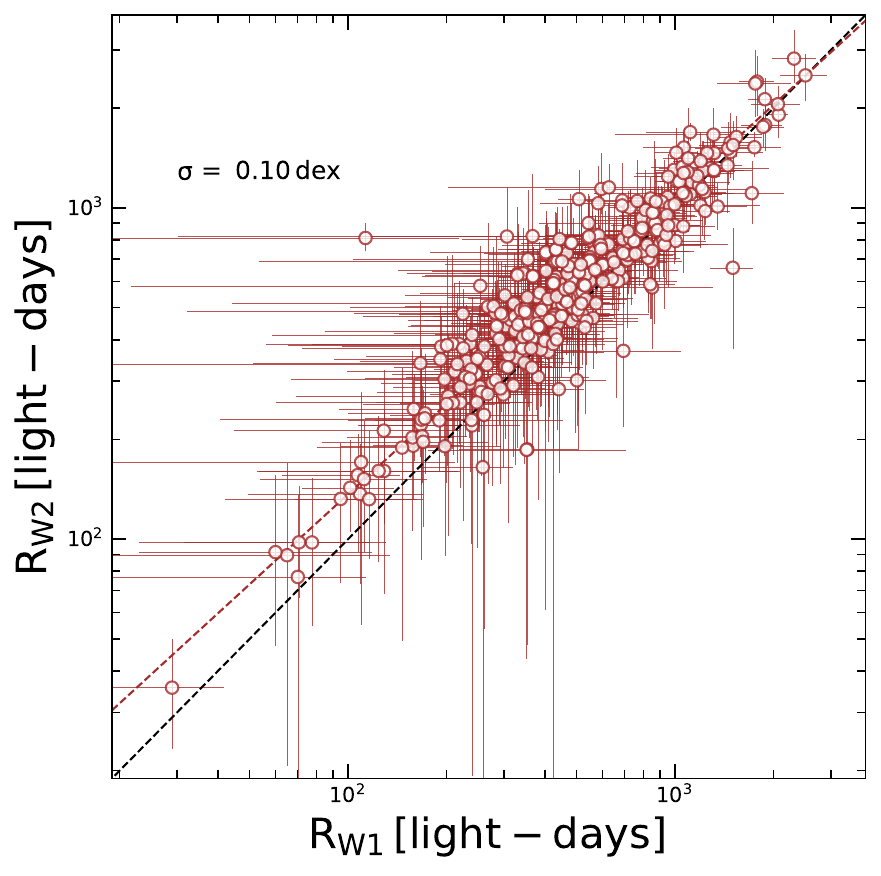}
\caption{Comparison between the torus sizes in W1 and W2- bands obtained from ICCF in the rest-frame of the targets based on their AD contamination corrected IR light curves.  The best-fit to the data points is shown by the dashed brown line, while the dashed black line represents the 1:1 relation.}
\label{fig:W1_W2_com}
\end{figure}

\begin{figure}
\centering
\includegraphics[scale=0.36]{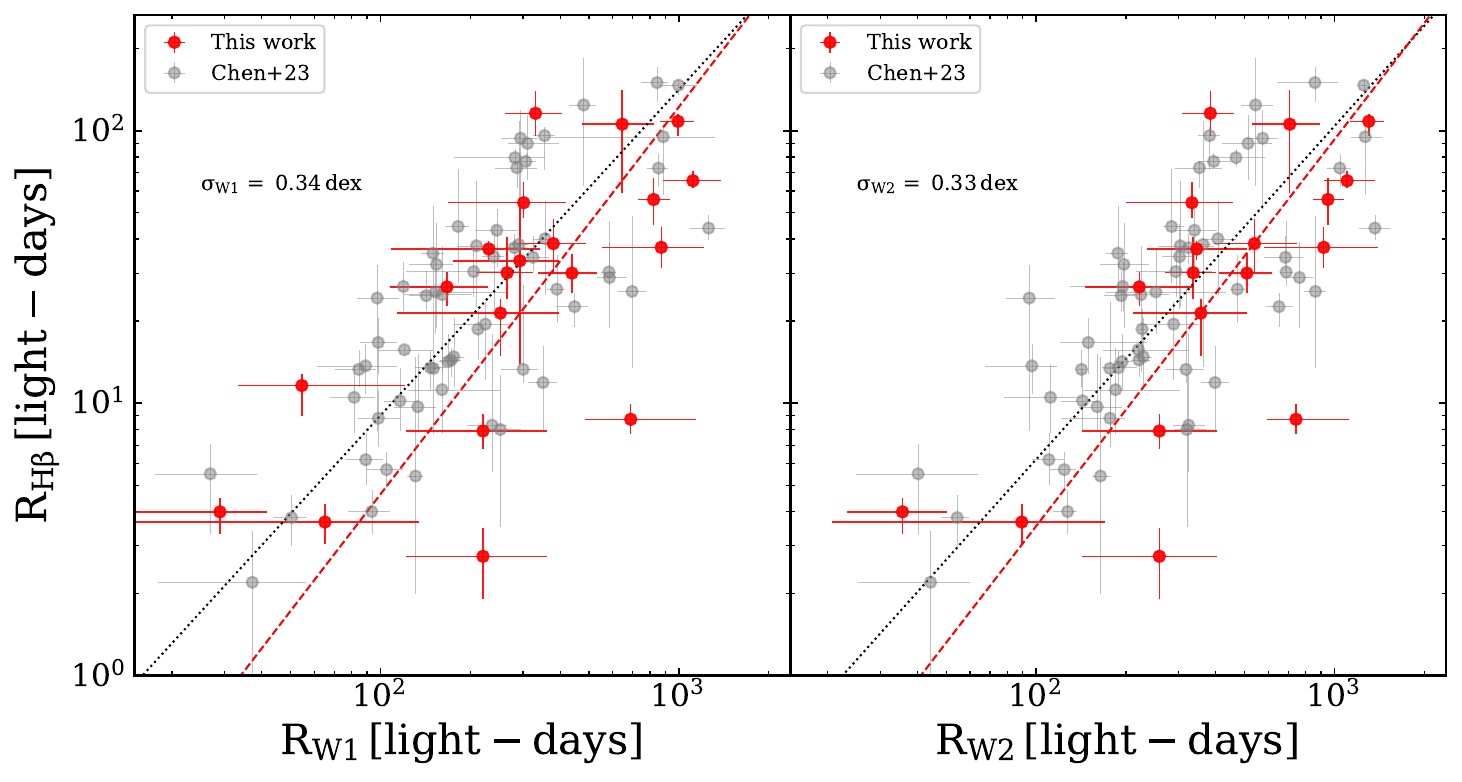}
%\resizebox{8.6cm}{5.0cm}{\includegraphics{MIR_RM_mod/FIG/com_BLR.pdf}}
\caption{Comparison between the BLR and torus sizes. The red circular points represent targets with torus sizes in W1 and W2- bands from this work. The dashed red lines are the linear fits between the BLR and the torus sizes in W1 and W2- bands, respectively. Measurements from \citet{2023MNRAS.tmp.1073C} are shown as grey points. The dotted black lines represent the best-fits for the total sample. The intrinsic scatter for the total sample is also mentioned at each panel.}
\label{fig:W_BLRcom}
\end{figure}

\section{Discussion}\label{discussion}

\subsection{Comparison between torus sizes in W1 and W2- bands}

In Figure \ref{fig:W1_W2_com}, we compare torus sizes in the W1 and W2- bands, revealing that $R_{dust, W2}$ exceeds that in the W1- band. This result aligns with our expectations, considering that lag generally increases with wavelength. Moreover, the mean torus size ratio between W2 and W1- bands is found to be $\sim$ 1.20. A linear fit to the torus sizes between W1 and W2 yields 

\begin{equation}
\begin{split}
\label{cmp1}
&\parbox{1.5\linewidth}{%
 \text{log}($R_{dust,W2}$/400 \, \text{light-day}) = ($0.90 \pm 0.02$) \, \text{log}($R_{dust,W1}$/ \\ 400 \, \text{light-day})  + ($0.09 \pm 0.01$)}
\end{split}
\end{equation}
and after correcting for the AD contamination, we found
\begin{equation}
\begin{split}
\label{cmp2}
&\parbox{1.5\linewidth}{%
 \text{log}($R_{dust,W2}$/400 \, \text{light-day}) = ($0.90 \pm 0.02$) \, \text{log}($R_{dust,W1}$/ \\ 400 \, \text{light-day})  + ($0.08 \pm 0.01$)}
\end{split}
\end{equation}
Indeed, the fitting results indicate that $R_{dust, W2}$ is statistically larger than that in the W1- band.

\begin{table*}
\caption{Details of BLR--RM sample}
\label{tab:BLR_RM}
\centering
\movetableright= -20mm
\begin{tabular}{lclcccr} \hline

Object name & ID no  & $R_{H\beta}$  & log$L_{5100 {\AA}}$  & $M_{BH}$  & Eddington ratio & Reference \\

 & &  (light-days) & ($\mathrm{erg \, s^{-1}}$) & ($\mathrm{\times 10^{7}M_{\odot}}$) & & \\

 (1) & (2) & (3) & (4) & (5) & (6) & (7) \\

\hline

PG 0923+201 & OB11 & $108.20^{+6.60}_{-12.30}$ & $45.31 \pm 0.06$ & $118.00^{+11.00}_{-16.00}$ & $-0.90 \pm 0.09$ & \citet{2021ApJ...920....9L} \\

PG 1001+291 & OB16 & $37.30^{+6.90}_{-6.00}$ & $45.59 \pm 0.03$ & $3.33^{+0.62}_{-0.54}$ & $0.93 \pm 0.11$  & \citet{2021ApJ...920....9L} \\

PG 1001+054 & OB17 & $65.50^{+5.60}_{-3.90}$ & $44.64 \pm 0.03^a$ & $10.73^{+1.15}_{-1.17}$ & $-0.52 \pm 0.07$ & \citet{2022ApJS..262...14B} \\

Mrk 142 & OB24 & $2.74^{+0.73}_{-0.83}$ & $43.54 \pm 0.04$ & $0.17^{+0.06}_{-0.07}$ & $0.18 \pm 0.24$  & \citet{2009ApJ...705..199B} \\
Mrk 142 & OB24 & $7.90^{+1.20}_{-1.10}$ & $43.57 \pm 0.05$ & $0.17^{+0.06}_{-0.07}$ & $0.21 \pm 0.24$  & \citet{2014ApJ...782...45D} \\

PG 1048+342 & OB27 & $36.80^{+2.40}_{-3.40}$ & $44.55 \pm 0.05^a$ & $4.44^{+0.31}_{-0.32}$ & $-0.23 \pm 0.07$ & \citet{2022ApJS..262...14B} \\

PG 1115+407 & OB31 & $54.40^{+10.20}_{-6.70}$ & $44.39 \pm 0.06$ & $7.76^{+2.23}_{-1.95}$ & $-0.63 \pm 0.18$ & \citet{2021ApJS..253...20H} \\

PG 1121+422 & OB32 & $115.80^{+24.20}_{-20.20}$ & $44.83 \pm 0.01$ & $10.10^{+3.00}_{-2.00}$ & $-0.31 \pm 0.16$ & \citet{2024ApJ...962...67W} \\

Mrk 40 & OB33 & $3.99^{+0.49}_{-0.68}$ & $42.48 \pm 0.11$ & $0.53^{+0.06}_{-0.07}$ & $-1.38 \pm 0.13$ & \citet{2009ApJ...705..199B} \\

Mrk 1310 & OB41 & $3.66^{+0.59}_{-0.61}$ & $42.23 \pm 0.17$ & $0.17^{+0.07}_{-0.07}$ & $-1.13 \pm 0.30$ & \citet{2009ApJ...705..199B} \\

PG 1202+281 & OB43 & $38.50^{+9.10}_{-8.50}$ & $44.43 \pm 0.01$ & $12.00^{+3.00}_{-3.00}$ & $-0.78 \pm 0.15$  & \citet{2024ApJ...962...67W} \\

PG 1307+085 & OB48 & $105.60^{+36.00}_{-46.60}$ & $44.79 \pm 0.02$ & $29.35^{+10.75}_{-13.53}$ & $-0.81 \pm 0.26$ & \citet{2000ApJ...533..631K} \\

PG 1404+226 & OB55 & $21.40^{+2.80}_{-6.50}$ & $44.11 \pm 0.03$ & $0.68^{+0.14}_{-0.23}$ & $0.14 \pm 0.17$ & \citet{2021ApJS..253...20H} \\

PG 1415+451 & OB56 & $30.00^{+5.40}_{-4.60}$ & $44.23 \pm 0.04$ & $1.75^{+0.36}_{-0.32}$ & $-0.15 \pm 0.13$ & \citet{2021ApJS..253...20H} \\

PG 1427+480 & OB59 & $33.30^{+20.60}_{-19.40}$ & $44.79 \pm 0.01$ & $5.00^{+3.00}_{-3.00}$ & $-0.04 \pm 0.37$ & \citet{2024ApJ...962...67W} \\

PG 1448+273 & OB62 & $30.10^{+10.70}_{-5.90}$ & $44.24 \pm 0.03$ & $1.01^{+0.38}_{-0.23}$ & $0.10 \pm 0.19$ & \citet{2021ApJS..253...20H} \\

Mrk 1392 & OB65 & $26.70^{+3.50}_{-3.90}$ & $43.20 \pm 0.16$ & $6.30^{+0.80}_{-0.90}$ & $-0.17 \pm 0.18$ & \citet{2022ApJ...925...52U} \\

%PG 1519+226 & OB82 & $73.10^{+4.00}_{-11.40}$ & $44.49 \pm 0.06$ & $4.87^{+0.49}_{-0.86}$ & $-0.33 \pm 0.11$  & \citet{2021ApJS..253...20H} \\

Mrk 290 & OB70 & $8.72^{+1.21}_{-1.02}$ & $43.11 \pm 0.06$ & $1.97^{+0.30}_{-0.26}$ & $-1.32 \pm 0.11$ & \citet{2010ApJ...721..715D} \\

Mrk 493 & OB75 & $11.60^{+1.20}_{-2.60}$ & $43.11 \pm 0.08$ & $0.15^{+0.04}_{-0.15}$ & $-0.20 \pm 0.46$  & \citet{2014ApJ...793..108W} \\

PG 2349-014 & OB83 & $55.90^{+11.00}_{-10.80}$ & $44.66 \pm 0.02$ & $63.00^{+14.00}_{-14.00}$ & $-1.27 \pm 0.14$ & \citet{2024ApJ...962...67W} \\

\hline

\end{tabular}

\vspace{0.15cm}

\raggedright Note: Columns are (1) object name, (2) object ID number used in this work, (3) $H\beta$ BLR size (rest-frame) obtained from BLR--RM, (4) host-corrected AGN optical luminosity at 5100 {\AA} collected from the literature with BLR--RM measurements, (5) $M_{BH}$ obtained from RM, (6) Eddington ratio in log scale derived from reverberation-mapped $M_{BH}$, and (7) references.  
$L_{5100{\AA}}$ marked with $a$ is not corrected for host-starlight contamination.
\end{table*}

\subsection{Comparison between torus and BLR sizes}

We have identified a total of 19 AGNs within our sample, all of which possess prior measurements of BLR size and $M_{BH}$ based on $H\beta$ line obtained through BLR--RM analysis available in the existing literature (See Table \ref{tab:BLR_RM}).

Figure \ref{fig:W_BLRcom} illustrates a comparative analysis between the torus sizes observed in the W1 and W2- bands and their corresponding  $H\beta$ BLR sizes ($R_{H\beta}$) as determined by BLR--RM in the left and right panel, respectively for those 19 AGNs as shown by the red points.   Notably, our examination reveals a discernible correlation between BLR sizes and our obtained torus sizes with a Spearman's rank correlation coefficient of $\mathrm{\rho = 0.73}$ with a p-value of $2.56\times10^{-4}$ for the W1- band, and $\mathrm{\rho = 0.71}$ with a p-value of $1.04\times10^{-3}$ for the W2- band. We employ the following linear equation to parameterize the relationship between $R_{H\beta}$ and $R_{dust, IR}$

\begin{equation}
%\begin{split}
\label{bl}
%&\parbox{1.0\linewidth}{%
\text{log}(R_{H\beta}/ 1 \, \text{light-day}) = \epsilon \, \text{log}(R_{dust, IR}/ 1 \, \text{light-day}) + K
%\end{split}
\end{equation}
where $\epsilon$ denotes the slope and $K$ is the intercept. We find that the BLR size scales with the torus size as, $R_{H\beta} \, \propto \, R_{dust, W1}^{1.42 \pm 0.32}$ with $K=-2.18 \pm 0.88$, $\sigma = 0.37$ dex, and $R_{H\beta} \, \propto \, R_{dust, W2}^{1.42 \pm 0.31}$ with $K=-2.29 \pm 0.87$, $\sigma = 0.39$ dex. The median size ratios between torus and BLR are approximately $\sim$ 9.5 and 15.1 for W1 and W2- bands, respectively. In addition, \citet{2014ApJ...788..159K} reported a significant four to five-fold disparity in the mean reverberation radii between the innermost dust torus based on K- band lag and the BLR.  A median size ratio between the torus and BLR, $R_{dust,K}/R_{H\beta}$, was determined to be approximately $\sim$6.5 based on 10 objects \citep{2015ApJ...812..113G}. \citet{2015ApJ...806...22D} reported $R_{dust,K}/R_{H\beta}$ $\sim$4 for 10 low-accreting AGNs, and $R_{dust,K}/R_{H\beta}$ $\sim$7 for high-accreting AGNs with a smaller sample size (4 objects). \citet{2020MNRAS.491.4615K} also reported $R_{dust,K}/R_{H\beta}$ $\sim$4 based on measurements from 15 Seyfert galaxies. Moreover, \citet{2018MNRAS.475.5330M} found about 11.5 times larger torus size (based on K-band) than BLR size in H0507+164. Recently, \citet{2023MNRAS.tmp.1073C} found $R_{dust,K}/R_{H\beta}$ $\sim$6.2, $R_{dust,W1}/R_{H\beta}$  $\sim$9.2, and $R_{dust,W2}/R_{H\beta}$  $\sim$11.2 based on a significantly larger sample of 78 AGNs with $H\beta$-RM measurements. Our findings indicate that the ratio of $R_{dust,W1}/R_{H\beta}$ is $\sim$ $0.98\pm0.36$ dex, suggesting a torus size roughly 9.5 times larger than that of the BLR, which closely aligns with the results of \citet{2023MNRAS.tmp.1073C}.  In contrast, we observe a slightly higher ratio of $R_{dust,W2}/R_{H\beta}$ at around $1.18\pm0.36$ dex, indicating a torus size approximately 15.1 times larger than the BLR size based on W2-band lags, derived from a sample of 19 AGNs.  The discrepancy observed in W2-band data may be attributed to the smaller number of AGNs with $H\beta$-RM data in our sample compared to their sample. Nonetheless, taking into account the rms scatter of approximately 0.36 dex, our measurements are consistent with previous findings reported in the literature. It is worth highlighting a consistent trend observed across these 19 AGNs: the torus sizes consistently surpass the corresponding BLR sizes.

Since, the above $R_{H\beta}$ -- $R_{dust, IR}$ relations are based on limited number of AGNs, to expand the sample size we incorporate measurements from \citet{2023MNRAS.tmp.1073C}, who reported torus sizes for 78 AGNs with $R_{H\beta}$ measurements. Their sample does not cover the S82 region, and only 28 among the 76 targets from their sample overlap with our initial Sample 1 before applying the lag-quality assessment. After implementing our lag-quality assessment (see Section \ref{qual}), only 12 targets remain, that overlap with their sample. Note that, \citet{2023MNRAS.tmp.1073C} did not correct for AD contamination in their measured lags. A direct comparison of the torus sizes based on W1- (W2) band lags for those 12 targets between our measurements and that from \citet{2023MNRAS.tmp.1073C} is discussed in Section \ref{lit_com}, which shows an offset of 0.16 (0.14) dex between the two sets of measurements. Additionally, we find that, among the remaining 66 (78-12) targets, 60 have torus size measurements with $r_{max} \, \geq \, 0.6$. Thus, we include these additional 60 targets from \citet{2023MNRAS.tmp.1073C} in the $R_{H\beta}$ -- $R_{dust, IR}$ plane represented by the grey points in Figure \ref{fig:W_BLRcom} and obtain the best fit relations using Equation \ref{bl}. We use same redshift correction factor of $(1+z)^{-0.38}$ to get the rest-frame torus size, while disregarding the extra uncertainties caused by the AD contamination correction which was not performed for the additional 60 targets. For the total BLR--RM sample (19+60),  $R_{H\beta}$ varies with $R_{dust, IR}$ as, $R_{H\beta} \, \propto \, R_{dust, W1}^{1.20 \pm 0.09}$ with $K=-1.44 \pm 0.23$, $\sigma = 0.34$ dex, and $R_{H\beta} \, \propto \, R_{dust, W2}^{1.22 \pm 0.08}$ with $K=-1.65 \pm 0.21$, $\sigma = 0.33$ dex.

While \citet{2023MNRAS.tmp.1073C} reported a linear relationship between $R_{H\beta}$ and $R_{dust, IR}$, our findings reveal a super-linear relationship characterized by a slope exceeding unity. This aligns with the findings of \citet{2023A&A...669A..14G}, who also identified a slope greater than unity of $\sim$ 1.12 and 1.10 in the $R_{H\beta}$ -- $R_{dust, IR}$ relationship using K-band torus sizes derived from RM and optical/NIR interferometry, respectively. According to the photoionization model, the relationship between $R_{H\beta}$ and $R_{dust, IR}$ can be expressed as \citep{2023MNRAS.tmp.1073C}:

\begin{equation}
    \frac{R_{H\beta}}{R_{dust, IR}} = \left(\frac{L_{ion}}{L_{OUV}}\right)^{1/2} \left(\frac{\sigma_{s}T_{IR}^4}{cn_{e}kT_{c}}\right)^{1/2}
\end{equation}
where, $L_{ion}$ and $L_{OUV}$ represent the ionizing and optical-UV luminosity, respectively. $\sigma_{s}$, $n_{e}$, $k$, $T_{c}$, and $T_{IR}$ denote the Stefan-Boltzmann constant, electron number density, Boltzmann constant, temperature of the ionized clouds, and temperature emitting IR photons, respectively. The deviation from a linear relationship between $R_{H\beta}$ and $R_{dust, IR}$ may occur for two primary reasons. First, within AGNs, BLR gas clouds could exhibit a range of $n_{e}$  \citep{1995ApJ...455L.119B}. Second, the optical luminosity measured at 5100 Å, utilized to derive $L_{bol}$, might not precisely represent the ionizing luminosity, implying that $L_{bol}$ is not directly proportional to $L_{ion}$ \citep[refer to the discussion by][]{2019ApJ...870...84C, 2020ApJ...899...73F}. Consequently, the observed super-linear relationship between $R_{H\beta}$ and $R_{dust, IR}$ is not unexpected.

Moreover, when considering the relationship between torus size in W1- (W2) band and AGN bolometric luminosity, we find that the BLR size scales with $L_{bol}$ as $R_{H\beta} \, \propto \, L_{bol}^{0.47}$ ($R_{H\beta} \, \propto \, L_{bol}^{0.40}$). Thus, relying on torus sizes in W1 and W2- bands, the slope of $R_{H\beta}$--$L_{bol}$ relation ranges between 0.47 and 0.40 with a mean value of $\sim 0.43$, which is  smaller than the expected value of 0.5 from photoionization model \citep{1972ApJ...171..213D}. However, it closely aligns with the observed slope of 0.402 obtained from BLR--RM, as reported by \citet{2024ApJ...962...67W} within the uncertainty.

Consequently, these findings imply that the BLR may either constitute an inner extension of the torus or originate from infalling material associated with the torus.

\begin{figure*}
\centering
\includegraphics[scale=0.48]{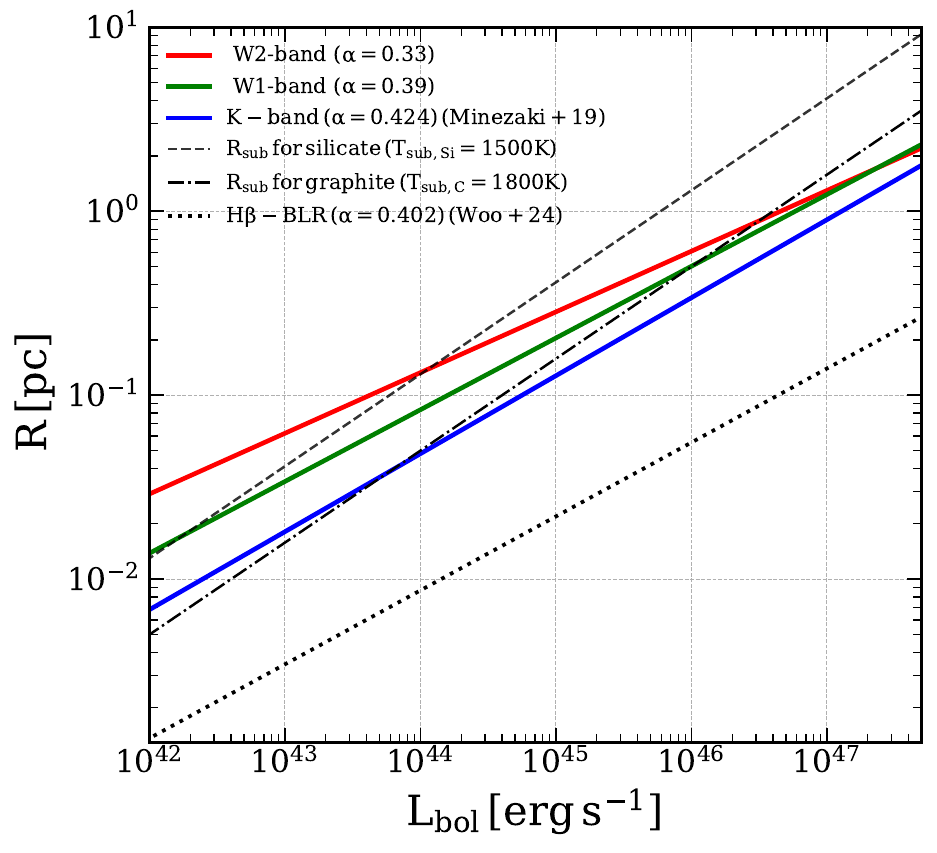}
\includegraphics[scale=0.5]{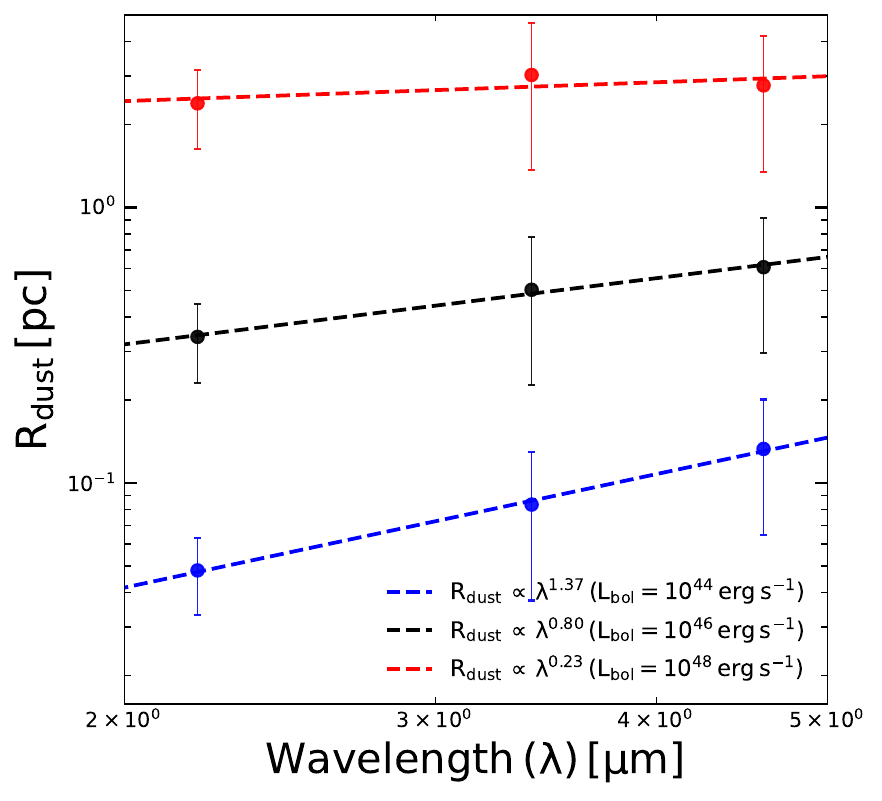}

\caption{Left: $R_{dust}$--$L_{bol}$ relations based on W1, W2 and K-band lags, which are corrected for the AD contamination. Theoretical estimates are calculated with
Equations \ref{sub_e1} and \ref{sub_e2}, respectively, for silicate with a sublimation temperature $T_{sub}$ = 1500 K (black dashed line) and for graphite grains with $T_{sub}$ = 1800 K (black dotted-dashed line). The H$\beta$ BLR size -luminosity relation adopted from \citet{2024ApJ...962...67W} is also presented. Right: Torus size measured at K, W1, and W2- band for three different values of $L_{bol}$ = $10^{44}$, $10^{46}$ and $10^{48} \, erg \, s^{-1}$, and the best-fit relation for each luminosity case. The error bars represent the intrinsic scatter in the corresponding torus size - luminosity relation.
%The dependency of torus size on wavelength at different $L_{bol}$ are also mentioned.
}
\label{fig:RL_all}
\end{figure*}

\subsection{Dependency of torus size on wavelength}

The availability of nearly simultaneous monitoring observations across various IR- bands, coupled with optical data, enables us to determine torus sizes across distinct IR wavelengths and, in turn, examine whether there exists any wavelength-dependent trend in torus sizes. However, in our sample, only two targets have previous DRM measurements based on K-band data in the literature: ID2149795 (J213818.96$+$011222.3, with W1 and W2)  and ID392158 (J232640.01$-$003041.4, with W1). Thus, we cannot establish a one-to-one relationship between the lags in the K- band and those in the other IR W1 and W2- bands.

Instead, we present the torus size -- AGN bolometric luminosity relations based on our work using W1 and W2-band, along with the work by \citet{2019ApJ...886..150M} using K-band in Figure \ref{fig:RL_all} (left). All these $R_{dust}$--$L_{bol}$ relations have been adjusted for the AD contamination in their respective IR (W1, W2 and K- band) light curves. The $R_{dust, K}$--$L_{V}$ relation is based on K-band lag and V-band luminosity of a sample of 36 quasars with relatively moderate to low luminosities ($10^{42.7} \, erg \, s^{-1} < L_{bol} < 10^{46.4} \, erg \, s^{-1}$), the majority of which have bolometric luminosities below $10^{46} \, erg \, s^{-1}$. We converted the V- band luminosity of the $R_{dust, K}$--$L_{V}$ relation to bolometric luminosity using a correction factor of 10 \citep{2020ApJ...900...58Y}. If we consider the same $R_{dust, K}$--$L_{bol}$ relation to hold for high luminosity AGNs, Figure \ref{fig:RL_all} (left) illustrates wavelength-dependent torus sizes: $R_{dust, W2}$ $>$ $R_{dust, W1}$ $>$ $R_{dust, K}$ for a given luminosity. We notice that as luminosity increases, the torus sizes in different IR bands gradually converge, suggesting that higher luminosity objects tend to exhibit a more compact torus structure, a notion in alignment with the findings of \citet{2011A&A...536A..78K} from interferometric observations.

The  innermost radius of the torus composed of silicate grains is defined by \citep{1987ApJ...320..537B, 2008ApJ...685..160N}

\begin{equation}
\label{sub_e1}
    R_{dust,Si} \simeq 1.3\left(\frac{L_{bol}}{10^{46} erg \, s^{-1}}\right)^{1/2} \left(\frac{1500 K}{T_{sub}}\right)^{2.6} pc,
\end{equation}
where $T_{sub}$ represents the dust sublimation temperature. Similarly, the sublimation radius for pure graphite grains is given by \citep{2012MNRAS.420..526M}

\begin{equation}
\label{sub_e2}
    R_{dust,C} \simeq 0.5\left(\frac{L_{bol}}{10^{46} erg \, s^{-1}}\right)^{1/2} \left(\frac{1800 K}{T_{sub}}\right)^{2.8} pc.
\end{equation}
The hot dust spectrum obtained by \citet{2012MNRAS.420..526M} suggests that the existence of hot graphite dust makes a substantial contribution to the observed luminosity at 3.6 and 4.5 $\mu m$. On the other hand, using $T_{sub}$ = 1700 K and a dust grain size of $\sim$ 0.1 $\mu m$, \citet{2014ApJ...788..159K} found that the sublimation radius is close to the $K$-band dust reverberation radius. For comparision we over-plot the theoretical prediction of the $R_{dust}$--$L_{bol}$ relations as defined by the Equations \ref{sub_e1} and \ref{sub_e2} for two distinct sublimation temperatures $T_{sub}$ = 1500 K and 1800 K, respectively.

 Using Equations \ref{sub_e1} and \ref{sub_e2}, we estimate the sublimation radius of a low-luminosity AGN with $L_{bol} = 10^{44} \, erg \, s^{-1}$. For silicate dust with $T_{sub}$ = 1500 K, $R_{dust,Si}$ is 0.13 pc, while for pure graphite with $T_{sub}$ = 1800 K, $R_{dust,C}$ is 0.05 pc. Whereas, for a high-luminosity AGN with $L_{bol} = 10^{48} \, erg \, s^{-1}$, the sublimation radius is significantly larger as $\sim$ 13.0 pc and 5.0 pc, respectively, for silicate with $T_{sub}$ = 1500 K and pure graphite grains with $T_{sub}$ = 1800 K.

We also empirically determine the torus size from the observed $R_{dust}$--$L_{bol}$ relationships (Equation \ref{eq_rL}). The derived radii are $R_{dust,W1} = 0.08$ pc ($R_{dust,W2} = 0.13$ pc), and $R_{dust,W1} = 3.0$ pc ($R_{dust,W2} = 2.8$ pc), respectively, for AGNs with $L_{bol} = 10^{44}$ and $10^{48} \, erg \, s^{-1}$. By comparing the theoretical estimates and the empirically determined torus sizes, we find that for low-luminosity AGNs ($\sim 10^{43.4}-10^{45} \, erg \, s^{-1}$), the primary source of the 3.4 and 4.6 $\mu m$ emission lies within the region bound by the graphite and silicate sublimation radii, respectively. These results align with the previous findings with the Spitzer Space Telescope data. For example, \citet{2015ApJ...801..127V} derived the torus size of NGC 6418 ($L_{bol} \geq 10^{43.3} \, erg \, s^{-1}$) as  $R_{dust,3.6\mu m} = 0.031$ pc with 3.6 $\mu m$ data, and $R_{dust,4.5\mu m} = 0.040$ pc with 4.5 $\mu m$ data.

In contrast, for very high luminosity AGNs ($L_{bol} \geq 10^{47} \, erg \, s^{-1}$), W1 and W2-band emissions originate from a similar region, which is roughly consistent with the sublimation radius of graphite grains with $T_{sub}=1800$ K (See left panel of Figure \ref{fig:RL_all}). This alignment is well-supported by the computed emission spectrum of graphite dust in the model by \citet{2012MNRAS.420..526M}, which peaks in the 2–3 $\mu m$ range and exhibits substantial emission in the 3.4$-$4.6 $\mu m$ range as well. Note that the observed $R_{dust}$--$L_{bol}$ relations show shallower slopes than expected from theoretical relations, resulting in a deviation of the observed torus sizes from theoretical predictions as luminosity increases. In addition, we present the H$\beta$ BLR size-luminosity relation from \citet{2024ApJ...962...67W}, demonstrating that the torus lies outside the BLR in AGNs.

There are only a few cases with all three (i.e., K, W1, and W2-band) lag measurements. For example, a wavelength dependent torus size is observed in  ID2149795 (J213818.96$+$011222.3, $L_{bol} = 10^{45.69} \, erg \, s^{-1}$) from our sample as $R_{dust,K}$ = $0.16\pm0.01$ pc \citep{2019ApJ...886..150M}, $R_{dust,W1}$ = $0.20^{+0.10}_{-0.15}$ pc, and $R_{dust,W2}$ = $0.35^{+0.10}_{-0.16}$ pc,  indicating a stratified torus structure. In contrast, Z229-15($L_{bol} \sim 10^{44.11} \, erg \, s^{-1}$) shows somewhat different trend. \citet{2022MNRAS.516.4898G} reported $R_{dust,3.6\mu m} \sim 0.009$ pc, and $R_{dust,4.5\mu m} \sim 0.023$ pc, while \citet{2021MNRAS.501.3905M} reported K-band lag as $R_{dust,K} \sim 0.017$ pc.  Similarly, ID392158 (J232640.01$-$003041.4) with $L_{bol} = 10^{45.64} \, erg \, s^{-1}$ from our sample shows comparable torus sizes in K and W1-bands with $R_{dust,K} = 0.19\pm0.06$ pc \citep{2019ApJ...886..150M} and $R_{dust,W1}=0.18^{+0.11}_{-0.13}$ pc. However, considering the relatively large uncertainty of the measured lags and the scatter in the $R_{dust}$--$L_{bol}$ relationships, we expect such exceptions.

%{\color{red}
% Hence, the resemblances among the K-band, 3.6 $\mu m$, and 4.5 $\mu m$ reverberation lags in Z229$-$15 suggest that, they collectively trace identical regions corresponding to the hottest dust near the inner radius of the dust-emitting region \citep{2022MNRAS.516.4898G}, despite it being a relatively low-luminosity AGN.
%}

By combining the best-fit parameters obtained by \citet{2019ApJ...886..150M} using K- band lags and our measurements in the W1 and W2- bands, we determine torus sizes from the best-fit torus size -- AGN bolometric luminosity relations at K, W1 and W2- bands at luminosities $L_{bol}$ = $10^{44}$, $10^{46}$, and $10^{48} \, erg \, s^{-1}$ spanning the entire dynamical range used in this study (see right panel of Figure \ref{fig:RL_all}). At $L_{bol}$ = $10^{46} \, erg \, s^{-1}$, nearly the median luminosity of our sample, we obtain the ratios of the dust torus sizes: $R_{dust,K}:R_{dust,W1}:R_{dust,W2}$ = 1.0:1.5:1.8. We find that the torus size scales with the wavelength as $R_{dust} \, \propto \, \lambda^{0.80}$ at $L_{bol}$ = $10^{46} \, erg \, s^{-1}$, indicating a stratified torus structure. Note that the dependence of $R_{dust}$ on wavelength becomes weaker as luminosity increases as indicated by the shallower slopes observed in the $R_{dust}$--$L_{bol}$ relations.

 Extensive efforts, comprising both observational studies \citep{2006ApJ...652L..13T, 2014ApJ...784L..11Y, 2019ApJ...886..150M} and theoretical investigations \citep{1987ApJ...320..537B, 2011ApJ...737..105K}, have been dedicated to gain insights into the structure of the dust torus in AGNs. The presence of a radial temperature gradient in the dust torus was reported by \citet{2006ApJ...652L..13T}. Additionally, a clumpy dust torus model \citep{2008ApJ...685..147N, 2011ApJ...737..105K} proposes a vertical stratification of rotating dust cloud distributions within galactic nuclei \citep{2019A&A...632A..61G}. Furthermore, \citet{2017ApJ...843....3A, 2020ApJ...891...26A} through simulations of the reverberation response of a clumpy torus, found that the torus response exhibits a strong wavelength dependency. This phenomenon arises from the temperature gradient within the cloud surfaces  of the torus and the highly anisotropic cloud emission at shorter wavelengths. Additionally, our findings support the idea that different IR emissions originate from distinct regions of the torus, thus lending support to the notion of a stratified torus structure.

%Moreover, several high-resolution observations, including those from the Atacama Large Millimeter Array (ALMA), have provided evidence for density radial stratification and complex gas kinematics in the torus and its surroundings \citep{2019A&A...632A..61G}. From high-angular-resolution IR and submillimeter interferometric observations, \citet{2019ApJ...884..171H} demonstrated that the vertical distribution of gas exhibits a clear density and temperature stratification. A clumpy dust torus model \citep{2008ApJ...685..147N} also suggests a vertical stratification of rotating dust cloud distributions in galactic nuclei \citep{2019A&A...632A..61G}. Additionally, our findings support the idea that different IR emissions originate from distinct regions of the torus, thus lending support to the notion of a stratified torus structure.

\begin{figure*}
\centering

%\resizebox{16cm}{9cm}{\includegraphics{Scatter_free.pdf}}
\includegraphics[scale=0.64]{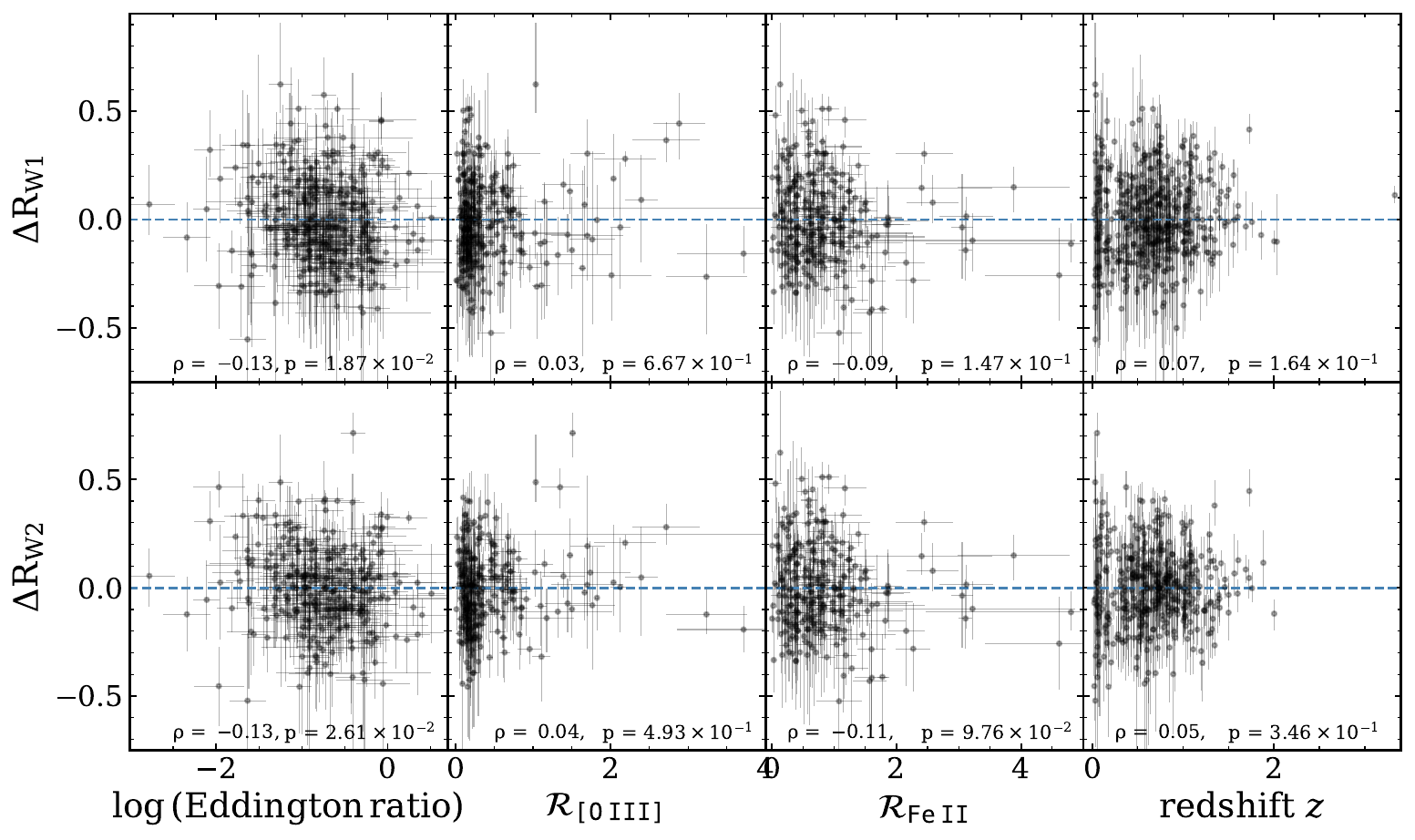}

\caption{The deviations of the dust-torus sizes from the corresponding best-fitting relationships ($\Delta R_{W1}$, and $\Delta R_{W2}$) as a function of Eddington ratio, $\mathcal{R}_{[O III]}$, $\mathcal{R}_{Fe II}$, and redshift $z$. The black points represent DRM sample corrected for AD contamination in the IR W1, W2 light curves. The Spearman's rank correlation coefficients and the corresponding probabilities are also mentioned in each panel. The blue dashed lines represent $\Delta R_{IR} = 0$.}

\label{fig:edd}
\end{figure*}

\begin{figure*}
\centering

%\resizebox{16cm}{9cm}{\includegraphics{Scatter_fixed.pdf}}
\includegraphics[scale=0.64]{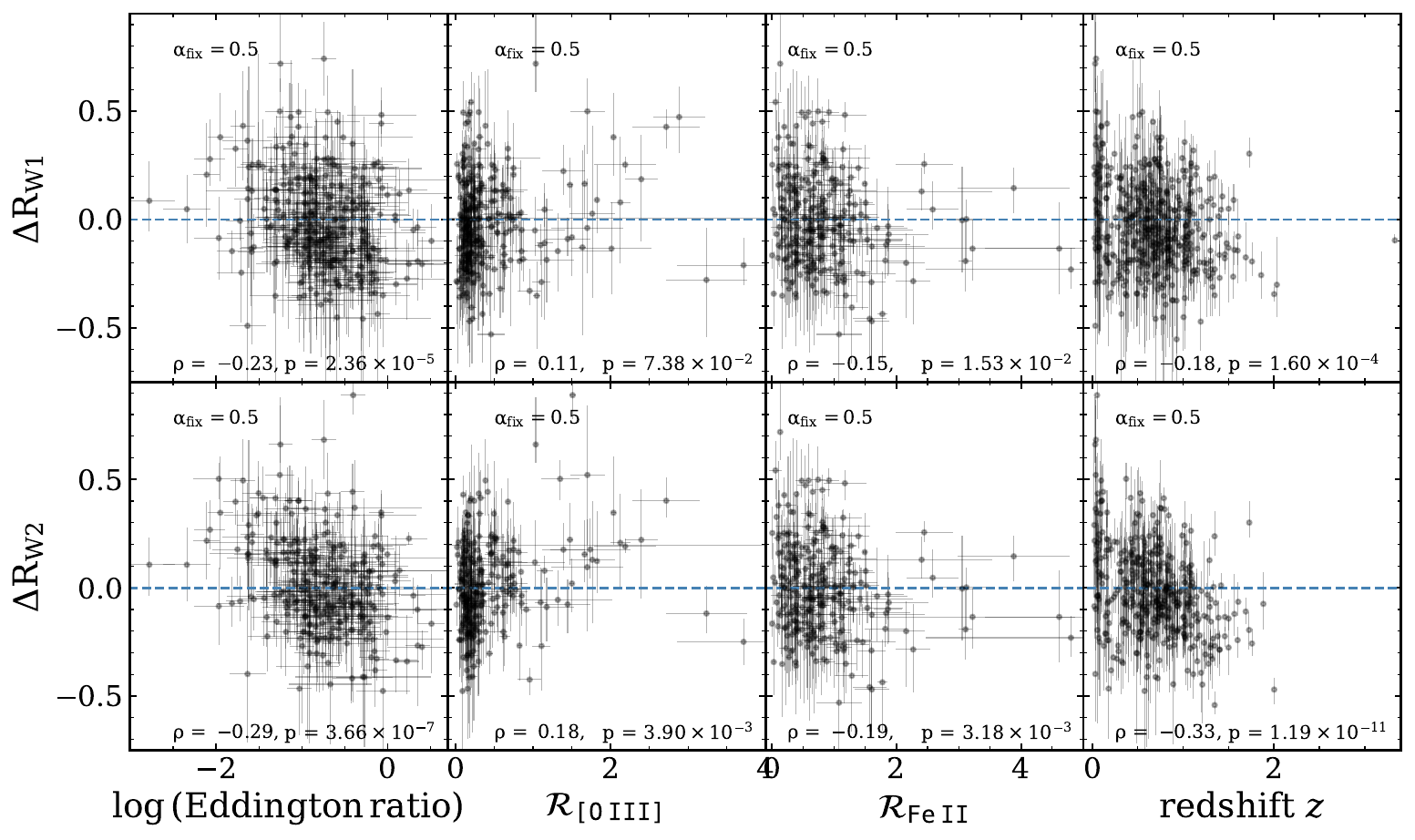}

\caption{Same as Figure \ref{fig:edd}, but for a fixed slope $\alpha_{fix}=0.5$.}

\label{fig:edd_fix}
\end{figure*}

\subsection{Scatter in the $R_{dust,IR}$--$L_{bol}$}\label{sec:scatter}

In this section, we explore the potential correlation between the scatter around the $R_{dust,IR}-L_{bol}$ relations and various other AGN parameters, including the Eddington ratio, the flux ratios {\OIII}/$H\beta$ ($\mathcal{R}_{[O III]}$), {\FeII}/$H\beta$ ($\mathcal{R}_{Fe II}$), and redshift $z$. We define the deviations of the measured torus sizes from the corresponding best-fitting relationships as $\Delta R_{IR}$ =  $\mathrm{log}R_{IR}$ - $\mathrm{log}R_{best-fit}$.

We compare $\Delta R_{IR}$ with the Eddington ratio in Figure \ref{fig:edd} (top left) considering the slope in the $R_{dust,IR}$--$L_{bol}$ relation as free parameter. It is important to note that the Eddington ratio in this analysis is derived from a single-epoch $M_{BH}$ (see Section \ref{sec:AD_cor}) using the BLR size -- luminosity relation \citep{2024ApJ...962...67W}, which is based on AGNs in the luminosity range $10^{42} < L_{5100{\AA}}  < 10^{46} \, erg \, s^{-1}$ with the majority of the AGNs having luminosities below $10^{44.5} \, erg \, s^{-1}$. This limitation imposes constraints on the precision of $M_{BH}$ measurements at higher redshift, where high-luminosity AGNs are more commonly observed. As a result, the canonical single-epoch $M_{BH}$ can lead to substantial over-estimations and, consequently, underestimations in the Eddington ratio, especially for AGNs with high accretion rates frequently found in the high-luminosity regime (also see section \ref{comp_edd}).

Nonetheless, despite these limitations, we identify a mild negative correlation between $\Delta R_{IR}$ and the Eddington ratio, with Spearman's rank correlation coefficients of $\mathrm{\rho = -0.13}$ and a probability of p = $1.87\times10^{-2}$ in the W1- band, and $\mathrm{\rho = -0.13}$ with a p-value of $2.61\times10^{-2}$ in the W2- band.

We also compare $\Delta R_{IR}$ with the Eddington ratio in Figure \ref{fig:edd_fix} considering a fixed slope of $\alpha_{fix} = 0.5$ in the $R_{dust,IR}-L_{bol}$ relation as predicted by dust radiation equilibrium theory. This time, we find a moderate negative correlation between them, with $\mathrm{\rho = -0.23}$, p = $2.36\times10^{-5}$ and $\mathrm{\rho = -0.29}$, p = $3.66\times10^{-7}$ in W1 and W2- bands, respectively.

However, it is important to note that the Eddington ratio scales with $L/R_{H\beta}$ $\sim$ $L/R_{dust,IR}$ due to the linear correlation present between $R_{H\beta}$ and $R_{dust,IR}$ \citep{2023MNRAS.tmp.1073C}. Further, the change in $\Delta R_{IR}$ varies as $R_{dust,IR}/L^{0.5}$. As a result, the negative correlation observed between $\Delta R_{IR}$ and the Eddington ratio may be attributed to the self-correlation between the Eddington ratios and the deviation from the best-fit relation \citep{2020ApJ...899...73F}.

Instead of Eddington ratio, we also utilized $\mathcal{R}_{[O III]}$ and $\mathcal{R}_{Fe II}$, as indicators of Eddington ratio \citep{1992ApJS...80..109B, 2000ApJ...536L...5S, Ludwig_2009, 2003MNRAS.345.1133M, Sun_2015, 2019ApJ...886...42D}. We collected $H\beta$, {\OIII}, and {\FeII} luminosities from \citet{2019ApJS..243...21L} for Sample 1. For Sample 2, the $H\beta$, and {\OIII} luminosities were retrieved from \citet{2011ApJS..194...45S}, whereas {\FeII} luminosities were obtained from \citet{2020ApJS..249...17R}. Hence, we  present $\Delta R_{IR}$ as a function of $\mathcal{R}_{[O III]}$ and $\mathcal{R}_{Fe II}$ in Figure \ref{fig:edd}, and \ref{fig:edd_fix} for varying $\alpha$ and a fixed $\alpha_{fix} = 0.5$, respectively. Notably, when $\alpha$ is treated as a free parameter, $\Delta R_{IR}$ exhibits relatively weak correlations with $\mathcal{R}_{[O III]}$ and $\mathcal{R}_{Fe II}$—positive and negative, respectively. However, these correlations become more pronounced when $\alpha$ is held constant at 0.5, as seen in Figures \ref{fig:edd} and \ref{fig:edd_fix} for comparison.

In addition, we examine the dependence of $\Delta R_{IR}$ with redshift $z$ to investigate any residual redshift dependence. We find no significant dependence when $\alpha$ is treated as a free parameter. This suggests that the redshift dependence of dust lags is negligible after the $z$ correction applied to our sample as detailed in Section \ref{cor_redshift}. However, a noticeable correlation between $\Delta R_{IR}$ and $z$ emerges when we use $\alpha_{fix} = 0.5$, indicating a reduction in torus sizes for AGNs at higher redshifts. However, this is mainly due to the luminosity effect since on average higher luminosity AGNs are selected at higher $z$.

Our findings are consistent with the previous studies by \citet{2023MNRAS.tmp.1073C}. Notably, \citet{2020ApJ...900...58Y} reported no dependency of $\Delta R_{W1}$ on Eddington ratio. Note that they used Eddington ratio measured from single epoch $M_{BH}$. In contrast, \citet{2023MNRAS.tmp.1073C} found that $\Delta R_{W1}$ and $\Delta R_{W2}$  systematically decrease with increasing accretion rate, suggesting the shortening of time lags for both BLRs and the dusty torus. They explained such behaviors by the self-shadowing effect of the slim disk model. Similarly, \citet{2020ApJ...899...73F} found that AGNs have a shorter BLR size with lower $\mathcal{R}_{[O III]}$. According to them, the deviation from the radius--luminosity relation depends on the distribution of the UV/optical SED and the relative contribution of ionizing radiation \citep[for the case of H$\beta$ BLR, see][]{2024ApJ...962...67W}. All of these findings align with our own results, implying that lag shortening of both BLRs and torus is caused by the same physical mechanism, i.e., the accretion rate. Additionally, the weakening of the overall correlation between $\Delta R_{IR}$ and the Eddington ratio, $\mathcal{R}_{[O III]}$, and $\mathcal{R}_{Fe II}$ as the slope becomes flatter suggests that the flattening of the $R_{dust,IR}$--$L_{bol}$ relation from the dust radiation equilibrium model is influenced by the self-shadowing effect of the accretion disk as accretion rate increases.

% It is important to note that \citet{2023MNRAS.tmp.1073C} obtained the accretion rate using BH mass measurements from RM, which are considered more accurate than single epoch BH mass measurements. Eddington ratios obtained from single epoch BH mass can be underestimated for high accreting AGNs, see section \ref{comp_edd} for more details. Thus, the dependence of the scatter in the $\mathrm{R_{dust,MIR}-L_{bol}}$ relation on accretion rate can be properly investigated using AGNs that have both MIR and BLR-RM measurements.                                                                                              

However, the obtained $R_{dust,IR}$--$L_{AGN}$ relations may suffer from several selection biases, such as, the IR lags may be underestimated for high luminosity sources because of the limited optical and IR baseline. \citet{2020ApJ...900...58Y} mentioned that  IR lags below 7 yr in the observed-frame can be detected within the base-line, whereas the lag detection probability will decrease toward longer lags in the observed-frame. 

Utilizing high cadence light curve data in optical with strong features present in both optical and IR light curves, our analysis has allowed us to detect lags that are shorter than the cadence of the IR light curves ($\sim$ 180 days) for a few targets. This lag detection was achieved through the application of both ICCF and {\tt JAVELIN} techniques. It is worth noting that {\tt JAVELIN} can detect lags shorter than the observing cadence \citep{2019ApJ...884..119L}. However, it is important to consider that the IR lags might be prone to overestimation when the observed-frame lags are significantly shorter than the cadence of the IR light curves, for instance, falling below 100 days.

To ensure the reliability of our lag measurements, we employed a comprehensive validation process in addition to assessing the quality of measured lags based on $r_{max}$ and $p(r_{max})_{\tau>0}$ values, as described in Section \ref{qual}. This validation procedure involved a systematic comparison of each lag measurement by shifting the optical light curves according to the calculated lag value and overlaying them onto the observed IR light curve. The results of this comparison demonstrated a consistent agreement between the shifted optical light curves and the observed IR light curves. As an illustrative example, please refer to Figure \ref{fig:light} for the target OB76 and Figure \ref{fig:light_app} for OB45 and OB60, all of which exhibit observed-frame W1 lags of less than 100 days. These figures provide clear visual evidence of the robustness of our lag measurements in such cases. The combined effect of these selection biases on the $R_{dust,IR}$--$L_{bol}$ relation can be demonstrated properly by detail simulation, which is beyond the scope of this paper.

\section{Summary}\label{summary}

We present DRM results for a total of $\sim$ 446 (416) AGNs with W1- (W2) band light curves, covering a large dynamic range in luminosities from $10^{43.4} \, \mathrm{to} \, 10^{47.6} \, erg \, s^{-1}$ with redshift $< 2$ using 16--20 yr optical light curve data from different ground based optical surveys and IR data in W1 and W2- bands from WISE during 2010--2020. We explored the $R_{dust,IR}$--$L_{bol}$ relations for different combinations of data. Our main findings are summarized below.

\begin{itemize}
%\begin{enumerate}

    \item We find that the torus size scales with the AGN bolometric luminosity as $R_{dust, W1} \propto L_{bol}^{0.38}$ and $R_{dust, W2} \propto L_{bol}^{0.35}$ in the W1 and W2- bands, respectively. Our best-fit slopes are much shallower than that expected from the thermal equilibrium model of $R_{dust} \propto L_{bol}^{0.5}$. However, our results are consistent with that obtained from the optical/IR interferometric observations, which also predict shallower slope than expected from the radiation equilibrium model.
                        
    \item We remove the AD contamination from the observed IR (W1 and W2- bands) fluxes and obtained the $R_{dust, IR}$--$ L_{bol}$ relation. The correction of the AD contamination in IR fluxes results slightly different slope with $R_{dust, W1} \propto L_{bol}^{0.39}$ and $R_{dust, W2} \propto L_{bol}^{0.33}$. As a result of this correction for AD contamination in the IR light curves, we observe larger torus sizes compared to cases where AD contamination is not accounted for in the analysis.

    \item Utilizing IR luminosity in the W2- band as the indicator for AGN luminosity, we identified analogous slopes in the $R_{dust, IR}$--$L_{W2}$ relationships as those obtained for $R_{dust,IR}$--$L_{bol}$ associations. This observation implies that both unobscured and obscured AGNs exhibit a consistent trend within the torus size--luminosity plane.

    \item The torus sizes in W1 and W2- bands are found to be strongly correlated with $R_{dust,W2}$ larger than $R_{dust,W1}$. We also compare our obtained IR $R_{dust,IR}$--$L_{bol}$ relations in W1 and W2- bands with the $R_{dust, K}$--$L_{bol}$ relation based on K- band lags available in the literature. We find a wavelength dependent torus size that increases with wavelengths as $R_{dust,K}:R_{dust,W1}:R_{dust,W2}$ = 1.0:1.5:1.8 ($R_{dust} \, \propto \, \lambda^{0.80}$) at $L_{bol}$ = $10^{46} \, erg \, s^{-1}$. This suggests a stratified dust torus. Additionally, a noteworthy observation is that objects with higher luminosities tend to display a more compact torus structure.

    \item The sizes of torus are consistently determined to be larger than that of the BLR, indicating that the dust torus exists outside of the BLR. Our analysis reveals a super-linear relationship between BLR size and torus size, as $R_{H\beta} \, \propto \, R_{dust, W1}^{1.20}$ and $R_{H\beta} \, \propto \, R_{dust, W2}^{1.22}$, demonstrating a consistent slope observed in the $R_{H\beta}$--$L_{5100 {\AA}}$ relationship available in the literature.

    \item We find that the scatter observed in the $R_{dust,IR}$--$L_{bol}$ relation depends on Eddington ratio, while a mild correlation is evident with $\mathcal{R}_{[O III]}$ and $\mathcal{R}_{Fe II}$. These findings collectively suggest that as the accretion rate increases, the torus size tends to shorten because of the self-shadowing effect of the slim disk, and the $R_{dust,IR}$--$L_{bol}$ relation flattens compared to the predictions from the dust radiation equilibrium model. Nevertheless, we cannot definitively establish this correlation, except for the one arising due to the inherent relationship between AGN luminosity and Eddington ratio.

%\end{enumerate}    
\end{itemize}
The uncertainty in the measured lags and the scatter in the obtained torus size -- luminosity relations can be improved using dedicated AGN monitoring survey, simultaneously, in optical and IR wavelengths in future with proper cadence. In this case the Vera C. Rubin Observatory Legacy Survey of Space and Time \citep[LSST;][]{2019ApJ...873..111I} and the Nancy Grace Roman Space Telescope (NGRST), which was previously known as WFIRST \citep{2012arXiv1208.4012G} can play an important role. The LSST+NGRST data set can be used to measure the size of the dust torus in local AGNs with redshift $z$ $<$ 0.5 since NGRST can observe up to $\sim$ 2 $\mu m$ \citep{2020ApJ...900...58Y}. In order to employ dust lag as a standard candle for cosmology and constrain cosmological parameters, \citet{2017MNRAS.464.1693H} started a large DRM program, namely 'VEILS' (VISTA Extragalactic Infrared Legacy Survey) that will observe about 1350 Seyfert 1 galaxies in the redshift range of 0.1 $<$ $z$ $<$ 1.2. The Spectro-Photometer for the History of the Universe, Epoch of Reionization and Ices Explorer \citep[SPHEREx;][]{2021JKAS...54...37K}  mission aims to provide optical and NIR multi-epoch spectroscopic data, which can be utilized in RM studies for bright AGNs. These RM studies can generate a unique data set that, when combined with complementary optical observations, can aid in comprehending the physical properties of the dust torus as well as the central structures of AGNs.

%\section{Software and third party data repository citations} \label{sec:cite}

%% Also note that the akcnowlodgment environment does not support long amounts of text. If you have a lot of people and institutions to acknowledge, do not use this command. Instead, create a new \section{Acknowledgments}.

%\section*{Acknowledgments}.

\begin{acknowledgments}

 We thank the anonymous referee for valuable comments and suggestions. This publication uses data products from the WISE, a joint project between the University of California, Los Angeles, and the Jet Propulsion Laboratory/California Institute of Technology, and is funded by the National Aeronautics and Space Administration. The publication also leverages data products from NEOWISE, a Jet Propulsion Laboratory/California Institute of Technology project, funded by the Planetary Science Division of the National Aeronautics and Space Administration. The CRTS survey is conducted by the U.S. National Science Foundation under grants AST-0909182 and AST-1313422. The ASAS-SN is supported by the Gordon and Betty Moore Foundation through grant GBMF5490 to the Ohio State University and NSF grant AST-1515927. The Development of ASAS-SN was supported by the NSF grant AST- 0908816, the Mt. Cuba Astronomical Foundation, the Center for Cosmology and AstroParticle Physics at the Ohio State University, the Chinese Academy of Science South America Center for Astronomy (CASSACA), the Villum Foundation, and George Skestos. ZTF is funded by the National Science Foundation through grant number AST-1440341 and a partnership with Caltech, IPAC, the Weizmann Institute for Science, the Oskar Klein Center at Stockholm University, the University of Maryland, the University of Washington, Deutsches Elektronen-Synchrotron and Humboldt University, Los Alamos National Laboratories, the TANGO Consortium of Taiwan, the University of Wisconsin at Milwaukee, and Lawrence Berkeley National Laboratories. COO, IPAC, and UW manage operations. This paper also used data observed with the Samuel Oschin Telescope at the Palomar Observatory as part of the Palomar Transient Factory project, a scientific collaboration between the California Institute of Technology, Columbia University, Las Cumbres Observatory, the Lawrence Berkeley National Laboratory, the National Energy Research Scientific Computing Center, the University of Oxford, and the Weizmann Institute of Science.

This work has been supported by the Basic Science Research Program through the National Research Foundation of the Korean Government (grant No. NRF-2021R1A2C3008486) and Samsung Science and Technology Foundation under Project Number SSTF --BA1501--05. S.W. acknowledges the support from the National Research Foundation of Korea (NRF) grant funded by the Korean government (MEST) (No. 2019R1A6A1A10073437). 

{Software:}  {\tt JAVELIN} \citep{2011ApJ...735...80Z}, {\tt PyCALI} \citep{2014ApJ...786L...6L}\\
%    {\bf Facilities:} WISE, NEOWISE, CRTS, ASAS-SN, PTF, ZTF

\facilities{WISE, NEOWISE, CRTS, ASAS-SN, PTF, ZTF}

\end{acknowledgments}
%\vspace{-13mm}


%\vspace{5mm}
%\facilities{HST(STIS), Swift(XRT and UVOT), AAVSO, CTIO:1.3m,


%\software{astropy \citep{2013A&A...558A..33A,2018AJ....156..123A},  
%          Cloudy \citep{2013RMxAA..49..137F}, 
%          Source Extractor \citep{1996A&AS..117..393B}
 %         }



\appendix \label{app}

\section{Examples of light curves and their lag measurements} 

\restartappendixnumbering

\begin{figure*}[!h]

\includegraphics[scale=0.5]{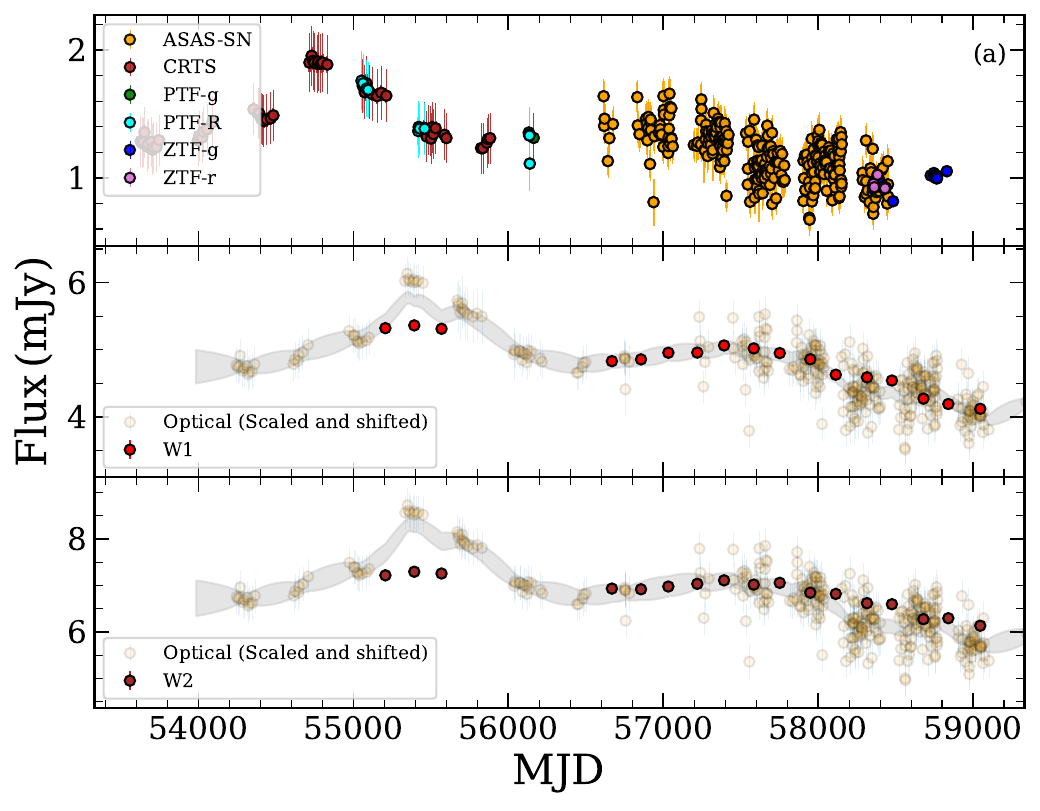}
\includegraphics[scale=0.5]{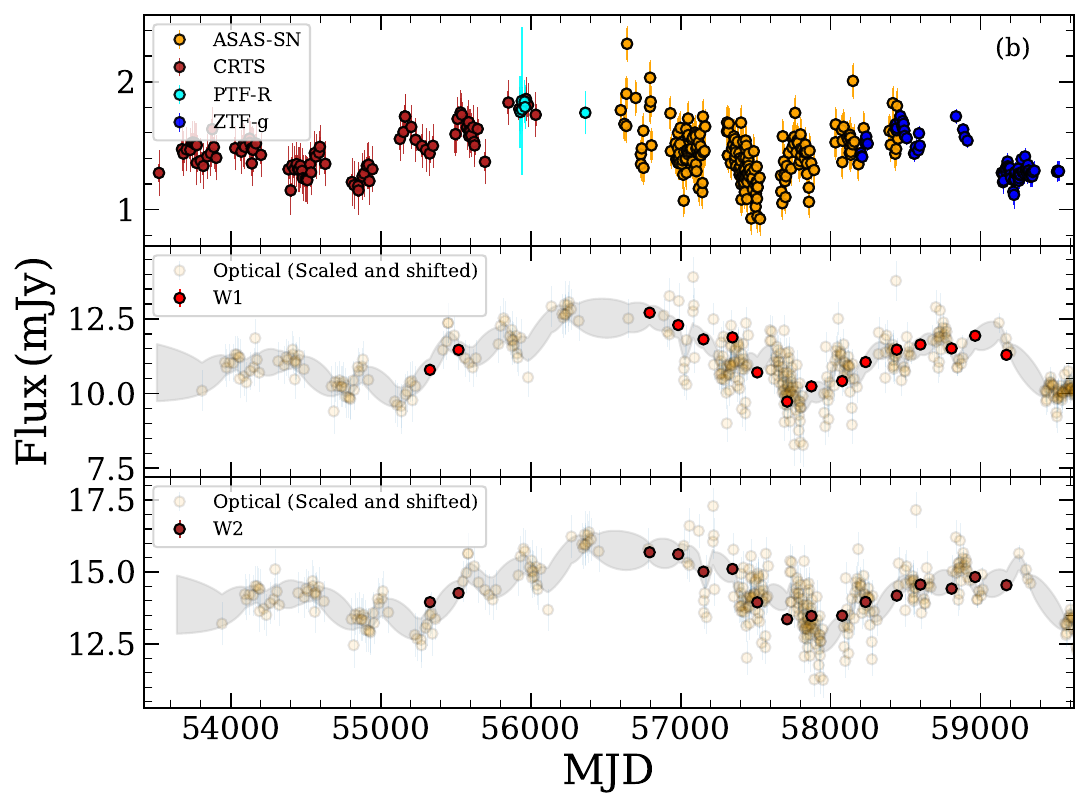}
\includegraphics[scale=0.5]{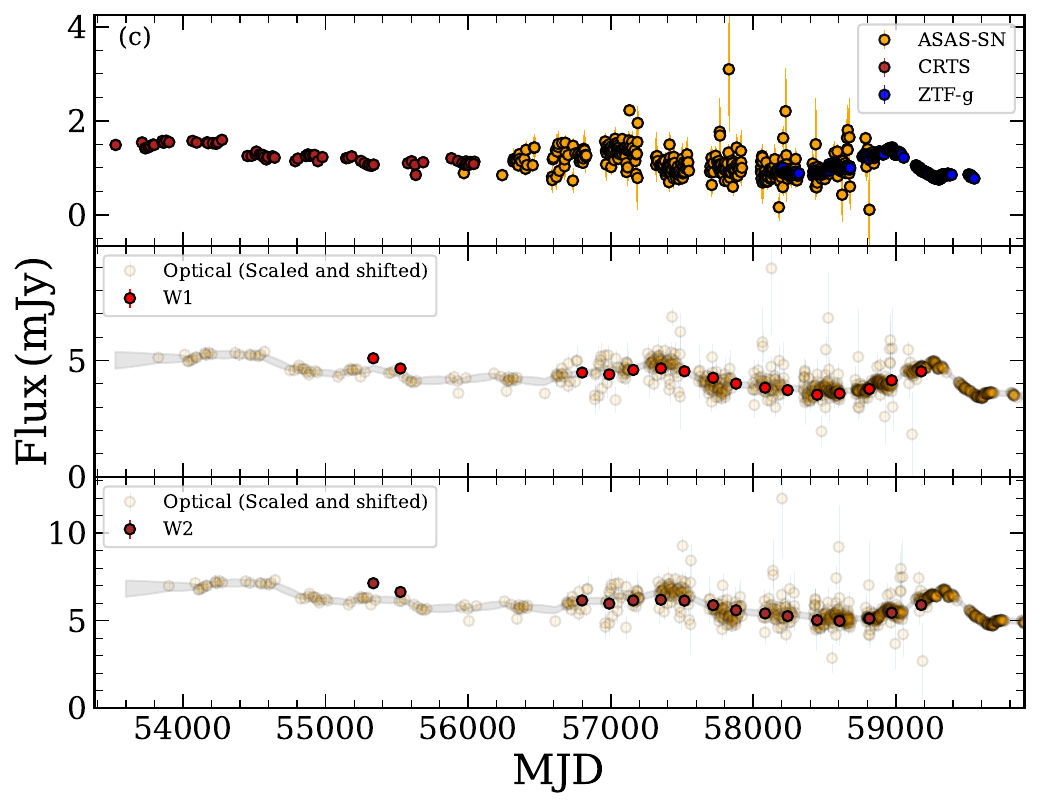}
\includegraphics[scale=0.5]{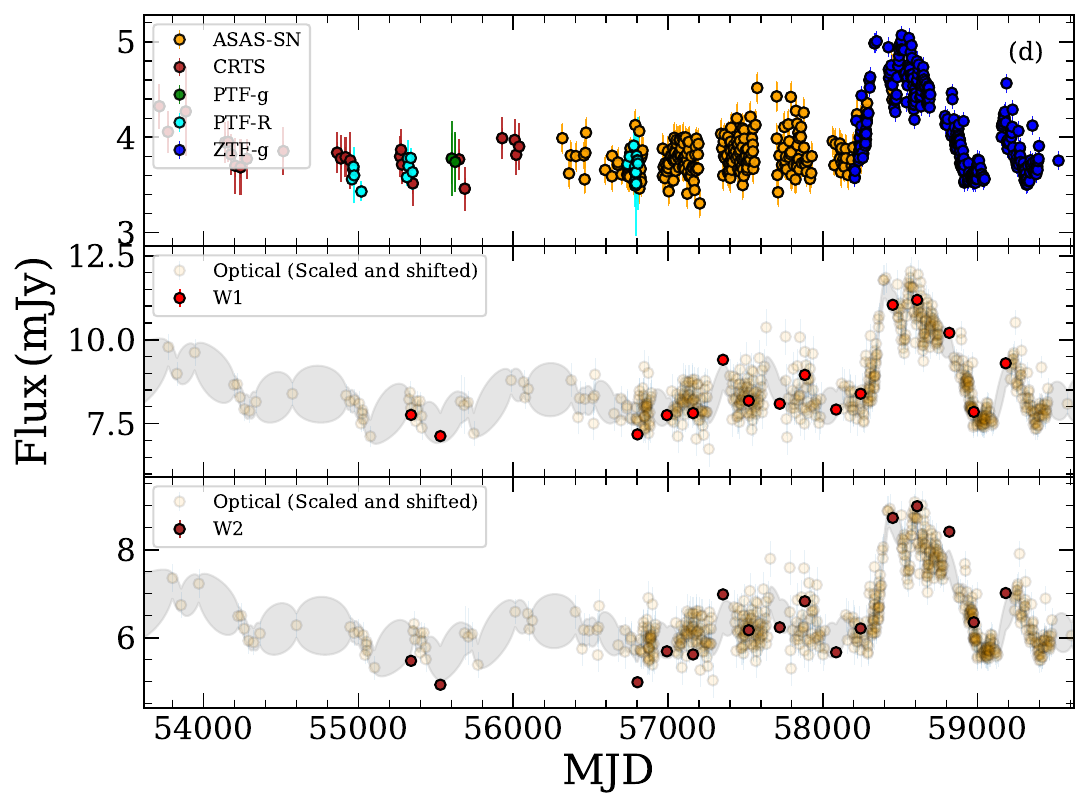}
\includegraphics[scale=0.5]{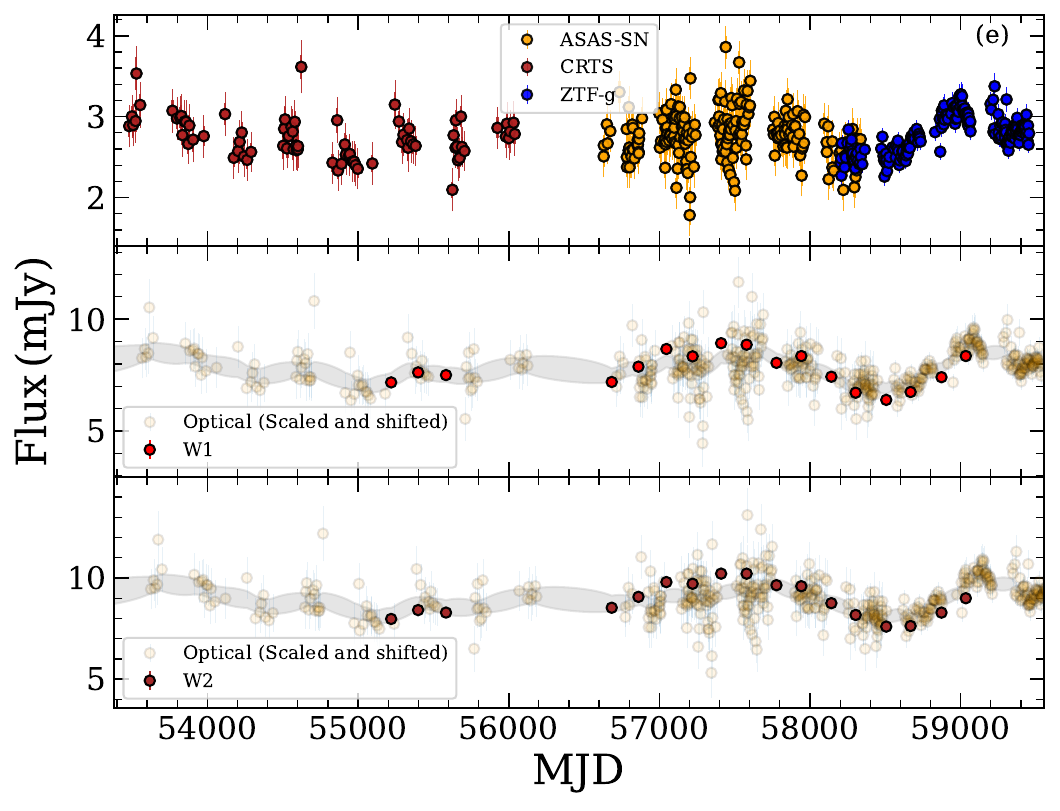}
\includegraphics[scale=0.5]{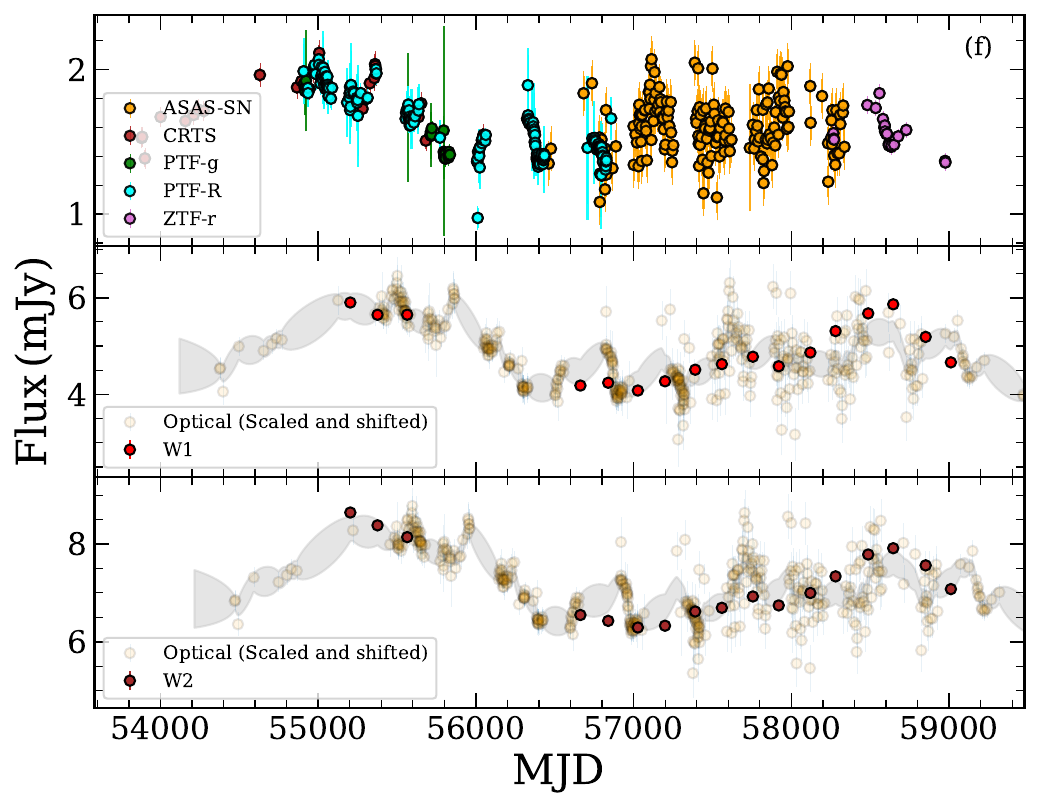}

%\resizebox{6cm}{5cm}{\includegraphics{grt_curve.pdf}}
\caption{Examples of light curves from Sample 1 in optical and IR  W1- and W2- bands for the targets (a) OB01, (b) OB14, (c) OB32, (d) OB45, (e) OB60, and (f) OB68. The IR W1- and W2- band light curves represented in the plot are not corrected for the AD contamination. Same notations are used as Figure \ref{fig:light}.}
\label{fig:light_app}
\end{figure*}

\begin{figure*}[!h]
\includegraphics[scale=0.48]{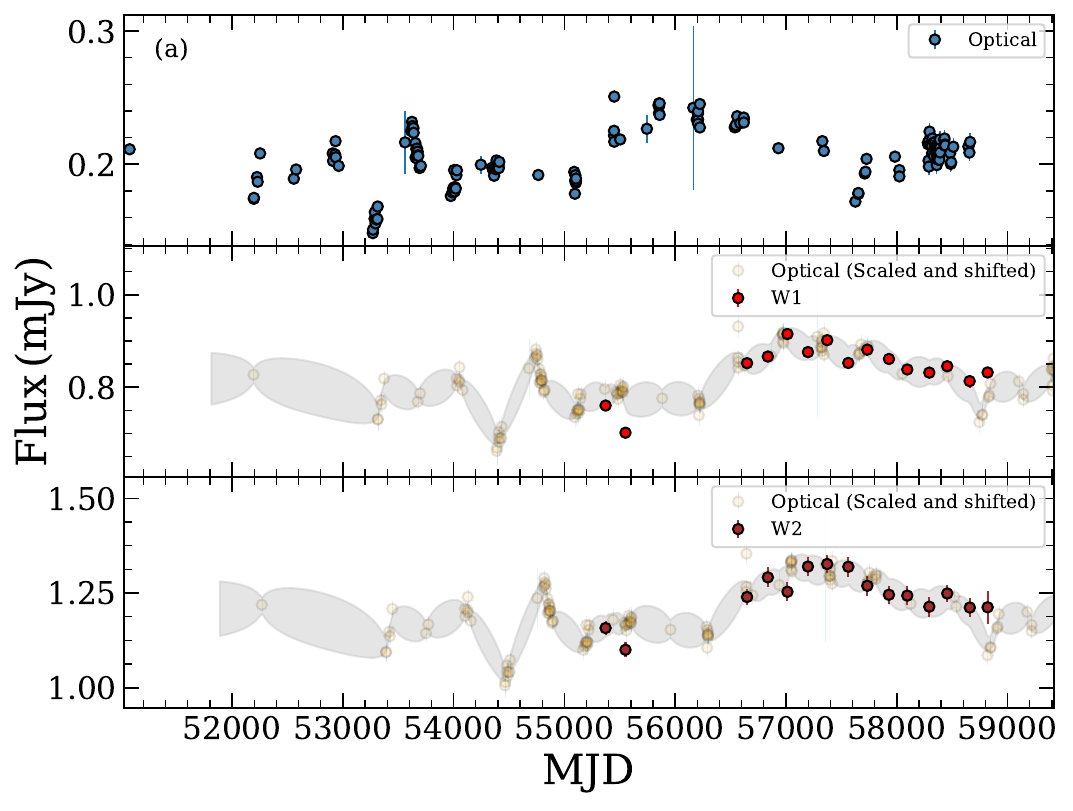}
\includegraphics[scale=0.48]{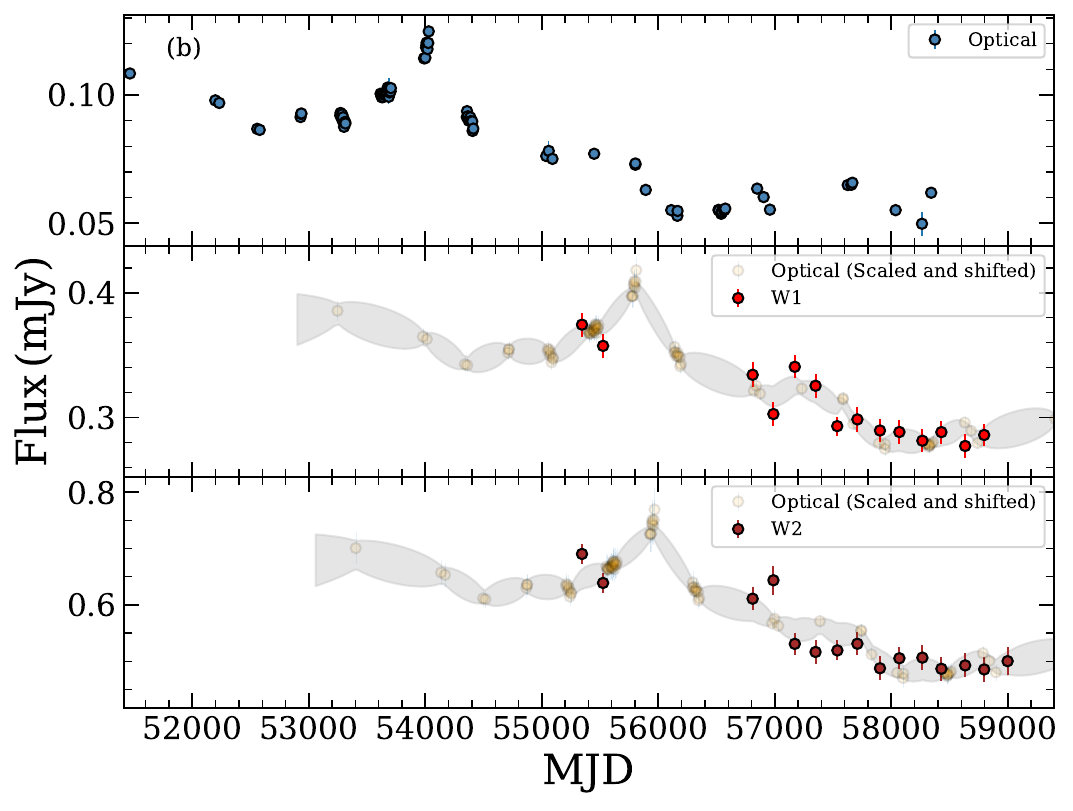}
\includegraphics[scale=0.48]{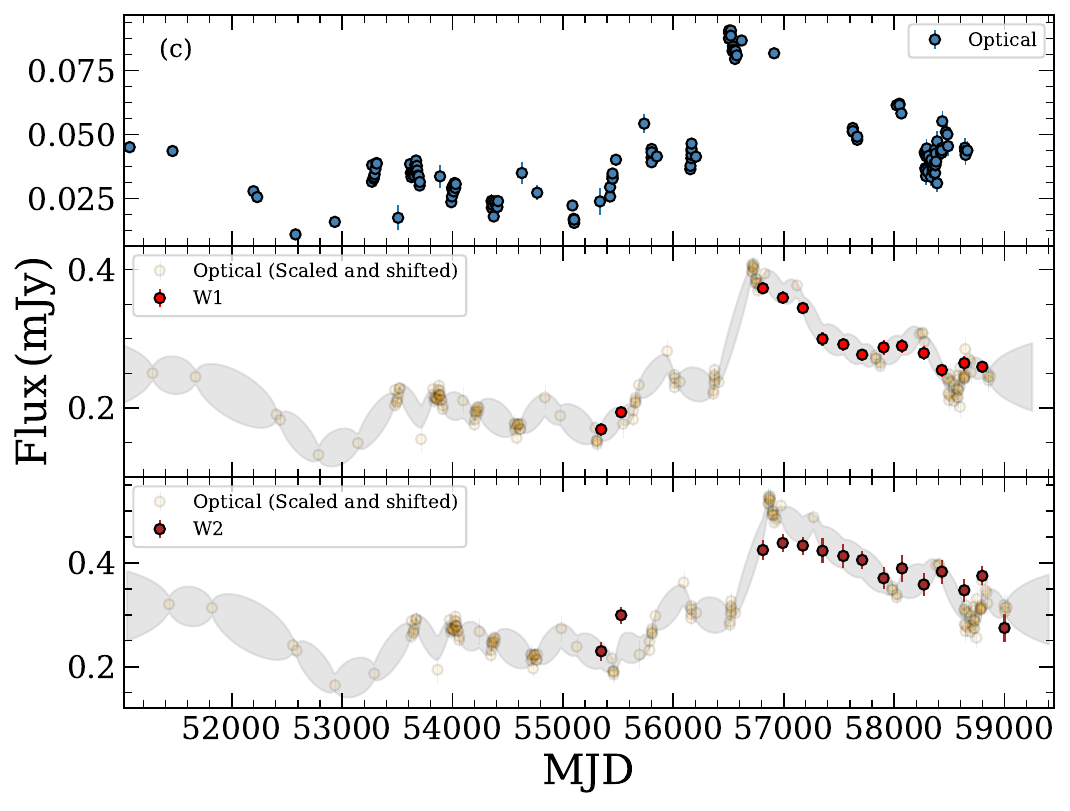}
\includegraphics[scale=0.48]{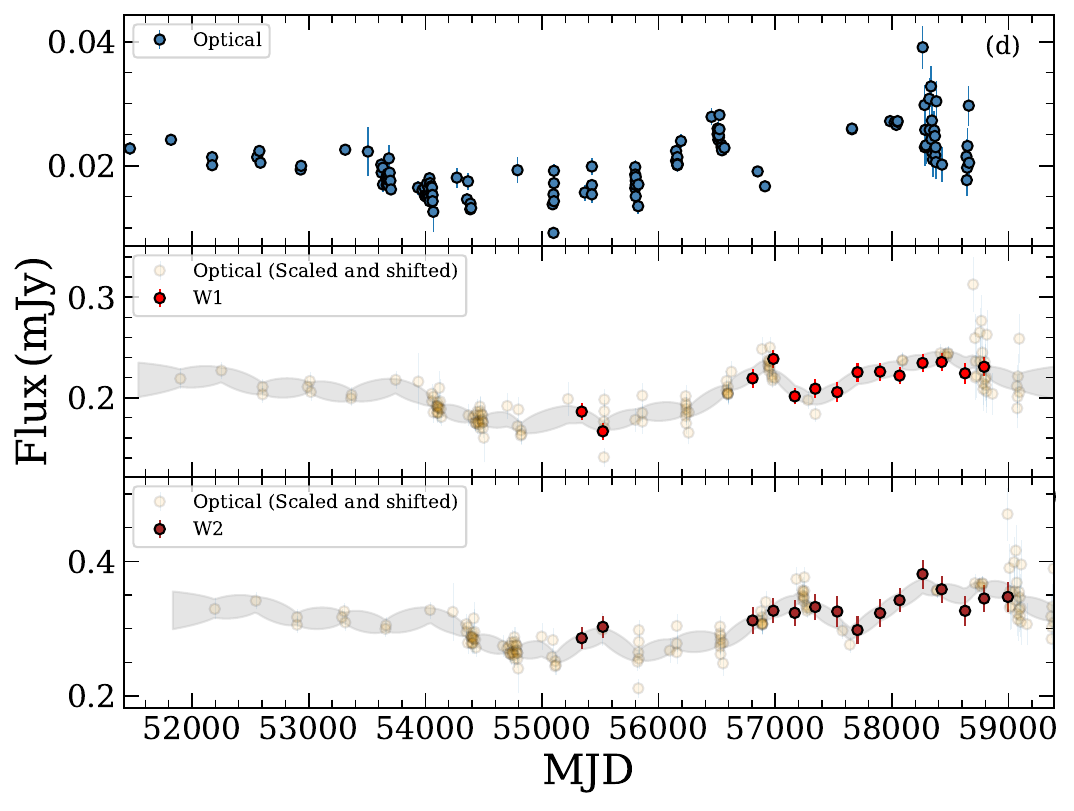}
\includegraphics[scale=0.495]{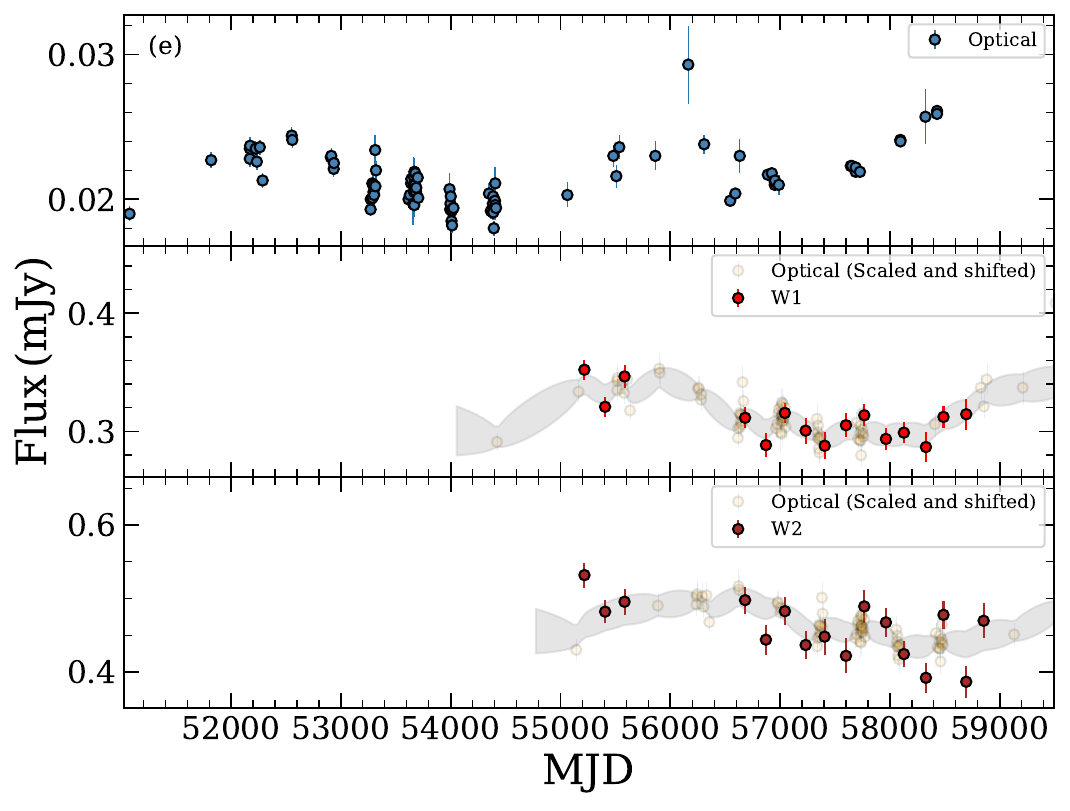}
\includegraphics[scale=0.495]{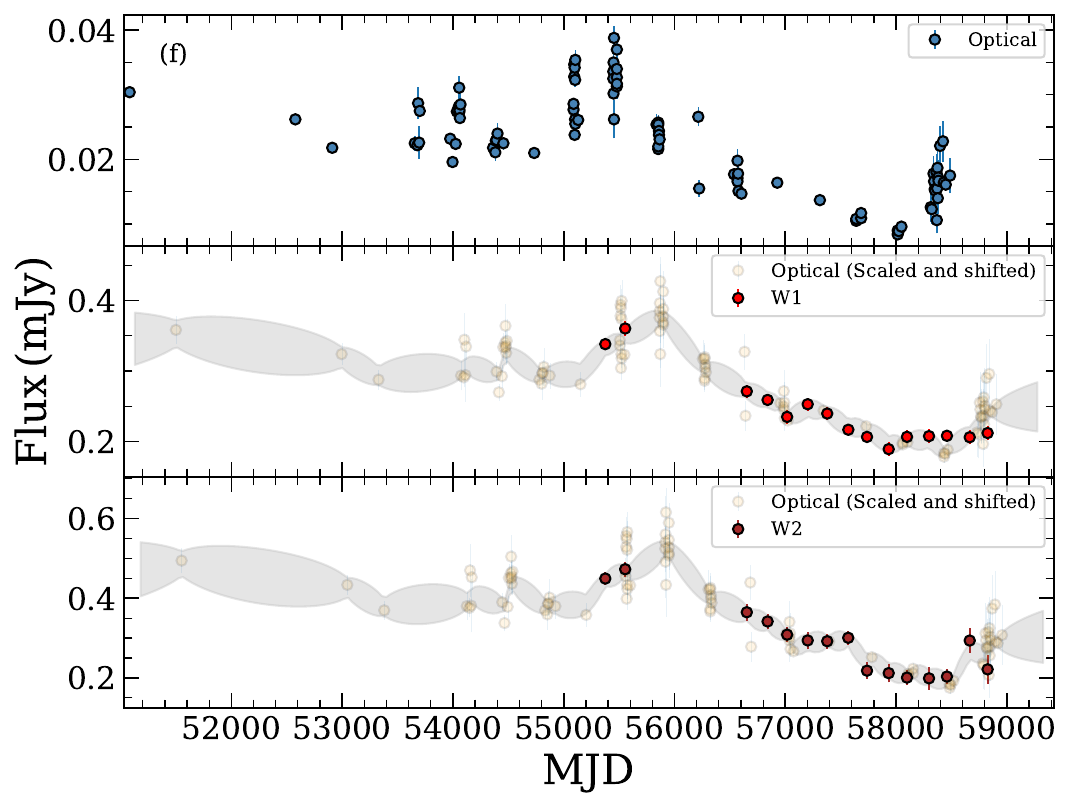}

\caption{Examples of light curves from Sample 2 in optical and IR  W1- and W2-  bands for the targets (a) ID47417, (b) ID1124174, (c) ID1189298, (d) ID1524735, (e) ID2023374, and (f) ID4065339.  The IR W1- and W2- band light curves represented in the plot are not corrected for the AD contamination. Same notations are used as Figure \ref{fig:light}.}
\label{fig:light_appS2}
\end{figure*}

\begin{figure*}[!h]

\resizebox{9.3cm}{4.8cm}{\includegraphics{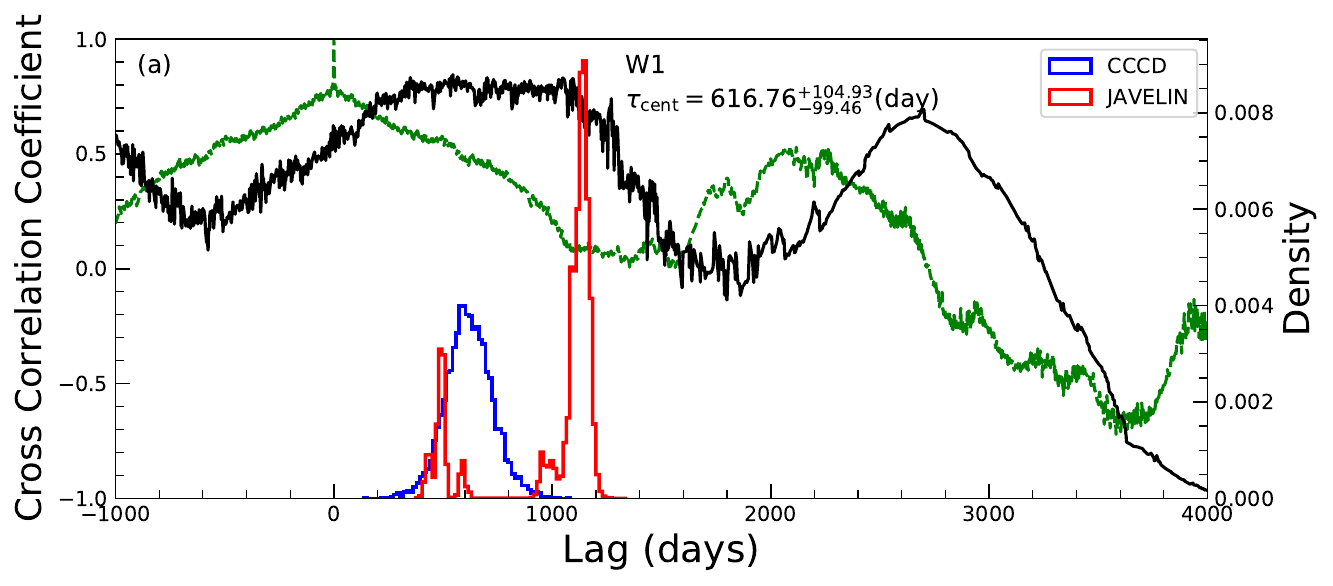}}
\resizebox{9.3cm}{4.8cm}{\includegraphics{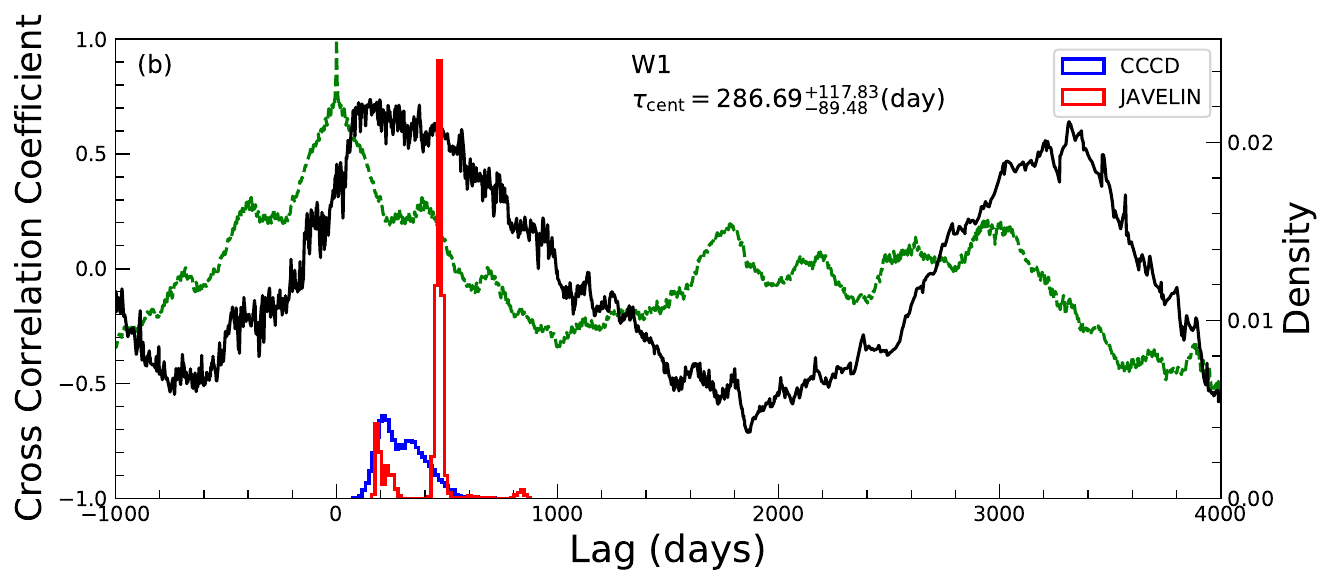}}
\resizebox{9.3cm}{4.8cm}{\includegraphics{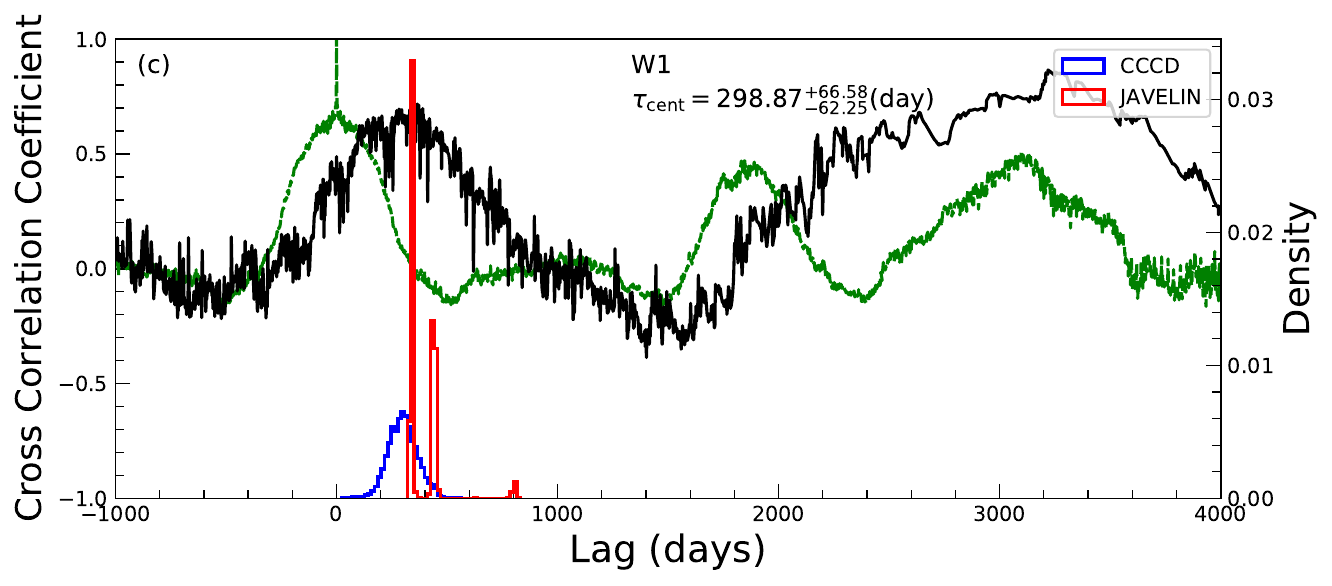}}
\resizebox{9.3cm}{4.8cm}{\includegraphics{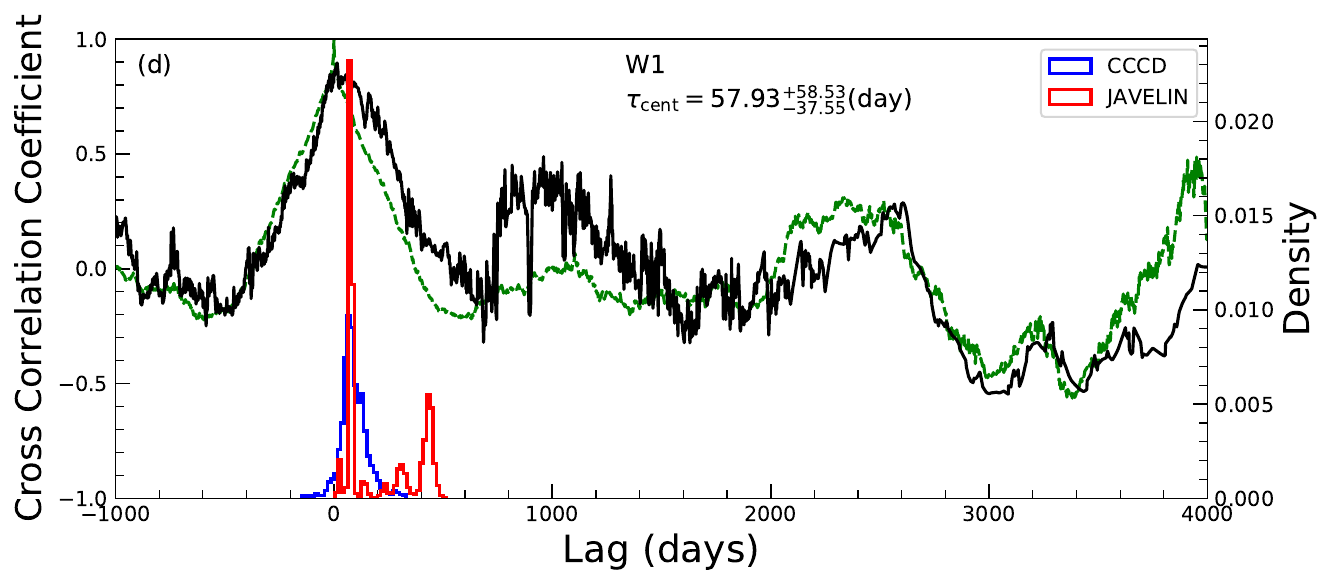}}
\resizebox{9.3cm}{4.8cm}{\includegraphics{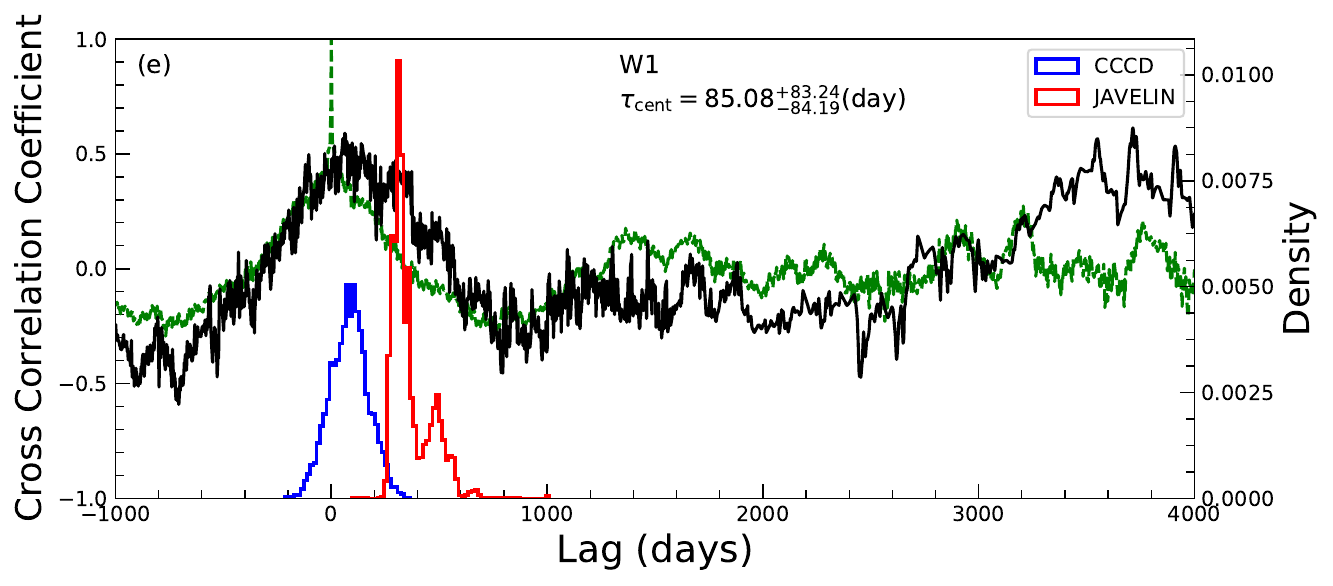}}
\resizebox{9.3cm}{4.8cm}{\includegraphics{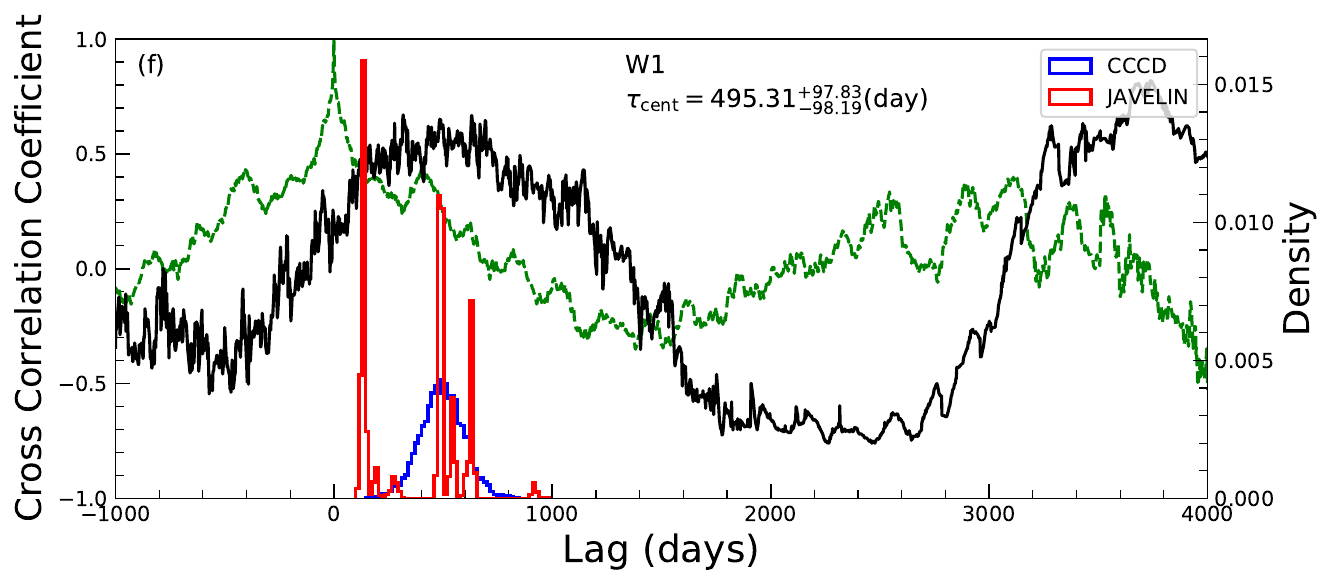}}

\caption{The ICCFs and ${\tt JAVELIN}$ analysis between optical and W1- band light curves of the example targets (a) OB01, (b) OB14, (c) OB32, (d) OB45, (e) OB60, and (f) OB68 from Sample 1. Same notations are used as Figure \ref{fig:ccf}.}
\label{fig:ccf_app}
\end{figure*}

\begin{figure*}[!h]

\resizebox{9.3cm}{4.8cm}{\includegraphics{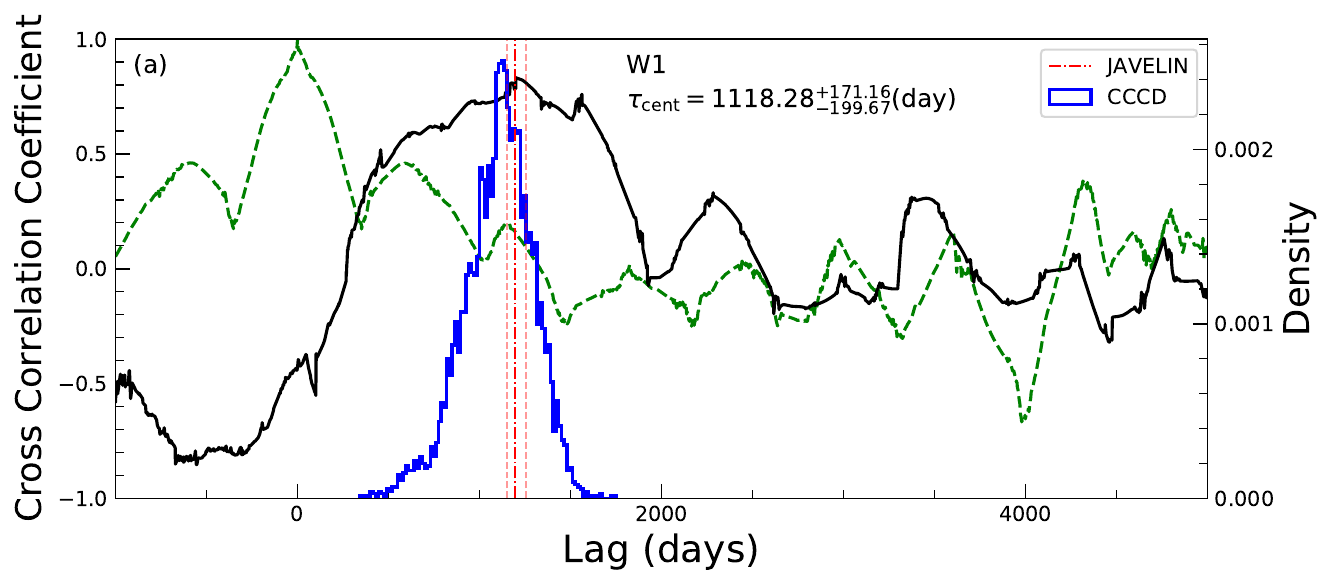}}
\resizebox{9.3cm}{4.8cm}{\includegraphics{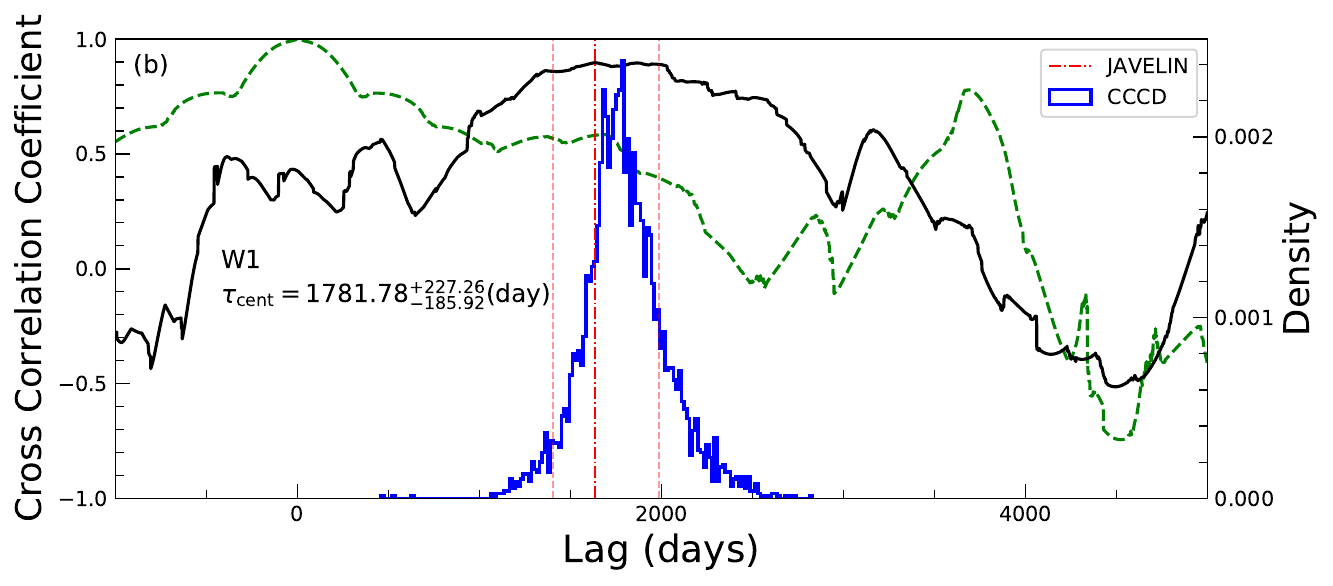}}
\resizebox{9.3cm}{4.8cm}{\includegraphics{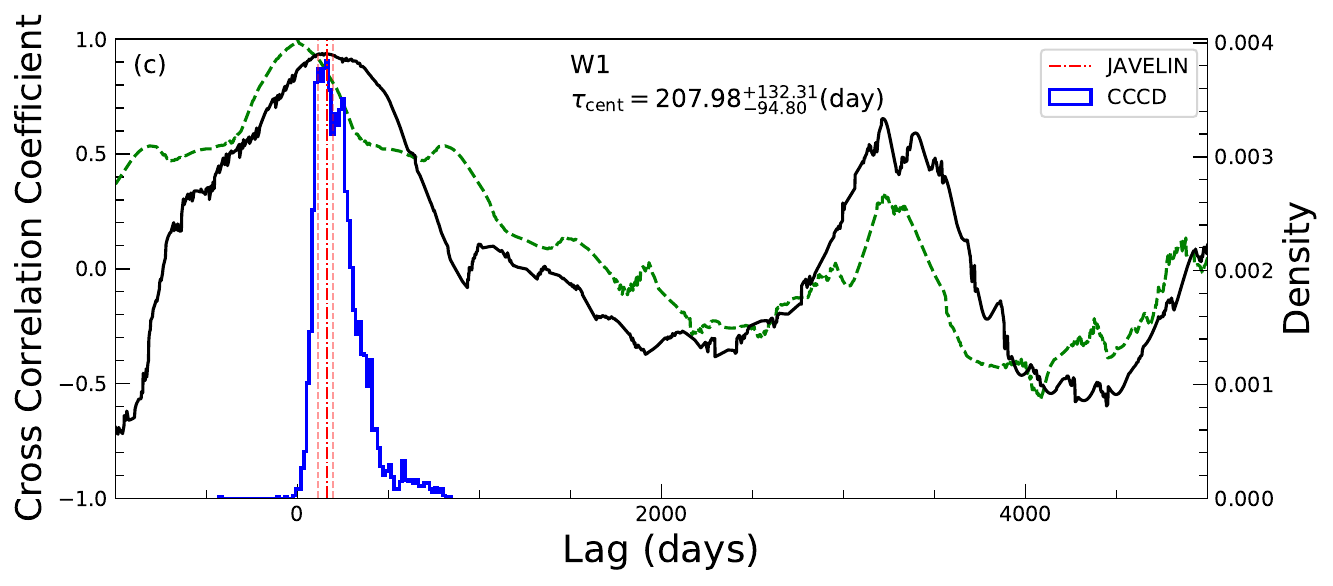}}
\resizebox{9.3cm}{4.8cm}{\includegraphics{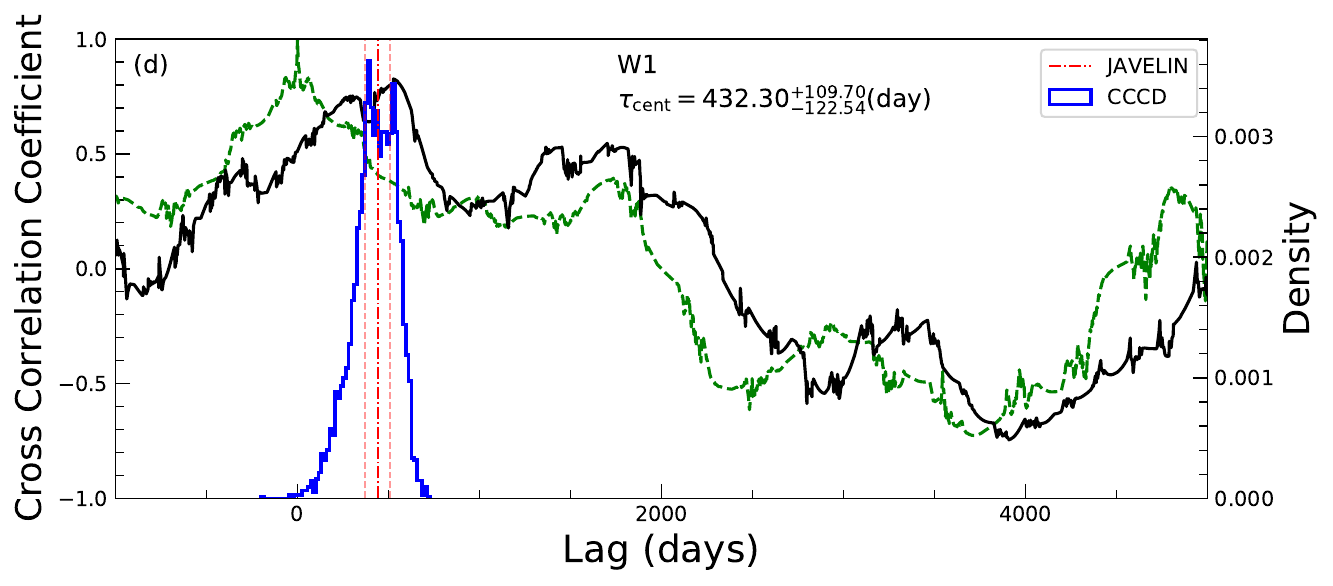}}
\resizebox{9.3cm}{4.8cm}{\includegraphics{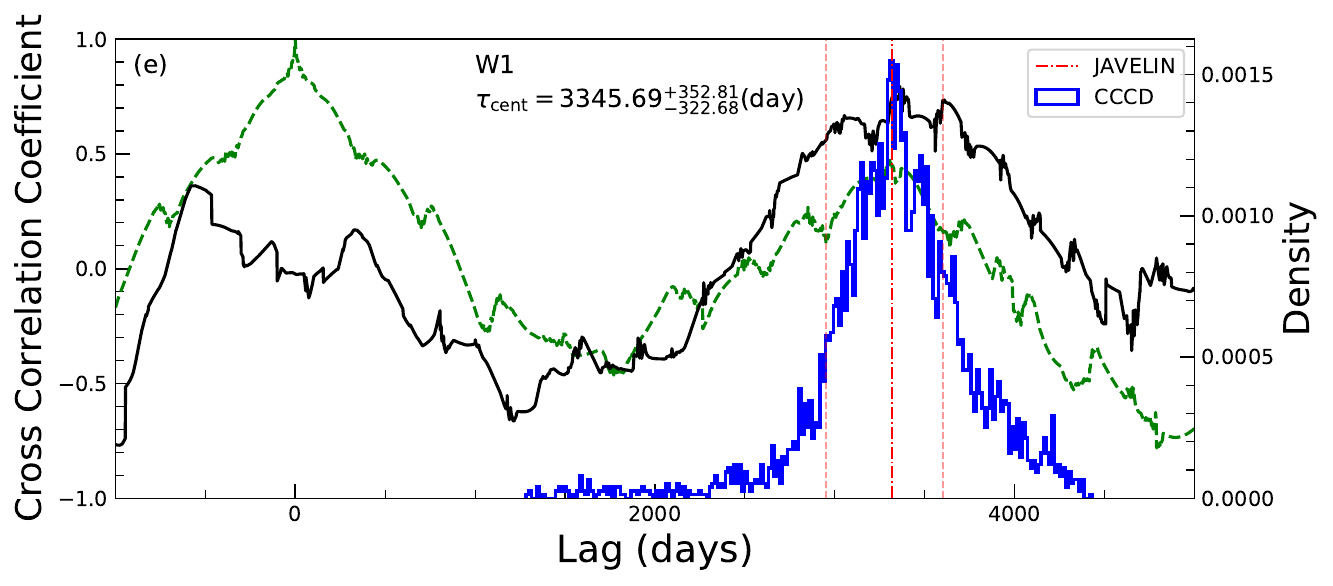}}
\resizebox{9.3cm}{4.8cm}{\includegraphics{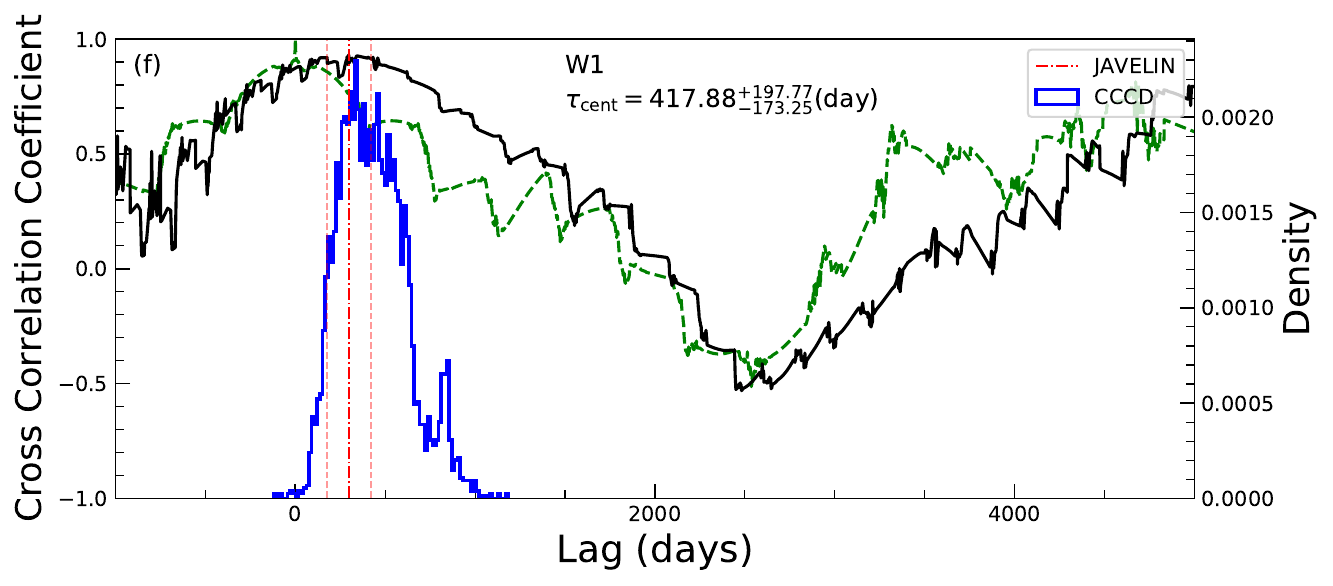}}

\caption{The ICCFs and ${\tt JAVELIN}$ analysis between optical and W1- band light curves of the example targets (a) ID47417, (b) ID1124174, (c) ID1189298,  (d) ID1524735, (e) ID2023374, and (f) ID4065339 from Sample 2. Same notations are used as Figure \ref{fig:ccf}.}
\label{fig:ccf_appS2}
\end{figure*}

We present examples of the light curves from a selection of 12 randomly chosen targets in Sample 1 and Sample 2, showcasing their wide range of bolometric luminosities ($10^{44.3}-10^{47.4} \, erg \, s^{-1}$). These examples, displayed in Figures \ref{fig:light_app} and \ref{fig:light_appS2}, include the light curves  in optical and IR W1 and W2- bands for Sample 1 and Sample 2, respectively. They serve as representative illustrations of our total sample. Furthermore, the lag measurements analyses between the optical and W1- band light curves for these targets, utilizing ICCF and {\tt JAVELIN}, are depicted in Figure \ref{fig:ccf_app} and \ref{fig:ccf_appS2}.

\section{Comparison of our lag measurements with earlier work} \label{lit_com}

\restartappendixnumbering

\begin{figure*}[!h]
\centering
\includegraphics[scale=0.6]{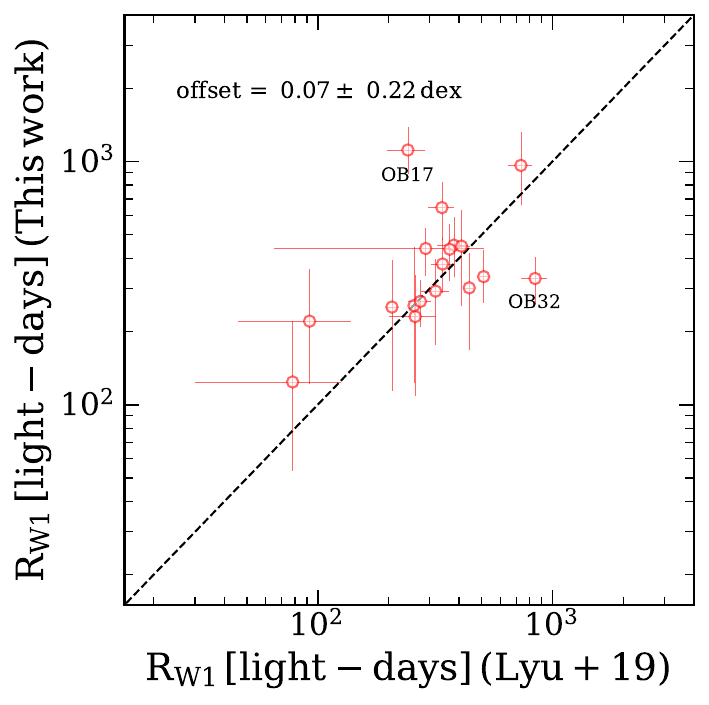}
\includegraphics[scale=0.6]{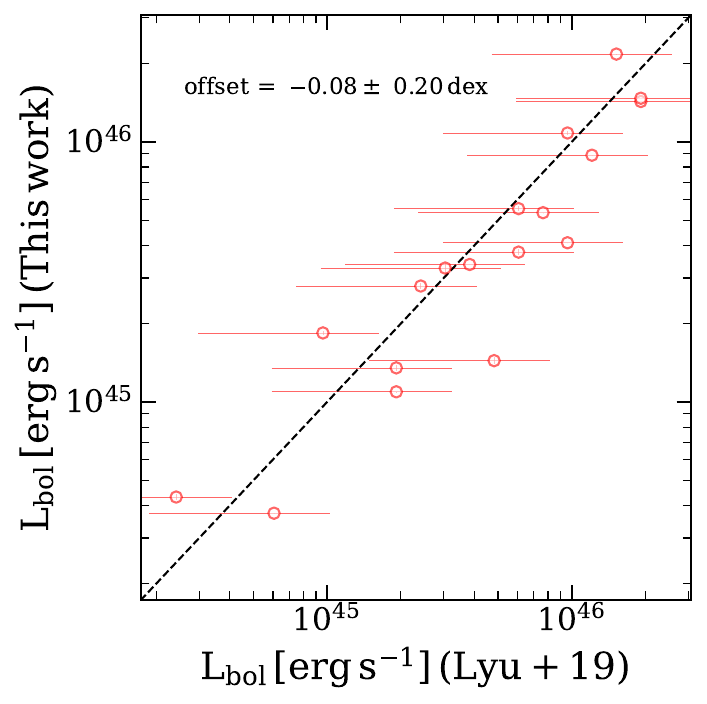}

\includegraphics[scale=0.57]{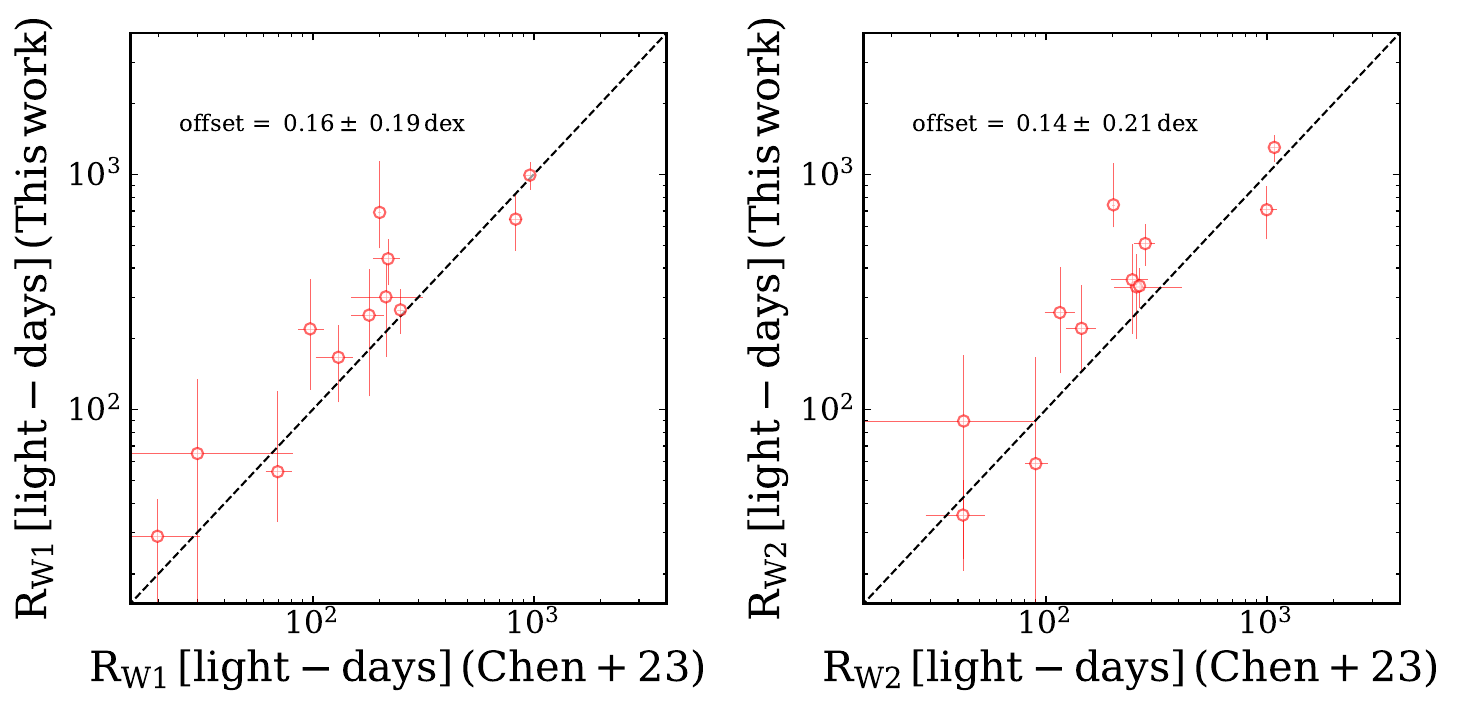}
\includegraphics[scale=0.5]{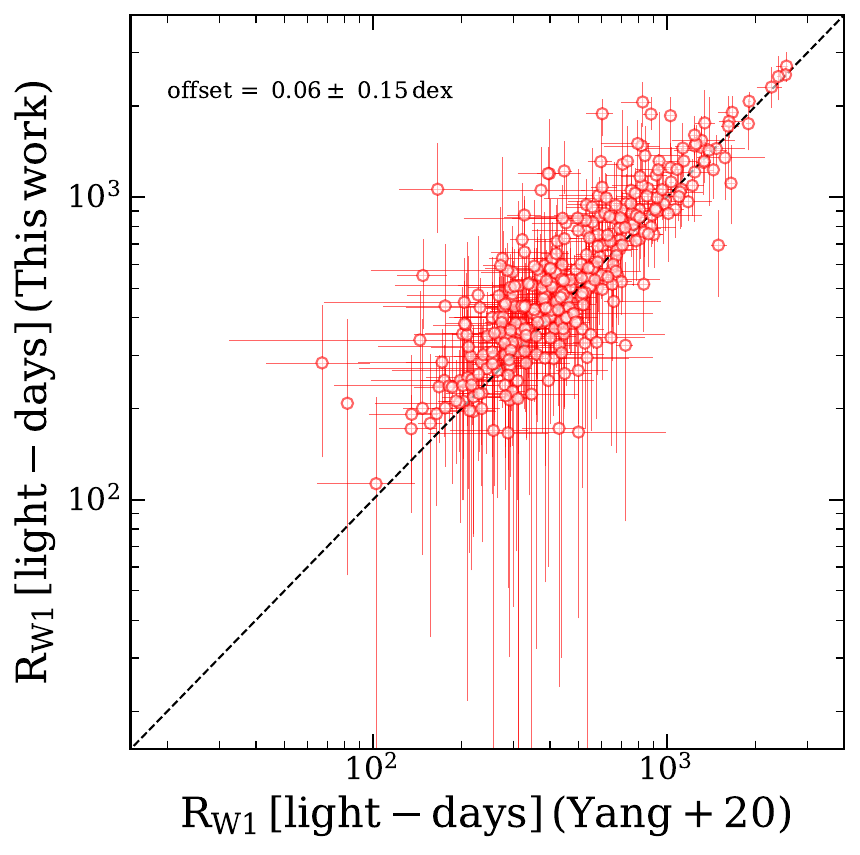}

\caption {Top panel: (Left) Comparison of the torus sizes in W1- band between our measurements  and that obtained by \citet{2019ApJ...886...33L}. These two measurements are found to be consistent with each other, except for two targets, namely OB17 and OB32, shown by individually marking them in the figure. (Right) Comparison of our obtained bolometric luminosities with that from  \citet{2019ApJ...886...33L} who performed optical-to-IR SED fitting to derive their bolometric luminosities.  Middle panel: a comparison of the torus sizes in both the W1 and W2- bands of our measurements (AD contamination corrected)  with the findings of  \citet{2023MNRAS.tmp.1073C} (AD contamination not corrected). Bottom panel: a comparison of the torus sizes in the W1-band between our measurements, corrected for AD contamination, and those from \citet{2020ApJ...900...58Y} without AD contamination correction.  The black dashed lines represent the 1:1 relation. The  offset values, along with their rms scatters as $1\sigma$ uncertainties are also shown at each panel in the figure.}
\label{fig:com_TL}
\end{figure*}

In the context of our study, it is important to note that \citet{2019ApJ...886...33L} previously reported lag measurements for 67 PG quasars. We identified 18 targets from our selected targets that overlap with the sample reported by \citet{2019ApJ...886...33L}. To assess the consistency between our new measurements and those obtained by \citet{2019ApJ...886...33L} using their independent $\chi^2-$fit method, we compare the torus sizes in W1- band in Figure \ref{fig:com_TL} (top left). With the exception of two targets, OB17 and OB32, our obtained torus sizes exhibit consistency (within 2$\sigma$) with those reported by \citet{2019ApJ...886...33L}, displaying an offset of 0.07 dex.

 In Figure \ref{fig:light_app}, we present the light curves of one of the most deviant targets, OB32, along with the corresponding results from the ICCF analysis shown in Figure \ref{fig:ccf_app}. Notably, the optical light curve of OB32, shifted by our recovered time lag, aligns well with the W1-band light curve. For determining AGN bolometric luminosity, \citet{2019ApJ...886...33L} employed optical-to-IR SED fitting and adopted an uncertainty of 0.3 dex in the luminosity. Figure \ref{fig:com_TL} (top right) demonstrates the comparable bolometric luminosities we obtained for these 18 targets, in line with those derived from optical-to-IR SED fitting by \citet{2019ApJ...886...33L}.

In addition, from our analysis, we identified a subset of 12 targets that are common to both our sample and the sample reported by \citet{2023MNRAS.tmp.1073C}. They used {\tt MICA} \citep{2016ApJ...831..206L} in their time-lag measurements. Despite the robust agreement observed between the measurements obtained through our approach and those reported by \citet{2023MNRAS.tmp.1073C} (See the middle panels of Figure \ref{fig:com_TL}), our derived torus sizes consistently appear larger with an offset of 0.16 (0.14) dex for the W1- (W2) bands. This discrepancy arises from the fact that \citet{2023MNRAS.tmp.1073C} did not perform an AD contamination correction in their analysis, whereas our obtained torus sizes are corrected for AD contamination in the W1 and W2- band light curves.

 Finally, we compare our measured torus sizes which are corrected for the AD contamination with those reported by \citet{2020ApJ...900...58Y}, who used {\tt JAVELIN} without employing AD contamination correction for Sample 2. Despite most of our measurements lying above the 1:1 line due to the AD contamination correction, the two sets of measurements exhibit good consistency, showing an offset of 0.06 dex and a scatter of 0.15 dex.

Note that to ensure consistency in our comparison, we obtained the corresponding rest-frame time lags by multiplying the estimated observed-frame lags both from the literature and our own analysis, by the redshift correction factor (1+$z$)$^{-0.38}$.

\section{Comparison between Eddington ratios derived from single epoch and reverberation mapped $M_{BH}$} \label{comp_edd}

\begin{figure}[!h]
\centering
\includegraphics[scale=0.7]{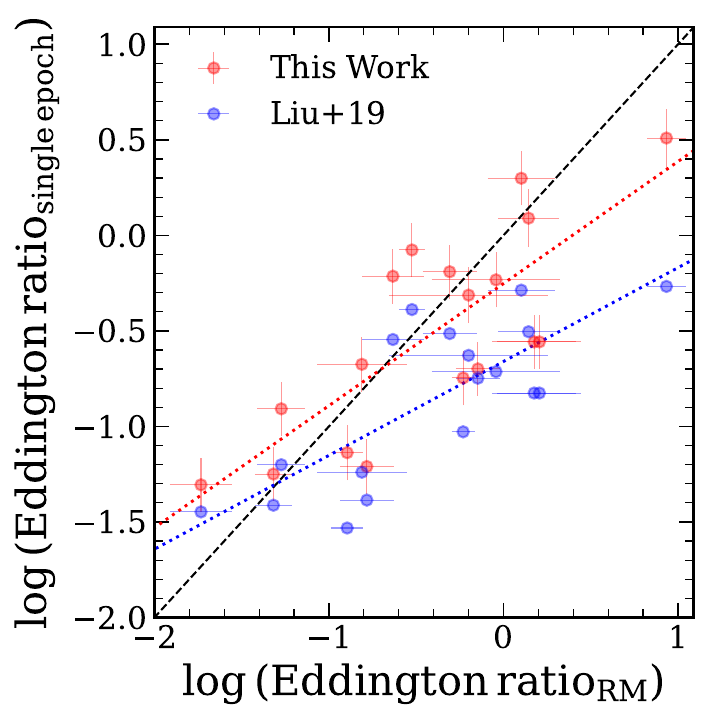}

\caption{Comparison between the Eddington ratios obtained from single epoch $M_{BH}$ and $M_{BH}$ from BLR--RM. The red and blue circles represent the objects with $\mathrm{Eddington \, ratio_{single \, epoch}}$ calculated in this work (see Section \ref{sec:AD_cor}) and retrieved from \citet{2019ApJS..243...21L}, respectively, with the linear-fits shown by the dotted lines. The black dashed line is the 1:1 relation.}
\label{fig:com_eddfig}
\end{figure}

We calculated the Eddington ratio for our sample by employing the single epoch $M_{BH}$ estimation method outlined in Section \ref{sec:AD_cor}. In Figure \ref{fig:com_eddfig}, we present a comparative analysis of the Eddington ratios derived from the single epoch $M_{BH}$ estimates in our study (represented by red circles) and those obtained by \citet{2019ApJS..243...21L} (depicted by blue circles) with those obtained by using $M_{BH}$ measurements from BLR--RM. Since the black hole mass derived from RM is considered to be more accurate, Figure \ref{fig:com_eddfig} clearly illustrates that the single epoch Eddington ratios tend to be underestimated as the accretion rate increases. This observation provides evidence for the need to account for the accurate measurement of $M_{BH}$ from BLR--RM to derive Eddington ratio.

\restartappendixnumbering

\bibliography{sample631}{}

\begin{thebibliography}{}
\expandafter\ifx\csname natexlab\endcsname\relax\def\natexlab#1{#1}\fi
\providecommand{\url}[1]{\href{#1}{#1}}
\providecommand{\dodoi}[1]{doi:~\href{http://doi.org/#1}{\nolinkurl{#1}}}
\providecommand{\doeprint}[1]{\href{http://ascl.net/#1}{\nolinkurl{http://ascl.net/#1}}}
\providecommand{\doarXiv}[1]{\href{https://arxiv.org/abs/#1}{\nolinkurl{https://arxiv.org/abs/#1}}}

\bibitem[{{Abbott} {et~al.}(2018){Abbott}, {Abdalla}, {Allam}, {Amara}, {Annis}, {Asorey}, {Avila}, {Ballester}, {Banerji}, {Barkhouse}, {Baruah}, {Baumer}, {Bechtol}, {Becker}, {Benoit-L{\'e}vy}, {Bernstein}, {Bertin}, {Blazek}, {Bocquet}, {Brooks}, {Brout}, {Buckley-Geer}, {Burke}, {Busti}, {Campisano}, {Cardiel-Sas}, {Carnero Rosell}, {Carrasco Kind}, {Carretero}, {Castander}, {Cawthon}, {Chang}, {Chen}, {Conselice}, {Costa}, {Crocce}, {Cunha}, {D'Andrea}, {da Costa}, {Das}, {Daues}, {Davis}, {Davis}, {De Vicente}, {DePoy}, {DeRose}, {Desai}, {Diehl}, {Dietrich}, {Dodelson}, {Doel}, {Drlica-Wagner}, {Eifler}, {Elliott}, {Evrard}, {Farahi}, {Fausti Neto}, {Fernandez}, {Finley}, {Flaugher}, {Foley}, {Fosalba}, {Friedel}, {Frieman}, {Garc{\'\i}a-Bellido}, {Gaztanaga}, {Gerdes}, {Giannantonio}, {Gill}, {Glazebrook}, {Goldstein}, {Gower}, {Gruen}, {Gruendl}, {Gschwend}, {Gupta}, {Gutierrez}, {Hamilton}, {Hartley}, {Hinton}, {Hislop}, {Hollowood}, {Honscheid}, {Hoyle}, {Huterer}, {Jain}, {James}, {Jeltema},
  {Johnson}, {Johnson}, {Kacprzak}, {Kent}, {Khullar}, {Klein}, {Kovacs}, {Koziol}, {Krause}, {Kremin}, {Kron}, {Kuehn}, {Kuhlmann}, {Kuropatkin}, {Lahav}, {Lasker}, {Li}, {Li}, {Liddle}, {Lima}, {Lin}, {L{\'o}pez-Reyes}, {MacCrann}, {Maia}, {Maloney}, {Manera}, {March}, {Marriner}, {Marshall}, {Martini}, {McClintock}, {McKay}, {McMahon}, {Melchior}, {Menanteau}, {Miller}, {Miquel}, {Mohr}, {Morganson}, {Mould}, {Neilsen}, {Nichol}, {Nogueira}, {Nord}, {Nugent}, {Nunes}, {Ogando}, {Old}, {Pace}, {Palmese}, {Paz-Chinch{\'o}n}, {Peiris}, {Percival}, {Petravick}, {Plazas}, {Poh}, {Pond}, {Porredon}, {Pujol}, {Refregier}, {Reil}, {Ricker}, {Rollins}, {Romer}, {Roodman}, {Rooney}, {Ross}, {Rykoff}, {Sako}, {Sanchez}, {Sanchez}, {Santiago}, {Saro}, {Scarpine}, {Scolnic}, {Serrano}, {Sevilla-Noarbe}, {Sheldon}, {Shipp}, {Silveira}, {Smith}, {Smith}, {Smith}, {Soares-Santos}, {Sobreira}, {Song}, {Stebbins}, {Suchyta}, {Sullivan}, {Swanson}, {Tarle}, {Thaler}, {Thomas}, {Thomas}, {Troxel}, {Tucker}, {Vikram}, {Vivas},
  {Walker}, {Wechsler}, {Weller}, {Wester}, {Wolf}, {Wu}, {Yanny}, {Zenteno}, {Zhang}, {Zuntz}, {DES Collaboration}, {Juneau}, {Fitzpatrick}, {Nikutta}, {Nidever}, {Olsen}, {Scott}, \& {NOAO Data Lab}}]{2018ApJS..239...18A}
{Abbott}, T.~M.~C., {Abdalla}, F.~B., {Allam}, S., {et~al.} 2018, \apjs, 239, 18, \dodoi{10.3847/1538-4365/aae9f0}

\bibitem[{{Akritas} \& {Bershady}(1996)}]{1996ApJ...470..706A}
{Akritas}, M.~G., \& {Bershady}, M.~A. 1996, \apj, 470, 706, \dodoi{10.1086/177901}

\bibitem[{{Almeyda} {et~al.}(2020){Almeyda}, {Robinson}, {Richmond}, {Nikutta}, \& {McDonough}}]{2020ApJ...891...26A}
{Almeyda}, T., {Robinson}, A., {Richmond}, M., {Nikutta}, R., \& {McDonough}, B. 2020, \apj, 891, 26, \dodoi{10.3847/1538-4357/ab6aa1}

\bibitem[{{Almeyda} {et~al.}(2017){Almeyda}, {Robinson}, {Richmond}, {Vazquez}, \& {Nikutta}}]{2017ApJ...843....3A}
{Almeyda}, T., {Robinson}, A., {Richmond}, M., {Vazquez}, B., \& {Nikutta}, R. 2017, \apj, 843, 3, \dodoi{10.3847/1538-4357/aa7687}

\bibitem[{{Baldwin} {et~al.}(1995){Baldwin}, {Ferland}, {Korista}, \& {Verner}}]{1995ApJ...455L.119B}
{Baldwin}, J., {Ferland}, G., {Korista}, K., \& {Verner}, D. 1995, \apjl, 455, L119, \dodoi{10.1086/309827}

\bibitem[{{Bao} {et~al.}(2022){Bao}, {Brotherton}, {Du}, {McLane}, {Zastrocky}, {Olson}, {Fang}, {Zhai}, {Huang}, {Wang}, {Zhao}, {Li}, {Yang}, {Chen}, {Liu}, {Yao}, {Peng}, {Guo}, {Songsheng}, {Li}, {Jiang}, {Kasper}, {Chick}, {Nguyen}, {Maithil}, {Kobulnicky}, {Dale}, {Hand}, {Adelman}, {Carter}, {Murphree}, {Oeur}, {Schonsberg}, {Roth}, {Winkler}, {Marziani}, {D'Onofrio}, {Hu}, {Xiao}, {Xue}, {Czerny}, {Aceituno}, {Ho}, {Bai}, {Wang}, \& {MAHA Collaboration}}]{2022ApJS..262...14B}
{Bao}, D.-W., {Brotherton}, M.~S., {Du}, P., {et~al.} 2022, \apjs, 262, 14, \dodoi{10.3847/1538-4365/ac7beb}

\bibitem[{{Barvainis}(1987)}]{1987ApJ...320..537B}
{Barvainis}, R. 1987, \apj, 320, 537, \dodoi{10.1086/165571}

\bibitem[{{Bellm} {et~al.}(2019){Bellm}, {Kulkarni}, {Graham}, {Dekany}, {Smith}, {Riddle}, {Masci}, {Helou}, {Prince}, {Adams}, {Barbarino}, {Barlow}, {Bauer}, {Beck}, {Belicki}, {Biswas}, {Blagorodnova}, {Bodewits}, {Bolin}, {Brinnel}, {Brooke}, {Bue}, {Bulla}, {Burruss}, {Cenko}, {Chang}, {Connolly}, {Coughlin}, {Cromer}, {Cunningham}, {De}, {Delacroix}, {Desai}, {Duev}, {Eadie}, {Farnham}, {Feeney}, {Feindt}, {Flynn}, {Franckowiak}, {Frederick}, {Fremling}, {Gal-Yam}, {Gezari}, {Giomi}, {Goldstein}, {Golkhou}, {Goobar}, {Groom}, {Hacopians}, {Hale}, {Henning}, {Ho}, {Hover}, {Howell}, {Hung}, {Huppenkothen}, {Imel}, {Ip}, {Ivezi{\'c}}, {Jackson}, {Jones}, {Juric}, {Kasliwal}, {Kaspi}, {Kaye}, {Kelley}, {Kowalski}, {Kramer}, {Kupfer}, {Landry}, {Laher}, {Lee}, {Lin}, {Lin}, {Lunnan}, {Giomi}, {Mahabal}, {Mao}, {Miller}, {Monkewitz}, {Murphy}, {Ngeow}, {Nordin}, {Nugent}, {Ofek}, {Patterson}, {Penprase}, {Porter}, {Rauch}, {Rebbapragada}, {Reiley}, {Rigault}, {Rodriguez}, {van Roestel}, {Rusholme}, {van
  Santen}, {Schulze}, {Shupe}, {Singer}, {Soumagnac}, {Stein}, {Surace}, {Sollerman}, {Szkody}, {Taddia}, {Terek}, {Van Sistine}, {van Velzen}, {Vestrand}, {Walters}, {Ward}, {Ye}, {Yu}, {Yan}, \& {Zolkower}}]{2019PASP..131a8002B}
{Bellm}, E.~C., {Kulkarni}, S.~R., {Graham}, M.~J., {et~al.} 2019, \pasp, 131, 018002, \dodoi{10.1088/1538-3873/aaecbe}

\bibitem[{{Bentz} {et~al.}(2009){Bentz}, {Walsh}, {Barth}, {Baliber}, {Bennert}, {Canalizo}, {Filippenko}, {Ganeshalingam}, {Gates}, {Greene}, {Hidas}, {Hiner}, {Lee}, {Li}, {Malkan}, {Minezaki}, {Sakata}, {Serduke}, {Silverman}, {Steele}, {Stern}, {Street}, {Thornton}, {Treu}, {Wang}, {Woo}, \& {Yoshii}}]{2009ApJ...705..199B}
{Bentz}, M.~C., {Walsh}, J.~L., {Barth}, A.~J., {et~al.} 2009, \apj, 705, 199, \dodoi{10.1088/0004-637X/705/1/199}

\bibitem[{{Bertin} \& {Arnouts}(1996)}]{1996A&AS..117..393B}
{Bertin}, E., \& {Arnouts}, S. 1996, \aaps, 117, 393, \dodoi{10.1051/aas:1996164}

\bibitem[{{Bessell} {et~al.}(1998){Bessell}, {Castelli}, \& {Plez}}]{1998A&A...333..231B}
{Bessell}, M.~S., {Castelli}, F., \& {Plez}, B. 1998, \aap, 333, 231

\bibitem[{{Blandford} \& {McKee}(1982)}]{1982ApJ...255..419B}
{Blandford}, R.~D., \& {McKee}, C.~F. 1982, \apj, 255, 419, \dodoi{10.1086/159843}

\bibitem[{{Boroson} \& {Green}(1992)}]{1992ApJS...80..109B}
{Boroson}, T.~A., \& {Green}, R.~F. 1992, \apjs, 80, 109, \dodoi{10.1086/191661}

\bibitem[{{Burtscher} {et~al.}(2013){Burtscher}, {Meisenheimer}, {Tristram}, {Jaffe}, {H{\"o}nig}, {Davies}, {Kishimoto}, {Pott}, {R{\"o}ttgering}, {Schartmann}, {Weigelt}, \& {Wolf}}]{2013A&A...558A.149B}
{Burtscher}, L., {Meisenheimer}, K., {Tristram}, K.~R.~W., {et~al.} 2013, \aap, 558, A149, \dodoi{10.1051/0004-6361/201321890}

\bibitem[{{Chambers} {et~al.}(2016){Chambers}, {Magnier}, {Metcalfe}, {Flewelling}, {Huber}, {Waters}, {Denneau}, {Draper}, {Farrow}, {Finkbeiner}, {Holmberg}, {Koppenhoefer}, {Price}, {Rest}, {Saglia}, {Schlafly}, {Smartt}, {Sweeney}, {Wainscoat}, {Burgett}, {Chastel}, {Grav}, {Heasley}, {Hodapp}, {Jedicke}, {Kaiser}, {Kudritzki}, {Luppino}, {Lupton}, {Monet}, {Morgan}, {Onaka}, {Shiao}, {Stubbs}, {Tonry}, {White}, {Ba{\~n}ados}, {Bell}, {Bender}, {Bernard}, {Boegner}, {Boffi}, {Botticella}, {Calamida}, {Casertano}, {Chen}, {Chen}, {Cole}, {Deacon}, {Frenk}, {Fitzsimmons}, {Gezari}, {Gibbs}, {Goessl}, {Goggia}, {Gourgue}, {Goldman}, {Grant}, {Grebel}, {Hambly}, {Hasinger}, {Heavens}, {Heckman}, {Henderson}, {Henning}, {Holman}, {Hopp}, {Ip}, {Isani}, {Jackson}, {Keyes}, {Koekemoer}, {Kotak}, {Le}, {Liska}, {Long}, {Lucey}, {Liu}, {Martin}, {Masci}, {McLean}, {Mindel}, {Misra}, {Morganson}, {Murphy}, {Obaika}, {Narayan}, {Nieto-Santisteban}, {Norberg}, {Peacock}, {Pier}, {Postman}, {Primak}, {Rae}, {Rai},
  {Riess}, {Riffeser}, {Rix}, {R{\"o}ser}, {Russel}, {Rutz}, {Schilbach}, {Schultz}, {Scolnic}, {Strolger}, {Szalay}, {Seitz}, {Small}, {Smith}, {Soderblom}, {Taylor}, {Thomson}, {Taylor}, {Thakar}, {Thiel}, {Thilker}, {Unger}, {Urata}, {Valenti}, {Wagner}, {Walder}, {Walter}, {Watters}, {Werner}, {Wood-Vasey}, \& {Wyse}}]{2016arXiv161205560C}
{Chambers}, K.~C., {Magnier}, E.~A., {Metcalfe}, N., {et~al.} 2016, arXiv e-prints, arXiv:1612.05560, \dodoi{10.48550/arXiv.1612.05560}

\bibitem[{{Chen} {et~al.}(2023){Chen}, {Liu}, {Zhai}, {Yao}, {Li}, {Du}, {Hu}, {Guo}, {Xiao}, {Songsheng}, \& {Wang}}]{2023MNRAS.tmp.1073C}
{Chen}, Y.-J., {Liu}, J.-R., {Zhai}, S., {et~al.} 2023, \mnras, \dodoi{10.1093/mnras/stad1136}

\bibitem[{{Czerny} {et~al.}(2019){Czerny}, {Wang}, {Du}, {Hryniewicz}, {Karas}, {Li}, {Panda}, {Sniegowska}, {Wildy}, \& {Yuan}}]{2019ApJ...870...84C}
{Czerny}, B., {Wang}, J.-M., {Du}, P., {et~al.} 2019, \apj, 870, 84, \dodoi{10.3847/1538-4357/aaf396}

\bibitem[{{Davidson}(1972)}]{1972ApJ...171..213D}
{Davidson}, K. 1972, \apj, 171, 213, \dodoi{10.1086/151274}

\bibitem[{{Denney} {et~al.}(2010){Denney}, {Peterson}, {Pogge}, {Adair}, {Atlee}, {Au-Yong}, {Bentz}, {Bird}, {Brokofsky}, {Chisholm}, {Comins}, {Dietrich}, {Doroshenko}, {Eastman}, {Efimov}, {Ewald}, {Ferbey}, {Gaskell}, {Hedrick}, {Jackson}, {Klimanov}, {Klimek}, {Kruse}, {Lad{\'e}route}, {Lamb}, {Leighly}, {Minezaki}, {Nazarov}, {Onken}, {Petersen}, {Peterson}, {Poindexter}, {Sakata}, {Schlesinger}, {Sergeev}, {Skolski}, {Stieglitz}, {Tobin}, {Unterborn}, {Vestergaard}, {Watkins}, {Watson}, \& {Yoshii}}]{2010ApJ...721..715D}
{Denney}, K.~D., {Peterson}, B.~M., {Pogge}, R.~W., {et~al.} 2010, \apj, 721, 715, \dodoi{10.1088/0004-637X/721/1/715}

\bibitem[{{Drake} {et~al.}(2009){Drake}, {Djorgovski}, {Mahabal}, {Beshore}, {Larson}, {Graham}, {Williams}, {Christensen}, {Catelan}, {Boattini}, {Gibbs}, {Hill}, \& {Kowalski}}]{2009ApJ...696..870D}
{Drake}, A.~J., {Djorgovski}, S.~G., {Mahabal}, A., {et~al.} 2009, \apj, 696, 870, \dodoi{10.1088/0004-637X/696/1/870}

\bibitem[{{Du} \& {Wang}(2019)}]{2019ApJ...886...42D}
{Du}, P., \& {Wang}, J.-M. 2019, \apj, 886, 42, \dodoi{10.3847/1538-4357/ab4908}

\bibitem[{{Du} {et~al.}(2014){Du}, {Hu}, {Lu}, {Wang}, {Qiu}, {Li}, {Bai}, {Kaspi}, {Netzer}, {Wang}, \& {SEAMBH Collaboration}}]{2014ApJ...782...45D}
{Du}, P., {Hu}, C., {Lu}, K.-X., {et~al.} 2014, \apj, 782, 45, \dodoi{10.1088/0004-637X/782/1/45}

\bibitem[{{Du} {et~al.}(2015){Du}, {Hu}, {Lu}, {Huang}, {Cheng}, {Qiu}, {Li}, {Zhang}, {Fan}, {Bai}, {Bian}, {Yuan}, {Kaspi}, {Ho}, {Netzer}, {Wang}, \& {SEAMBH Collaboration}}]{2015ApJ...806...22D}
---. 2015, \apj, 806, 22, \dodoi{10.1088/0004-637X/806/1/22}

\bibitem[{{Du} {et~al.}(2018){Du}, {Zhang}, {Wang}, {Huang}, {Zhang}, {Lu}, {Hu}, {Li}, {Bai}, {Bian}, {Yuan}, {Ho}, {Wang}, \& {SEAMBH Collaboration}}]{2018ApJ...856....6D}
{Du}, P., {Zhang}, Z.-X., {Wang}, K., {et~al.} 2018, \apj, 856, 6, \dodoi{10.3847/1538-4357/aaae6b}

\bibitem[{{Elvis} {et~al.}(1994){Elvis}, {Wilkes}, {McDowell}, {Green}, {Bechtold}, {Willner}, {Oey}, {Polomski}, \& {Cutri}}]{1994ApJS...95....1E}
{Elvis}, M., {Wilkes}, B.~J., {McDowell}, J.~C., {et~al.} 1994, \apjs, 95, 1, \dodoi{10.1086/192093}

\bibitem[{{Feigelson} \& {Babu}(1992)}]{1992ApJ...397...55F}
{Feigelson}, E.~D., \& {Babu}, G.~J. 1992, \apj, 397, 55, \dodoi{10.1086/171766}

\bibitem[{{Fonseca Alvarez} {et~al.}(2020){Fonseca Alvarez}, {Trump}, {Homayouni}, {Grier}, {Shen}, {Horne}, {Li}, {Brandt}, {Ho}, {Peterson}, \& {Schneider}}]{2020ApJ...899...73F}
{Fonseca Alvarez}, G., {Trump}, J.~R., {Homayouni}, Y., {et~al.} 2020, \apj, 899, 73, \dodoi{10.3847/1538-4357/aba001}

\bibitem[{{Fukugita} {et~al.}(1996){Fukugita}, {Ichikawa}, {Gunn}, {Doi}, {Shimasaku}, \& {Schneider}}]{1996AJ....111.1748F}
{Fukugita}, M., {Ichikawa}, T., {Gunn}, J.~E., {et~al.} 1996, \aj, 111, 1748, \dodoi{10.1086/117915}

\bibitem[{{Gandhi} {et~al.}(2015){Gandhi}, {H{\"o}nig}, \& {Kishimoto}}]{2015ApJ...812..113G}
{Gandhi}, P., {H{\"o}nig}, S.~F., \& {Kishimoto}, M. 2015, \apj, 812, 113, \dodoi{10.1088/0004-637X/812/2/113}

\bibitem[{{Garc{\'\i}a-Burillo} {et~al.}(2019){Garc{\'\i}a-Burillo}, {Combes}, {Ramos Almeida}, {Usero}, {Alonso-Herrero}, {Hunt}, {Rouan}, {Aalto}, {Querejeta}, {Viti}, {van der Werf}, {Vives-Arias}, {Fuente}, {Colina}, {Mart{\'\i}n-Pintado}, {Henkel}, {Mart{\'\i}n}, {Krips}, {Gratadour}, {Neri}, \& {Tacconi}}]{2019A&A...632A..61G}
{Garc{\'\i}a-Burillo}, S., {Combes}, F., {Ramos Almeida}, C., {et~al.} 2019, \aap, 632, A61, \dodoi{10.1051/0004-6361/201936606}

\bibitem[{{Gaskell} \& {Peterson}(1987)}]{1987ApJS...65....1G}
{Gaskell}, C.~M., \& {Peterson}, B.~M. 1987, \apjs, 65, 1, \dodoi{10.1086/191216}

\bibitem[{{Gaskell} \& {Sparke}(1986)}]{1986ApJ...305..175G}
{Gaskell}, C.~M., \& {Sparke}, L.~S. 1986, \apj, 305, 175, \dodoi{10.1086/164238}

\bibitem[{{Gravity Collaboration} {et~al.}(2020){Gravity Collaboration}, {Dexter}, {Shangguan}, {H{\"o}nig}, {Kishimoto}, {Lutz}, {Netzer}, {Davies}, {Sturm}, {Pfuhl}, {Amorim}, {Baub{\"o}ck}, {Brandner}, {Cl{\'e}net}, {de Zeeuw}, {Eckart}, {Eisenhauer}, {F{\"o}rster Schreiber}, {Gao}, {Garcia}, {Genzel}, {Gillessen}, {Gratadour}, {Jim{\'e}nez-Rosales}, {Lacour}, {Millour}, {Ott}, {Paumard}, {Perraut}, {Perrin}, {Peterson}, {Petrucci}, {Prieto}, {Rouan}, {Schartmann}, {Shimizu}, {Sternberg}, {Straub}, {Straubmeier}, {Tacconi}, {Tristram}, {Vermot}, {Waisberg}, {Widmann}, \& {Woillez}}]{2020A&A...635A..92G}
{Gravity Collaboration}, {Dexter}, J., {Shangguan}, J., {et~al.} 2020, \aap, 635, A92, \dodoi{10.1051/0004-6361/201936767}

\bibitem[{{Gravity Collaboration} {et~al.}(2023){Gravity Collaboration}, {Amorim}, {Bourdarot}, {Brandner}, {Cao}, {Cl{\'e}net}, {Davies}, {de Zeeuw}, {Dexter}, {Drescher}, {Eckart}, {Eisenhauer}, {Fabricius}, {F{\"o}rster Schreiber}, {Garcia}, {Genzel}, {Gillessen}, {Gratadour}, {H{\"o}nig}, {Kishimoto}, {Lacour}, {Lutz}, {Millour}, {Netzer}, {Ott}, {Paumard}, {Perraut}, {Perrin}, {Peterson}, {Petrucci}, {Pfuhl}, {Prieto}, {Rouan}, {Santos}, {Shangguan}, {Shimizu}, {Sternberg}, {Straubmeier}, {Sturm}, {Tacconi}, {Tristram}, {Widmann}, \& {Woillez}}]{2023A&A...669A..14G}
{Gravity Collaboration}, {Amorim}, A., {Bourdarot}, G., {et~al.} 2023, \aap, 669, A14, \dodoi{10.1051/0004-6361/202244655}

\bibitem[{{Green} {et~al.}(2012){Green}, {Schechter}, {Baltay}, {Bean}, {Bennett}, {Brown}, {Conselice}, {Donahue}, {Fan}, {Gaudi}, {Hirata}, {Kalirai}, {Lauer}, {Nichol}, {Padmanabhan}, {Perlmutter}, {Rauscher}, {Rhodes}, {Roellig}, {Stern}, {Sumi}, {Tanner}, {Wang}, {Weinberg}, {Wright}, {Gehrels}, {Sambruna}, {Traub}, {Anderson}, {Cook}, {Garnavich}, {Hillenbrand}, {Ivezic}, {Kerins}, {Lunine}, {McDonald}, {Penny}, {Phillips}, {Rieke}, {Riess}, {van der Marel}, {Barry}, {Cheng}, {Content}, {Cutri}, {Goullioud}, {Grady}, {Helou}, {Jackson}, {Kruk}, {Melton}, {Peddie}, {Rioux}, \& {Seiffert}}]{2012arXiv1208.4012G}
{Green}, J., {Schechter}, P., {Baltay}, C., {et~al.} 2012, arXiv e-prints, arXiv:1208.4012, \dodoi{10.48550/arXiv.1208.4012}

\bibitem[{{Grier} {et~al.}(2017){Grier}, {Trump}, {Shen}, {Horne}, {Kinemuchi}, {McGreer}, {Starkey}, {Brandt}, {Hall}, {Kochanek}, {Chen}, {Denney}, {Greene}, {Ho}, {Homayouni}, {I-Hsiu Li}, {Pei}, {Peterson}, {Petitjean}, {Schneider}, {Sun}, {AlSayyad}, {Bizyaev}, {Brinkmann}, {Brownstein}, {Bundy}, {Dawson}, {Eftekharzadeh}, {Fernandez-Trincado}, {Gao}, {Hutchinson}, {Jia}, {Jiang}, {Oravetz}, {Pan}, {Paris}, {Ponder}, {Peters}, {Rogerson}, {Simmons}, {Smith}, \& {Wang}}]{2017ApJ...851...21G}
{Grier}, C.~J., {Trump}, J.~R., {Shen}, Y., {et~al.} 2017, \apj, 851, 21, \dodoi{10.3847/1538-4357/aa98dc}

\bibitem[{{Grier} {et~al.}(2019){Grier}, {Shen}, {Horne}, {Brandt}, {Trump}, {Hall}, {Kinemuchi}, {Starkey}, {Schneider}, {Ho}, {Homayouni}, {I-Hsiu Li}, {McGreer}, {Peterson}, {Bizyaev}, {Chen}, {Dawson}, {Eftekharzadeh}, {Guo}, {Jia}, {Jiang}, {Kneib}, {Li}, {Li}, {Nie}, {Oravetz}, {Oravetz}, {Pan}, {Petitjean}, {Ponder}, {Rogerson}, {Vivek}, {Zhang}, \& {Zou}}]{2019ApJ...887...38G}
{Grier}, C.~J., {Shen}, Y., {Horne}, K., {et~al.} 2019, \apj, 887, 38, \dodoi{10.3847/1538-4357/ab4ea5}

\bibitem[{{Guise} {et~al.}(2022){Guise}, {H{\"o}nig}, {Gorjian}, {Barth}, {Almeyda}, {Pei}, {Cenko}, {Edelson}, {Filippenko}, {Joner}, {Laney}, {Li}, {Malkan}, {Nguyen}, \& {Zheng}}]{2022MNRAS.516.4898G}
{Guise}, E., {H{\"o}nig}, S.~F., {Gorjian}, V., {et~al.} 2022, \mnras, 516, 4898, \dodoi{10.1093/mnras/stac2529}

\bibitem[{{Guo} \& {Barth}(2021)}]{2021AAS...23722608G}
{Guo}, H., \& {Barth}, A.~J. 2021, in American Astronomical Society Meeting Abstracts, Vol.~53, American Astronomical Society Meeting Abstracts, 226.08

\bibitem[{{Guo} {et~al.}(2022){Guo}, {Barth}, \& {Wang}}]{2022ApJ...940...20G}
{Guo}, H., {Barth}, A.~J., \& {Wang}, S. 2022, \apj, 940, 20, \dodoi{10.3847/1538-4357/ac96ec}

\bibitem[{{Homayouni} {et~al.}(2020){Homayouni}, {Trump}, {Grier}, {Horne}, {Shen}, {Brandt}, {Dawson}, {Alvarez}, {Green}, {Hall}, {Hern{\'a}ndez Santisteban}, {Ho}, {Kinemuchi}, {Kochanek}, {Li}, {Peterson}, {Schneider}, {Starkey}, {Bizyaev}, {Pan}, {Oravetz}, \& {Simmons}}]{2020ApJ...901...55H}
{Homayouni}, Y., {Trump}, J.~R., {Grier}, C.~J., {et~al.} 2020, \apj, 901, 55, \dodoi{10.3847/1538-4357/ababa9}

\bibitem[{{H{\"o}nig} {et~al.}(2017){H{\"o}nig}, {Watson}, {Kishimoto}, {Gandhi}, {Goad}, {Horne}, {Shankar}, {Banerji}, {Boulderstone}, {Jarvis}, {Smith}, \& {Sullivan}}]{2017MNRAS.464.1693H}
{H{\"o}nig}, S.~F., {Watson}, D., {Kishimoto}, M., {et~al.} 2017, \mnras, 464, 1693, \dodoi{10.1093/mnras/stw2484}

\bibitem[{{Hovatta} {et~al.}(2014){Hovatta}, {Pavlidou}, {King}, {Mahabal}, {Sesar}, {Dancikova}, {Djorgovski}, {Drake}, {Laher}, {Levitan}, {Max-Moerbeck}, {Ofek}, {Pearson}, {Prince}, {Readhead}, {Richards}, \& {Surace}}]{2014MNRAS.439..690H}
{Hovatta}, T., {Pavlidou}, V., {King}, O.~G., {et~al.} 2014, \mnras, 439, 690, \dodoi{10.1093/mnras/stt2494}

\bibitem[{{Hu} {et~al.}(2021){Hu}, {Li}, {Yang}, {Yang}, {Guo}, {Bao}, {Jiang}, {Du}, {Li}, {Xiao}, {Songsheng}, {Yu}, {Bai}, {Ho}, {Brotherton}, {Aceituno}, {Winkler}, {Wang}, \& {Seambh Collaboration}}]{2021ApJS..253...20H}
{Hu}, C., {Li}, S.-S., {Yang}, S., {et~al.} 2021, \apjs, 253, 20, \dodoi{10.3847/1538-4365/abd774}

\bibitem[{{Isobe} {et~al.}(1990){Isobe}, {Feigelson}, {Akritas}, \& {Babu}}]{1990ApJ...364..104I}
{Isobe}, T., {Feigelson}, E.~D., {Akritas}, M.~G., \& {Babu}, G.~J. 1990, \apj, 364, 104, \dodoi{10.1086/169390}

\bibitem[{{Ivezi{\'c}} {et~al.}(2019){Ivezi{\'c}}, {Kahn}, {Tyson}, {Abel}, {Acosta}, {Allsman}, {Alonso}, {AlSayyad}, {Anderson}, {Andrew}, {Angel}, {Angeli}, {Ansari}, {Antilogus}, {Araujo}, {Armstrong}, {Arndt}, {Astier}, {Aubourg}, {Auza}, {Axelrod}, {Bard}, {Barr}, {Barrau}, {Bartlett}, {Bauer}, {Bauman}, {Baumont}, {Bechtol}, {Bechtol}, {Becker}, {Becla}, {Beldica}, {Bellavia}, {Bianco}, {Biswas}, {Blanc}, {Blazek}, {Blandford}, {Bloom}, {Bogart}, {Bond}, {Booth}, {Borgland}, {Borne}, {Bosch}, {Boutigny}, {Brackett}, {Bradshaw}, {Brandt}, {Brown}, {Bullock}, {Burchat}, {Burke}, {Cagnoli}, {Calabrese}, {Callahan}, {Callen}, {Carlin}, {Carlson}, {Chandrasekharan}, {Charles-Emerson}, {Chesley}, {Cheu}, {Chiang}, {Chiang}, {Chirino}, {Chow}, {Ciardi}, {Claver}, {Cohen-Tanugi}, {Cockrum}, {Coles}, {Connolly}, {Cook}, {Cooray}, {Covey}, {Cribbs}, {Cui}, {Cutri}, {Daly}, {Daniel}, {Daruich}, {Daubard}, {Daues}, {Dawson}, {Delgado}, {Dellapenna}, {de Peyster}, {de Val-Borro}, {Digel}, {Doherty}, {Dubois},
  {Dubois-Felsmann}, {Durech}, {Economou}, {Eifler}, {Eracleous}, {Emmons}, {Fausti Neto}, {Ferguson}, {Figueroa}, {Fisher-Levine}, {Focke}, {Foss}, {Frank}, {Freemon}, {Gangler}, {Gawiser}, {Geary}, {Gee}, {Geha}, {Gessner}, {Gibson}, {Gilmore}, {Glanzman}, {Glick}, {Goldina}, {Goldstein}, {Goodenow}, {Graham}, {Gressler}, {Gris}, {Guy}, {Guyonnet}, {Haller}, {Harris}, {Hascall}, {Haupt}, {Hernandez}, {Herrmann}, {Hileman}, {Hoblitt}, {Hodgson}, {Hogan}, {Howard}, {Huang}, {Huffer}, {Ingraham}, {Innes}, {Jacoby}, {Jain}, {Jammes}, {Jee}, {Jenness}, {Jernigan}, {Jevremovi{\'c}}, {Johns}, {Johnson}, {Johnson}, {Jones}, {Juramy-Gilles}, {Juri{\'c}}, {Kalirai}, {Kallivayalil}, {Kalmbach}, {Kantor}, {Karst}, {Kasliwal}, {Kelly}, {Kessler}, {Kinnison}, {Kirkby}, {Knox}, {Kotov}, {Krabbendam}, {Krughoff}, {Kub{\'a}nek}, {Kuczewski}, {Kulkarni}, {Ku}, {Kurita}, {Lage}, {Lambert}, {Lange}, {Langton}, {Le Guillou}, {Levine}, {Liang}, {Lim}, {Lintott}, {Long}, {Lopez}, {Lotz}, {Lupton}, {Lust}, {MacArthur}, {Mahabal},
  {Mandelbaum}, {Markiewicz}, {Marsh}, {Marshall}, {Marshall}, {May}, {McKercher}, {McQueen}, {Meyers}, {Migliore}, {Miller}, {Mills}, {Miraval}, {Moeyens}, {Moolekamp}, {Monet}, {Moniez}, {Monkewitz}, {Montgomery}, {Morrison}, {Mueller}, {Muller}, {Mu{\~n}oz Arancibia}, {Neill}, {Newbry}, {Nief}, {Nomerotski}, {Nordby}, {O'Connor}, {Oliver}, {Olivier}, {Olsen}, {O'Mullane}, {Ortiz}, {Osier}, {Owen}, {Pain}, {Palecek}, {Parejko}, {Parsons}, {Pease}, {Peterson}, {Peterson}, {Petravick}, {Libby Petrick}, {Petry}, {Pierfederici}, {Pietrowicz}, {Pike}, {Pinto}, {Plante}, {Plate}, {Plutchak}, {Price}, {Prouza}, {Radeka}, {Rajagopal}, {Rasmussen}, {Regnault}, {Reil}, {Reiss}, {Reuter}, {Ridgway}, {Riot}, {Ritz}, {Robinson}, {Roby}, {Roodman}, {Rosing}, {Roucelle}, {Rumore}, {Russo}, {Saha}, {Sassolas}, {Schalk}, {Schellart}, {Schindler}, {Schmidt}, {Schneider}, {Schneider}, {Schoening}, {Schumacher}, {Schwamb}, {Sebag}, {Selvy}, {Sembroski}, {Seppala}, {Serio}, {Serrano}, {Shaw}, {Shipsey}, {Sick}, {Silvestri},
  {Slater}, {Smith}, {Smith}, {Sobhani}, {Soldahl}, {Storrie-Lombardi}, {Stover}, {Strauss}, {Street}, {Stubbs}, {Sullivan}, {Sweeney}, {Swinbank}, {Szalay}, {Takacs}, {Tether}, {Thaler}, {Thayer}, {Thomas}, {Thornton}, {Thukral}, {Tice}, {Trilling}, {Turri}, {Van Berg}, {Vanden Berk}, {Vetter}, {Virieux}, {Vucina}, {Wahl}, {Walkowicz}, {Walsh}, {Walter}, {Wang}, {Wang}, {Warner}, {Wiecha}, {Willman}, {Winters}, {Wittman}, {Wolff}, {Wood-Vasey}, {Wu}, {Xin}, {Yoachim}, \& {Zhan}}]{2019ApJ...873..111I}
{Ivezi{\'c}}, {\v{Z}}., {Kahn}, S.~M., {Tyson}, J.~A., {et~al.} 2019, \apj, 873, 111, \dodoi{10.3847/1538-4357/ab042c}

\bibitem[{{Jarrett} {et~al.}(2011){Jarrett}, {Cohen}, {Masci}, {Wright}, {Stern}, {Benford}, {Blain}, {Carey}, {Cutri}, {Eisenhardt}, {Lonsdale}, {Mainzer}, {Marsh}, {Padgett}, {Petty}, {Ressler}, {Skrutskie}, {Stanford}, {Surace}, {Tsai}, {Wheelock}, \& {Yan}}]{2011ApJ...735..112J}
{Jarrett}, T.~H., {Cohen}, M., {Masci}, F., {et~al.} 2011, \apj, 735, 112, \dodoi{10.1088/0004-637X/735/2/112}

\bibitem[{{Jayasinghe} {et~al.}(2019){Jayasinghe}, {Stanek}, {Kochanek}, {Shappee}, {Holoien}, {Thompson}, {Prieto}, {Dong}, {Pawlak}, {Pejcha}, {Shields}, {Pojmanski}, {Otero}, {Hurst}, {Britt}, \& {Will}}]{2019MNRAS.485..961J}
{Jayasinghe}, T., {Stanek}, K.~Z., {Kochanek}, C.~S., {et~al.} 2019, \mnras, 485, 961, \dodoi{10.1093/mnras/stz444}

\bibitem[{{Jha} {et~al.}(2022){Jha}, {Joshi}, {Chand}, {Wu}, {Ho}, {Rastogi}, \& {Ma}}]{2022MNRAS.511.3005J}
{Jha}, V.~K., {Joshi}, R., {Chand}, H., {et~al.} 2022, \mnras, 511, 3005, \dodoi{10.1093/mnras/stac109}

\bibitem[{{Jiang} {et~al.}(2017){Jiang}, {Green}, {Greene}, {Morganson}, {Shen}, {Pancoast}, {MacLeod}, {Anderson}, {Brandt}, {Grier}, {Rix}, {Ruan}, {Protopapas}, {Scott}, {Burgett}, {Hodapp}, {Huber}, {Kaiser}, {Kudritzki}, {Magnier}, {Metcalfe}, {Tonry}, {Wainscoat}, \& {Waters}}]{2017ApJ...836..186J}
{Jiang}, Y.-F., {Green}, P.~J., {Greene}, J.~E., {et~al.} 2017, \apj, 836, 186, \dodoi{10.3847/1538-4357/aa5b91}

\bibitem[{{Kaspi} {et~al.}(2000){Kaspi}, {Smith}, {Netzer}, {Maoz}, {Jannuzi}, \& {Giveon}}]{2000ApJ...533..631K}
{Kaspi}, S., {Smith}, P.~S., {Netzer}, H., {et~al.} 2000, \apj, 533, 631, \dodoi{10.1086/308704}

\bibitem[{{Kawaguchi} \& {Mori}(2011)}]{2011ApJ...737..105K}
{Kawaguchi}, T., \& {Mori}, M. 2011, \apj, 737, 105, \dodoi{10.1088/0004-637X/737/2/105}

\bibitem[{{Kelly}(2007)}]{2007ApJ...665.1489K}
{Kelly}, B.~C. 2007, \apj, 665, 1489, \dodoi{10.1086/519947}

\bibitem[{{Kelly} {et~al.}(2009){Kelly}, {Bechtold}, \& {Siemiginowska}}]{2009ApJ...698..895K}
{Kelly}, B.~C., {Bechtold}, J., \& {Siemiginowska}, A. 2009, \apj, 698, 895, \dodoi{10.1088/0004-637X/698/1/895}

\bibitem[{{Kim} {et~al.}(2021){Kim}, {Jeong}, {Yang}, {Son}, {Ho}, {Woo}, {Im}, \& {Byun}}]{2021JKAS...54...37K}
{Kim}, M., {Jeong}, W.-S., {Yang}, Y., {et~al.} 2021, Journal of Korean Astronomical Society, 54, 37, \dodoi{10.5303/JKAS.2021.54.2.37}

\bibitem[{{Kishimoto} {et~al.}(2008){Kishimoto}, {Antonucci}, {Blaes}, {Lawrence}, {Boisson}, {Albrecht}, \& {Leipski}}]{2008Natur.454..492K}
{Kishimoto}, M., {Antonucci}, R., {Blaes}, O., {et~al.} 2008, \nat, 454, 492, \dodoi{10.1038/nature07114}

\bibitem[{{Kishimoto} {et~al.}(2011{\natexlab{a}}){Kishimoto}, {H{\"o}nig}, {Antonucci}, {Barvainis}, {Kotani}, {Tristram}, {Weigelt}, \& {Levin}}]{2011A&A...527A.121K}
{Kishimoto}, M., {H{\"o}nig}, S.~F., {Antonucci}, R., {et~al.} 2011{\natexlab{a}}, \aap, 527, A121, \dodoi{10.1051/0004-6361/201016054}

\bibitem[{{Kishimoto} {et~al.}(2009){Kishimoto}, {H{\"o}nig}, {Antonucci}, {Kotani}, {Barvainis}, {Tristram}, \& {Weigelt}}]{2009A&A...507L..57K}
---. 2009, \aap, 507, L57, \dodoi{10.1051/0004-6361/200913512}

\bibitem[{{Kishimoto} {et~al.}(2011{\natexlab{b}}){Kishimoto}, {H{\"o}nig}, {Antonucci}, {Millour}, {Tristram}, \& {Weigelt}}]{2011A&A...536A..78K}
---. 2011{\natexlab{b}}, \aap, 536, A78, \dodoi{10.1051/0004-6361/201117367}

\bibitem[{{Kishimoto} {et~al.}(2007){Kishimoto}, {H{\"o}nig}, {Beckert}, \& {Weigelt}}]{2007A&A...476..713K}
{Kishimoto}, M., {H{\"o}nig}, S.~F., {Beckert}, T., \& {Weigelt}, G. 2007, \aap, 476, 713, \dodoi{10.1051/0004-6361:20077911}

\bibitem[{{Kokubo} \& {Minezaki}(2020)}]{2020MNRAS.491.4615K}
{Kokubo}, M., \& {Minezaki}, T. 2020, \mnras, 491, 4615, \dodoi{10.1093/mnras/stz3397}

\bibitem[{{Koshida} {et~al.}(2009){Koshida}, {Yoshii}, {Kobayashi}, {Minezaki}, {Sakata}, {Sugawara}, {Enya}, {Suganuma}, {Tomita}, {Aoki}, \& {Peterson}}]{2009ApJ...700L.109K}
{Koshida}, S., {Yoshii}, Y., {Kobayashi}, Y., {et~al.} 2009, \apjl, 700, L109, \dodoi{10.1088/0004-637X/700/2/L109}

\bibitem[{{Koshida} {et~al.}(2014){Koshida}, {Minezaki}, {Yoshii}, {Kobayashi}, {Sakata}, {Sugawara}, {Enya}, {Suganuma}, {Tomita}, {Aoki}, \& {Peterson}}]{2014ApJ...788..159K}
{Koshida}, S., {Minezaki}, T., {Yoshii}, Y., {et~al.} 2014, \apj, 788, 159, \dodoi{10.1088/0004-637X/788/2/159}

\bibitem[{{Law} {et~al.}(2009){Law}, {Kulkarni}, {Dekany}, {Ofek}, {Quimby}, {Nugent}, {Surace}, {Grillmair}, {Bloom}, {Kasliwal}, {Bildsten}, {Brown}, {Cenko}, {Ciardi}, {Croner}, {Djorgovski}, {van Eyken}, {Filippenko}, {Fox}, {Gal-Yam}, {Hale}, {Hamam}, {Helou}, {Henning}, {Howell}, {Jacobsen}, {Laher}, {Mattingly}, {McKenna}, {Pickles}, {Poznanski}, {Rahmer}, {Rau}, {Rosing}, {Shara}, {Smith}, {Starr}, {Sullivan}, {Velur}, {Walters}, \& {Zolkower}}]{2009PASP..121.1395L}
{Law}, N.~M., {Kulkarni}, S.~R., {Dekany}, R.~G., {et~al.} 2009, \pasp, 121, 1395, \dodoi{10.1086/648598}

\bibitem[{{Li} \& {Shen}(2023)}]{2023arXiv230212437L}
{Li}, J., \& {Shen}, Y. 2023, arXiv e-prints, arXiv:2302.12437, \dodoi{10.48550/arXiv.2302.12437}

\bibitem[{{Li} {et~al.}(2019){Li}, {Shen}, {Brandt}, {Grier}, {Hall}, {Ho}, {Homayouni}, {Horne}, {Schneider}, {Trump}, \& {Starkey}}]{2019ApJ...884..119L}
{Li}, I-Hsiu, J., {Shen}, Y., {Brandt}, W.~N., {et~al.} 2019, \apj, 884, 119, \dodoi{10.3847/1538-4357/ab41fb}

\bibitem[{{Li} {et~al.}(2021){Li}, {Yang}, {Yang}, {Chen}, {Songsheng}, {Liu}, {Du}, {Luo}, {Yu}, {Hu}, {Jiang}, {Bao}, {Guo}, {Zhang}, {Li}, {Xiao}, {Lu}, {Ho}, {Bai}, {Bian}, {Aceituno}, {Minezaki}, {Horne}, {Kokubo}, \& {Wang}}]{2021ApJ...920....9L}
{Li}, S.-S., {Yang}, S., {Yang}, Z.-X., {et~al.} 2021, \apj, 920, 9, \dodoi{10.3847/1538-4357/ac116e}

\bibitem[{{Li} {et~al.}(2016){Li}, {Wang}, \& {Bai}}]{2016ApJ...831..206L}
{Li}, Y.-R., {Wang}, J.-M., \& {Bai}, J.-M. 2016, \apj, 831, 206, \dodoi{10.3847/0004-637X/831/2/206}

\bibitem[{{Li} {et~al.}(2014){Li}, {Wang}, {Hu}, {Du}, \& {Bai}}]{2014ApJ...786L...6L}
{Li}, Y.-R., {Wang}, J.-M., {Hu}, C., {Du}, P., \& {Bai}, J.-M. 2014, \apjl, 786, L6, \dodoi{10.1088/2041-8205/786/1/L6}

\bibitem[{{Lira} {et~al.}(2011){Lira}, {Ar{\'e}valo}, {Uttley}, {McHardy}, \& {Breedt}}]{2011MNRAS.415.1290L}
{Lira}, P., {Ar{\'e}valo}, P., {Uttley}, P., {McHardy}, I., \& {Breedt}, E. 2011, \mnras, 415, 1290, \dodoi{10.1111/j.1365-2966.2011.18774.x}

\bibitem[{{Liu} {et~al.}(2019){Liu}, {Liu}, {Dong}, {Zhou}, {Wang}, {Lu}, \& {Yuan}}]{2019ApJS..243...21L}
{Liu}, H.-Y., {Liu}, W.-J., {Dong}, X.-B., {et~al.} 2019, \apjs, 243, 21, \dodoi{10.3847/1538-4365/ab298b}

\bibitem[{Ludwig {et~al.}(2009)Ludwig, Wills, Greene, \& Robinson}]{Ludwig_2009}
Ludwig, R.~R., Wills, B., Greene, J.~E., \& Robinson, E.~L. 2009, The Astrophysical Journal, 706, 995, \dodoi{10.1088/0004-637X/706/2/995}

\bibitem[{{Lyu} {et~al.}(2019){Lyu}, {Rieke}, \& {Smith}}]{2019ApJ...886...33L}
{Lyu}, J., {Rieke}, G.~H., \& {Smith}, P.~S. 2019, \apj, 886, 33, \dodoi{10.3847/1538-4357/ab481d}

\bibitem[{{MacLeod} {et~al.}(2010){MacLeod}, {Ivezi{\'c}}, {Kochanek}, {Koz{\l}owski}, {Kelly}, {Bullock}, {Kimball}, {Sesar}, {Westman}, {Brooks}, {Gibson}, {Becker}, \& {de Vries}}]{2010ApJ...721.1014M}
{MacLeod}, C.~L., {Ivezi{\'c}}, {\v{Z}}., {Kochanek}, C.~S., {et~al.} 2010, \apj, 721, 1014, \dodoi{10.1088/0004-637X/721/2/1014}

\bibitem[{{MacLeod} {et~al.}(2012){MacLeod}, {Ivezi{\'c}}, {Sesar}, {de Vries}, {Kochanek}, {Kelly}, {Becker}, {Lupton}, {Hall}, {Richards}, {Anderson}, \& {Schneider}}]{2012ApJ...753..106M}
{MacLeod}, C.~L., {Ivezi{\'c}}, {\v{Z}}., {Sesar}, B., {et~al.} 2012, \apj, 753, 106, \dodoi{10.1088/0004-637X/753/2/106}

\bibitem[{{Mainzer} {et~al.}(2014){Mainzer}, {Bauer}, {Cutri}, {Grav}, {Masiero}, {Beck}, {Clarkson}, {Conrow}, {Dailey}, {Eisenhardt}, {Fabinsky}, {Fajardo-Acosta}, {Fowler}, {Gelino}, {Grillmair}, {Heinrichsen}, {Kendall}, {Kirkpatrick}, {Liu}, {Masci}, {McCallon}, {Nugent}, {Papin}, {Rice}, {Royer}, {Ryan}, {Sevilla}, {Sonnett}, {Stevenson}, {Thompson}, {Wheelock}, {Wiemer}, {Wittman}, {Wright}, \& {Yan}}]{2014ApJ...792...30M}
{Mainzer}, A., {Bauer}, J., {Cutri}, R.~M., {et~al.} 2014, \apj, 792, 30, \dodoi{10.1088/0004-637X/792/1/30}

\bibitem[{{Mandal} {et~al.}(2018){Mandal}, {Rakshit}, {Kurian}, {Stalin}, {Mathew}, {Hoenig}, {Gand hi}, {Sagar}, \& {Pandge}}]{2018MNRAS.475.5330M}
{Mandal}, A.~K., {Rakshit}, S., {Kurian}, K.~S., {et~al.} 2018, \mnras, 475, 5330, \dodoi{10.1093/mnras/sty200}

\bibitem[{{Mandal} {et~al.}(2021){Mandal}, {Rakshit}, {Stalin}, {Wylezalek}, {Patig}, {Sagar}, {Mathew}, {Muneer}, \& {Pal}}]{2021MNRAS.501.3905M}
{Mandal}, A.~K., {Rakshit}, S., {Stalin}, C.~S., {et~al.} 2021, \mnras, 501, 3905, \dodoi{10.1093/mnras/staa3828}

\bibitem[{{Marziani} {et~al.}(2003){Marziani}, {Zamanov}, {Sulentic}, \& {Calvani}}]{2003MNRAS.345.1133M}
{Marziani}, P., {Zamanov}, R.~K., {Sulentic}, J.~W., \& {Calvani}, M. 2003, \mnras, 345, 1133, \dodoi{10.1046/j.1365-2966.2003.07033.x}

\bibitem[{{Masci} {et~al.}(2019){Masci}, {Laher}, {Rusholme}, {Shupe}, {Groom}, {Surace}, {Jackson}, {Monkewitz}, {Beck}, {Flynn}, {Terek}, {Landry}, {Hacopians}, {Desai}, {Howell}, {Brooke}, {Imel}, {Wachter}, {Ye}, {Lin}, {Cenko}, {Cunningham}, {Rebbapragada}, {Bue}, {Miller}, {Mahabal}, {Bellm}, {Patterson}, {Juri{\'c}}, {Golkhou}, {Ofek}, {Walters}, {Graham}, {Kasliwal}, {Dekany}, {Kupfer}, {Burdge}, {Cannella}, {Barlow}, {Van Sistine}, {Giomi}, {Fremling}, {Blagorodnova}, {Levitan}, {Riddle}, {Smith}, {Helou}, {Prince}, \& {Kulkarni}}]{2019PASP..131a8003M}
{Masci}, F.~J., {Laher}, R.~R., {Rusholme}, B., {et~al.} 2019, \pasp, 131, 018003, \dodoi{10.1088/1538-3873/aae8ac}

\bibitem[{{Minezaki} {et~al.}(2019){Minezaki}, {Yoshii}, {Kobayashi}, {Sugawara}, {Sakata}, {Enya}, {Koshida}, {Tomita}, {Suganuma}, {Aoki}, \& {Peterson}}]{2019ApJ...886..150M}
{Minezaki}, T., {Yoshii}, Y., {Kobayashi}, Y., {et~al.} 2019, \apj, 886, 150, \dodoi{10.3847/1538-4357/ab4f7b}

\bibitem[{{Mor} \& {Netzer}(2012)}]{2012MNRAS.420..526M}
{Mor}, R., \& {Netzer}, H. 2012, \mnras, 420, 526, \dodoi{10.1111/j.1365-2966.2011.20060.x}

\bibitem[{{Mudd} {et~al.}(2018){Mudd}, {Martini}, {Zu}, {Kochanek}, {Peterson}, {Kessler}, {Davis}, {Hoormann}, {King}, {Lidman}, {Sommer}, {Tucker}, {Asorey}, {Hinton}, {Glazebrook}, {Kuehn}, {Lewis}, {Macaulay}, {Moeller}, {O'Neill}, {Zhang}, {Abbott}, {Abdalla}, {Allam}, {Banerji}, {Benoit-L{\'e}vy}, {Bertin}, {Brooks}, {Carnero Rosell}, {Carollo}, {Carrasco Kind}, {Carretero}, {Cunha}, {D'Andrea}, {da Costa}, {Davis}, {Desai}, {Doel}, {Fosalba}, {Garc{\'\i}a-Bellido}, {Gaztanaga}, {Gerdes}, {Gruen}, {Gruendl}, {Gschwend}, {Gutierrez}, {Hartley}, {Honscheid}, {James}, {Kuhlmann}, {Kuropatkin}, {Lima}, {Maia}, {Marshall}, {McMahon}, {Menanteau}, {Miquel}, {Plazas}, {Romer}, {Sanchez}, {Schindler}, {Schubnell}, {Smith}, {Smith}, {Soares-Santos}, {Sobreira}, {Suchyta}, {Swanson}, {Tarle}, {Thomas}, {Tucker}, {Walker}, \& {DES Collaboration}}]{2018ApJ...862..123M}
{Mudd}, D., {Martini}, P., {Zu}, Y., {et~al.} 2018, \apj, 862, 123, \dodoi{10.3847/1538-4357/aac9bb}

\bibitem[{{Nenkova} {et~al.}(2008{\natexlab{a}}){Nenkova}, {Sirocky}, {Ivezi{\'c}}, \& {Elitzur}}]{2008ApJ...685..147N}
{Nenkova}, M., {Sirocky}, M.~M., {Ivezi{\'c}}, {\v{Z}}., \& {Elitzur}, M. 2008{\natexlab{a}}, \apj, 685, 147, \dodoi{10.1086/590482}

\bibitem[{{Nenkova} {et~al.}(2008{\natexlab{b}}){Nenkova}, {Sirocky}, {Nikutta}, {Ivezi{\'c}}, \& {Elitzur}}]{2008ApJ...685..160N}
{Nenkova}, M., {Sirocky}, M.~M., {Nikutta}, R., {Ivezi{\'c}}, {\v{Z}}., \& {Elitzur}, M. 2008{\natexlab{b}}, \apj, 685, 160, \dodoi{10.1086/590483}

\bibitem[{{Oknyanskij} \& {Horne}(2001)}]{2001ASPC..224..149O}
{Oknyanskij}, V.~L., \& {Horne}, K. 2001, in Astronomical Society of the Pacific Conference Series, Vol. 224, Probing the Physics of Active Galactic Nuclei, ed. B.~M. {Peterson}, R.~W. {Pogge}, \& R.~S. {Polidan}, 149

\bibitem[{{Park} {et~al.}(2012){Park}, {Kelly}, {Woo}, \& {Treu}}]{2012ApJS..203....6P}
{Park}, D., {Kelly}, B.~C., {Woo}, J.-H., \& {Treu}, T. 2012, \apjs, 203, 6, \dodoi{10.1088/0067-0049/203/1/6}

\bibitem[{{Peterson}(1993)}]{1993PASP..105..247P}
{Peterson}, B.~M. 1993, \pasp, 105, 247, \dodoi{10.1086/133140}

\bibitem[{{Peterson} {et~al.}(1998){Peterson}, {Wanders}, {Horne}, {Collier}, {Alexander}, {Kaspi}, \& {Maoz}}]{1998PASP..110..660P}
{Peterson}, B.~M., {Wanders}, I., {Horne}, K., {et~al.} 1998, \pasp, 110, 660, \dodoi{10.1086/316177}

\bibitem[{{Peterson} {et~al.}(2004){Peterson}, {Ferrarese}, {Gilbert}, {Kaspi}, {Malkan}, {Maoz}, {Merritt}, {Netzer}, {Onken}, {Pogge}, {Vestergaard}, \& {Wandel}}]{2004ApJ...613..682P}
{Peterson}, B.~M., {Ferrarese}, L., {Gilbert}, K.~M., {et~al.} 2004, \apj, 613, 682, \dodoi{10.1086/423269}

\bibitem[{{Rakshit} {et~al.}(2020){Rakshit}, {Stalin}, \& {Kotilainen}}]{2020ApJS..249...17R}
{Rakshit}, S., {Stalin}, C.~S., \& {Kotilainen}, J. 2020, \apjs, 249, 17, \dodoi{10.3847/1538-4365/ab99c5}

\bibitem[{{Rees}(1984)}]{1984ARA&A..22..471R}
{Rees}, M.~J. 1984, \araa, 22, 471, \dodoi{10.1146/annurev.aa.22.090184.002351}

\bibitem[{{Richards} {et~al.}(2006){Richards}, {Lacy}, {Storrie-Lombardi}, {Hall}, {Gallagher}, {Hines}, {Fan}, {Papovich}, {Vanden Berk}, {Trammell}, {Schneider}, {Vestergaard}, {York}, {Jester}, {Anderson}, {Budav{\'a}ri}, \& {Szalay}}]{2006ApJS..166..470R}
{Richards}, G.~T., {Lacy}, M., {Storrie-Lombardi}, L.~J., {et~al.} 2006, \apjs, 166, 470, \dodoi{10.1086/506525}

\bibitem[{{Salpeter}(1964)}]{1964ApJ...140..796S}
{Salpeter}, E.~E. 1964, \apj, 140, 796, \dodoi{10.1086/147973}

\bibitem[{{Sanders} {et~al.}(1989){Sanders}, {Phinney}, {Neugebauer}, {Soifer}, \& {Matthews}}]{1989ApJ...347...29S}
{Sanders}, D.~B., {Phinney}, E.~S., {Neugebauer}, G., {Soifer}, B.~T., \& {Matthews}, K. 1989, \apj, 347, 29, \dodoi{10.1086/168094}

\bibitem[{{Seyfert}(1943)}]{1943ApJ....97...28S}
{Seyfert}, C.~K. 1943, \apj, 97, 28, \dodoi{10.1086/144488}

\bibitem[{{Shakura} \& {Sunyaev}(1973)}]{1973A&A....24..337S}
{Shakura}, N.~I., \& {Sunyaev}, R.~A. 1973, \aap, 500, 33

\bibitem[{{Shappee} {et~al.}(2014){Shappee}, {Prieto}, {Stanek}, {Kochanek}, {Holoien}, {Jencson}, {Basu}, {Beacom}, {Szczygiel}, {Pojmanski}, {Brimacombe}, {Dubberley}, {Elphick}, {Foale}, {Hawkins}, {Mullins}, {Rosing}, {Ross}, \& {Walker}}]{2014AAS...22323603S}
{Shappee}, B., {Prieto}, J., {Stanek}, K.~Z., {et~al.} 2014, in American Astronomical Society Meeting Abstracts, Vol. 223, American Astronomical Society Meeting Abstracts \#223, 236.03

\bibitem[{{Shen} {et~al.}(2011){Shen}, {Richards}, {Strauss}, {Hall}, {Schneider}, {Snedden}, {Bizyaev}, {Brewington}, {Malanushenko}, {Malanushenko}, {Oravetz}, {Pan}, \& {Simmons}}]{2011ApJS..194...45S}
{Shen}, Y., {Richards}, G.~T., {Strauss}, M.~A., {et~al.} 2011, \apjs, 194, 45, \dodoi{10.1088/0067-0049/194/2/45}

\bibitem[{{Sitko} {et~al.}(1993){Sitko}, {Sitko}, {Siemiginowska}, \& {Szczerba}}]{1993ApJ...409..139S}
{Sitko}, M.~L., {Sitko}, A.~K., {Siemiginowska}, A., \& {Szczerba}, R. 1993, \apj, 409, 139, \dodoi{10.1086/172649}

\bibitem[{{Sobrino Figaredo} {et~al.}(2020){Sobrino Figaredo}, {Haas}, {Ramolla}, {Chini}, {Blex}, {Hodapp}, {Murphy}, {Kollatschny}, {Chelouche}, \& {Kaspi}}]{2020AJ....159..259S}
{Sobrino Figaredo}, C., {Haas}, M., {Ramolla}, M., {et~al.} 2020, \aj, 159, 259, \dodoi{10.3847/1538-3881/ab89b1}

\bibitem[{{Suganuma} {et~al.}(2006){Suganuma}, {Yoshii}, {Kobayashi}, {Minezaki}, {Enya}, {Tomita}, {Aoki}, {Koshida}, \& {Peterson}}]{2006ApJ...639...46S}
{Suganuma}, M., {Yoshii}, Y., {Kobayashi}, Y., {et~al.} 2006, \apj, 639, 46, \dodoi{10.1086/499326}

\bibitem[{{Sulentic} {et~al.}(2000){Sulentic}, {Zwitter}, {Marziani}, \& {Dultzin-Hacyan}}]{2000ApJ...536L...5S}
{Sulentic}, J.~W., {Zwitter}, T., {Marziani}, P., \& {Dultzin-Hacyan}, D. 2000, \apjl, 536, L5, \dodoi{10.1086/312717}

\bibitem[{Sun \& Shen(2015)}]{Sun_2015}
Sun, J., \& Shen, Y. 2015, The Astrophysical Journal Letters, 804, L15, \dodoi{10.1088/2041-8205/804/1/L15}

\bibitem[{{Tomita} {et~al.}(2006){Tomita}, {Yoshii}, {Kobayashi}, {Minezaki}, {Enya}, {Suganuma}, {Aoki}, {Koshida}, \& {Yamauchi}}]{2006ApJ...652L..13T}
{Tomita}, H., {Yoshii}, Y., {Kobayashi}, Y., {et~al.} 2006, \apjl, 652, L13, \dodoi{10.1086/509878}

\bibitem[{{U} {et~al.}(2022){U}, {Barth}, {Vogler}, {Guo}, {Treu}, {Bennert}, {Canalizo}, {Filippenko}, {Gates}, {Hamann}, {Joner}, {Malkan}, {Pancoast}, {Williams}, {Woo}, {Abolfathi}, {Abramson}, {Armen}, {Bae}, {Bohn}, {Boizelle}, {Bostroem}, {Brandel}, {Brink}, {Channa}, {Cooper}, {Cosens}, {Donohue}, {Fillingham}, {Gonz{\'a}lez-Buitrago}, {Halevi}, {Halle}, {Hood}, {Horne}, {Horst}, {de Kouchkovsky}, {Kuhn}, {Kumar}, {Leonard}, {Loveland}, {Manzano-King}, {McHardy}, {Michel}, {Olaes}, {Park}, {Park}, {Pei}, {Ross}, {Runco}, {Samuel}, {S{\'a}nchez}, {Scott}, {Sexton}, {Shin}, {Shivvers}, {Spencer}, {Stahl}, {Stegman}, {Stomberg}, {Valenti}, {Villafa{\~n}a}, {Walsh}, {Yuk}, \& {Zheng}}]{2022ApJ...925...52U}
{U}, V., {Barth}, A.~J., {Vogler}, H.~A., {et~al.} 2022, \apj, 925, 52, \dodoi{10.3847/1538-4357/ac3d26}

\bibitem[{{Urry} \& {Padovani}(1995)}]{1995PASP..107..803U}
{Urry}, C.~M., \& {Padovani}, P. 1995, \pasp, 107, 803, \dodoi{10.1086/133630}

\bibitem[{{Vazquez} {et~al.}(2015){Vazquez}, {Galianni}, {Richmond}, {Robinson}, {Axon}, {Horne}, {Almeyda}, {Fausnaugh}, {Peterson}, {Bottorff}, {Gallimore}, {Eltizur}, {Netzer}, {Storchi-Bergmann}, {Marconi}, {Capetti}, {Batcheldor}, {Buchanan}, {Stirpe}, {Kishimoto}, {Packham}, {Perez}, {Tadhunter}, {Upton}, \& {Estrada-Carpenter}}]{2015ApJ...801..127V}
{Vazquez}, B., {Galianni}, P., {Richmond}, M., {et~al.} 2015, \apj, 801, 127, \dodoi{10.1088/0004-637X/801/2/127}

\bibitem[{{Wandel} {et~al.}(1999){Wandel}, {Peterson}, \& {Malkan}}]{1999ApJ...526..579W}
{Wandel}, A., {Peterson}, B.~M., \& {Malkan}, M.~A. 1999, \apj, 526, 579, \dodoi{10.1086/308017}

\bibitem[{{Wang} {et~al.}(2014{\natexlab{a}}){Wang}, {Qiu}, {Du}, \& {Ho}}]{2014ApJ...797...65W}
{Wang}, J.-M., {Qiu}, J., {Du}, P., \& {Ho}, L.~C. 2014{\natexlab{a}}, \apj, 797, 65, \dodoi{10.1088/0004-637X/797/1/65}

\bibitem[{{Wang} {et~al.}(2014{\natexlab{b}}){Wang}, {Du}, {Hu}, {Netzer}, {Bai}, {Lu}, {Kaspi}, {Qiu}, {Li}, {Wang}, \& {SEAMBH Collaboration}}]{2014ApJ...793..108W}
{Wang}, J.-M., {Du}, P., {Hu}, C., {et~al.} 2014{\natexlab{b}}, \apj, 793, 108, \dodoi{10.1088/0004-637X/793/2/108}

\bibitem[{{Woo} \& {Urry}(2002)}]{2002ApJ...579..530W}
{Woo}, J.-H., \& {Urry}, C.~M. 2002, \apj, 579, 530, \dodoi{10.1086/342878}

\bibitem[{{Woo} {et~al.}(2015){Woo}, {Yoon}, {Park}, {Park}, \& {Kim}}]{2015ApJ...801...38W}
{Woo}, J.-H., {Yoon}, Y., {Park}, S., {Park}, D., \& {Kim}, S.~C. 2015, \apj, 801, 38, \dodoi{10.1088/0004-637X/801/1/38}

\bibitem[{{Woo} {et~al.}(2024){Woo}, {Wang}, {Rakshit}, {Cho}, {Son}, {Bennert}, {Gallo}, {Hodges-Kluck}, {Treu}, {Barth}, {Cho}, {Foord}, {Geum}, {Guo}, {Jadhav}, {Jeon}, {Kabasares}, {Kang}, {Kim}, {Kim}, {Kim}, {Le}, {Malkan}, {Mandal}, {Park}, {Spencer}, {Shin}, {Sung}, {U}, {Williams}, \& {Yee}}]{2024ApJ...962...67W}
{Woo}, J.-H., {Wang}, S., {Rakshit}, S., {et~al.} 2024, \apj, 962, 67, \dodoi{10.3847/1538-4357/ad132f}

\bibitem[{{Wright} {et~al.}(2010){Wright}, {Eisenhardt}, {Mainzer}, {Ressler}, {Cutri}, {Jarrett}, {Kirkpatrick}, {Padgett}, {McMillan}, {Skrutskie}, {Stanford}, {Cohen}, {Walker}, {Mather}, {Leisawitz}, {Gautier}, {McLean}, {Benford}, {Lonsdale}, {Blain}, {Mendez}, {Irace}, {Duval}, {Liu}, {Royer}, {Heinrichsen}, {Howard}, {Shannon}, {Kendall}, {Walsh}, {Larsen}, {Cardon}, {Schick}, {Schwalm}, {Abid}, {Fabinsky}, {Naes}, \& {Tsai}}]{2010AJ....140.1868W}
{Wright}, E.~L., {Eisenhardt}, P. R.~M., {Mainzer}, A.~K., {et~al.} 2010, \aj, 140, 1868, \dodoi{10.1088/0004-6256/140/6/1868}

\bibitem[{{Yang} {et~al.}(2020){Yang}, {Shen}, {Liu}, {Aguena}, {Annis}, {Avila}, {Banerji}, {Bertin}, {Brooks}, {Burke}, {Carnero Rosell}, {Carrasco Kind}, {da Costa}, {De Vicente}, {Desai}, {Diehl}, {Doel}, {Flaugher}, {Fosalba}, {Frieman}, {Garcia-Bellido}, {Gerdes}, {Gruen}, {Gruendl}, {Gschwend}, {Gutierrez}, {Hinton}, {Hollowood}, {Honscheid}, {Kuropatkin}, {Maia}, {March}, {Marshall}, {Martini}, {Melchior}, {Menanteau}, {Miquel}, {Paz-Chinchon}, {Malag{\'o}n}, {Romer}, {Sanchez}, {Scarpine}, {Schubnell}, {Serrano}, {Sevilla}, {Smith}, {Suchyta}, {Tarle}, {Varga}, \& {Wilkinson}}]{2020ApJ...900...58Y}
{Yang}, Q., {Shen}, Y., {Liu}, X., {et~al.} 2020, \apj, 900, 58, \dodoi{10.3847/1538-4357/aba59b}

\bibitem[{{Yoshii} {et~al.}(2014){Yoshii}, {Kobayashi}, {Minezaki}, {Koshida}, \& {Peterson}}]{2014ApJ...784L..11Y}
{Yoshii}, Y., {Kobayashi}, Y., {Minezaki}, T., {Koshida}, S., \& {Peterson}, B.~A. 2014, \apjl, 784, L11, \dodoi{10.1088/2041-8205/784/1/L11}

\bibitem[{{Zu} {et~al.}(2011){Zu}, {Kochanek}, \& {Peterson}}]{2011ApJ...735...80Z}
{Zu}, Y., {Kochanek}, C.~S., \& {Peterson}, B.~M. 2011, \apj, 735, 80, \dodoi{10.1088/0004-637X/735/2/80}

\end{thebibliography}
%\bibliography{sample631}{}
\bibliographystyle{aasjournal}



\end{document}